\documentclass[bj]{imsart}

\RequirePackage{amsthm,amsmath,amsfonts,amssymb,mathrsfs,comment}
\RequirePackage[numbers]{natbib}

\RequirePackage[colorlinks,citecolor=blue,urlcolor=blue]{hyperref}
\RequirePackage{graphicx,enumerate,bm,xcolor,subfigure}
\usepackage{enumitem} 
\newlist{condenum}{enumerate}{1} 
\setlist[condenum]{label=\bfseries Condition \arabic*.,ref=\arabic*, wide}
\startlocaldefs
\theoremstyle{plain}

\newtheorem{theorem}{Theorem}[section]
\newtheorem{lem}[theorem]{Lemma}
\newtheorem{cor}[theorem]{Corollary}
\newtheorem{proposition}[theorem]{Proposition}
\theoremstyle{remark}   
\newtheorem{definition}[theorem]{Definition}
\newtheorem{remark}[theorem]{Remark}

\endlocaldefs

\begin{document}

\begin{frontmatter}
\title{Hypothesis testing for equality of latent positions in random graphs}
\runtitle{Hypothesis testing for equality of latent positions in random graphs}

\begin{aug}
\author[A]{\fnms{Xinjie} \snm{Du}\ead[label=e1,mark]{xdu22@ncsu.edu}}
\and
\author[A]{\fnms{Minh} \snm{Tang}\ead[label=e2,mark]{mtang8@ncsu.edu}}
\address[A]{Department of Statistics, North Carolina State University, Raleigh, North Carolina, USA. \printead{e1,e2}}
\end{aug}

\begin{abstract}
We consider the hypothesis testing problem that two
vertices $i$ and $j$ of a generalized random dot product graph have the same
latent positions, possibly up to scaling. Special cases of this
hypothesis test include testing whether two vertices in a
stochastic block model or degree-corrected stochastic block model
graph have the same block membership vectors, or testing whether two vertices in a popularity adjusted block model have the same community assignment. We propose several test
statistics based on the empirical Mahalanobis distances between
the $i$th and $j$th rows of either the adjacency or
the normalized Laplacian spectral embedding of the graph. We show that, under mild conditions,
these test statistics
have limiting chi-square distributions under both the null
and local alternative hypothesis, and we derived explicit expressions
for the non-centrality parameters under the local alternative. Using these limit results, we address the model selection problems including
choosing between the standard stochastic block model and its
degree-corrected variant, and choosing between the Erdős–Rényi model and stochastic block model. The effectiveness of our proposed tests are illustrated via both simulation studies and real data applications.
\end{abstract}


\end{frontmatter}

\section{Introduction}
\label{sec:intro}
A large number of real-world data across many fields of study such as
social science, computer science, and biology, can be modeled as
a network or a graph wherein the vertices or nodes represent objects of
interest and the edges represent the pairwise relationship between
different objects, e.g., in a social
network the nodes correspond to people and an edge between
two nodes indicates friendship. The prevalence of network data had, in turn, lead to the development of numerous statistical models for network data with perhaps the simplest and most
widely studied being the stochastic block model (SBM) of \citep{holland1983stochastic}. This model
provides a framework for generating graphs with underlying community
structures. More specifically, in a SBM graph each vertex is assigned to one out of $K$ possible communities and the
probability of an edge between two vertices depends only on their
community memberships. The assumption that a node belongs to
a single community or that the probabilities of connection depend
only on the community assignments is, however, too
restrictive for many applications;
several variants of stochastic block models were thus proposed to address
these limitations, with perhaps the two most well known being the
mixed membership SBM and the degree-corrected SBM. The mixed membership
stochastic block model (MMSBM) \citep{airoldi2008mixed} assumes that
each vertex $v_i$ is assigned to multiple communities simultaneously
and with {\em vertex specific} community membership vector $\pi_i$ while the
degree-corrected SBM \citep{karrer2011stochastic} 
incorporates a degree parameter $\theta_i$ for each vertex $v_i$, thereby allowing for
heterogeneous degrees for $v_i$ and $v_j$ even when they belong
to the same community.\par
The stochastic block model and its variants, as described above, are
themselves special cases of the 
latent position model \citep{hoff2002latent} wherein each
vertex $i$ is mapped to a latent position $X_i$, and that, conditional on the
collection of latent positions $\{X_i\}$, the edges are independent
Bernoulli random variables and the probability of a link
between any two nodes $i$ and $j$ is $f(X_i, X_j)$ for
some given link function or kernel $f$. Other special cases of latent
position models including the notion of a random dot product
graph (RDPG) \citep{nickel2008random} where
$f(x,y)=x^{\top}y$ or its generalised version (GRDPG) \citep{rubin2017statistical}
where $f(x,y)=x^{\top}\mathbf{I}_{a,b}y$; here $\mathbf{I}_{a,b}$ for
integers $a \geq 1$ and $b \geq 0$ is a diagonal matrix with $a$ ``$1$''
followed by $b$ ``$-1$''. The GRDPG is a
simple yet far reaching extension of RDPG and allows for modelling
disassortative connectivity behaviour, e.g., ``opposites
attract''; furthermore, the stochastic block model and its variants are
all special cases of the GRDPG (see Remark~\ref{rem:SBM_MMSBM} of this
paper) and any latent position model can be approximated
by a GRDPG for possibly large values of $a$ and $b$ \citep{graph_root}.

Hypothesis testing for graphs is a nascent research area, with a
significant portion of the existing literature being focused on either
goodness of fit test for graphs, e.g., whether or not a graph is an
instance of a stochastic block model graph with $K$ blocks, or
two-sample hypothesis testing, e.g., whether or not two collection of
graphs came from the same distribution.
For these hypothesis tests, the atomic object of interests are the
graphs themselves; see
\citep{bickel2016hypothesis,ghoshdastidar2017two,ginestet2017hypothesis,lei2016goodness,tang2017semiparametric}
for a few examples of these type of formulations.

In this paper we also consider hypothesis testing for graphs but we
take a different perspective wherein the atomic object of interests
are the individual vertices. More specifically, we consider the problem of determining whether or not
two nodes $i$ and $j$ in a generalized random dot product graph $\mathbf{A}$ have the same
latent positions, i.e., we test the null hypothesis $\mathbb{H}_0
\colon X_i=X_j$ against the
alternative hypothesis $\mathbb{H}_A \colon X_i \neq X_j$. This hypothesis
test includes, as a special case, the test that two nodes $i$ and $j$
in a mixed membership SBM have the same community membership vectors,
possibly up to scaling by some unknown degree heterogeneity parameters
$\theta_i$ and $\theta_j$.
These hypotheses arise naturally in many applications, including vertex nomination
\citep{fishkind2015vertex} and roles discovery \citep{eliassi}; in both of these
applications we are given a graph $G$ together with a notion of
``interesting'' vertices and our task
is to find vertices in $G$ that are most ``interesting''. Examples include
finding outliers in a stochastic block model with adversarial
outliers nodes \citep{agterberg2019vertex,cai_outliers} or finding
vertices that are most ``similar'' to a given subset of
vertices. 

Our test statistics are based on an estimate of the Mahalanobis
distance between the $i$th and $j$th row of the spectral embeddings for either the observed
adjacency matrix or the normalized Laplacian matrix for $\mathbf{A}$.
It is widely known that
 spectral embedding methods provide consistent estimates
for the latent position, see
\citep{lei2015consistency,rohe2011spectral,sussman12}, among
others. In particular, \citep{athreya2016limit,rubin2017statistical,tang2018limit} showed that
the spectral embeddings of either the adjacency or the normalized
Laplacian matrices yield estimates of the latent
positions that are both uniformly consistent and 
asymptotically normal. Leveraging these results, we derive the
limiting distributions for our proposed test statistics under both the
null hypothesis and under a {\em local} alternative hypothesis, i.e., 
they converge in
distribution to chi-square random variables with $d$ degrees of
freedom and non-centrality parameter $\mu$; here $d$ is the dimension of the latent
positions and $\mu$ is a Mahalanobis distance
between $X_i$ and $X_j$. In the degree-corrected case, in order
to eliminate the degree heterogeneity, we normalize the embedded
vectors by their norm before computing the test
statistic. For this setting the limiting distributions of our
test statistics under the null and local alternative hypothesis are both chi-square with
$d-1$ degrees of freedom and non-centrality parameter $\tilde{\mu}$
given by a Mahalanobis distance between the normalized $X_i$ and
$X_j$. The above limit results allow us to develop model selection
procedures for choosing between a stochastic block model and its 
degree-corrected variant and choosing between the Erd\H{o}s–Rényi and stochastic block models.

Our work is most similar to that in \citep{fan2019simple} wherein the authors
consider the problem of hypothesis testing for equality, up to
possible scaling due to degree-heterogeneity, of membership profiles
in large networks. They also propose a test statistic based on a Mahalanobis distance
between the $i$th and $j$th rows of $\hat{\mathbf{U}}$, the matrix
whose columns are eigenvectors corresponding to the $d$ largest eigenvalues of the observed graph.
We will show in Section~\ref{sec:related_work} that for the test
statistics constructed using the spectral embedding of the adjacency
matrix, our test statistics are closely related to that of
\citep{fan2019simple} and furthermore our results are
generalizations of
the corresponding results in \citep{fan2019simple}. In particular we relax three main assumptions made in
\citep{fan2019simple}, namely (1) we do not assume distinct
eigenvalues in the edge probabilities matrix (2) we do not assume that
the block probabilities matrix $\mathbf{B}$ is of full-rank and (3)
in the setting of the degree-corrected SBMs, we do not assume that
$\mathbf{B}$ is positive definite. Finally, for the hypothesis
test of equality up to scaling, we also obtain several related expressions for the
non-centrality parameter of the test statistic under the local
alternative; elucidating the subtle relationships between these
expressions, as is done in the current paper, is non-trivial.

The rest of this article is organized as follows. Section~\ref{sec2}
introduces the model setting and technical preparation. In
Section~\ref{sec:ase_test1} and Section~\ref{sec:degree_corrected} we present the proposed test statistics using adjacency spectral embedding and
establish their limiting distributions for the undirected graphs setting; extension of these results to the directed graphs setting is presented in Appendix~\ref{sec:directed}. Relationships with previous work and  proposed model selection procedure are presented in Section~\ref{sec:related_work} and Section~\ref{sec:model_selection}, respectively. Section~\ref{sec:LSE} introduces the test statistics using Laplacian spectral embedding. In Section~\ref{sec4},
simulations are conducted to illustrate the empirical performance of
the test statistic. Some applications of the proposed test on real
data are given in Section~\ref{sec5}. Conclusion and discussions are
presented in Section~\ref{sec6}. Proofs of the stated results and additional numerical experiments are provided
in the Appendix.
\section{Model Setting}
\label{sec2}
\subsection{The generalised random dot product graph}
\label{sec2.1}
We start by recalling the notion of a generalised random dot product
graph as introduced in \citep{rubin2017statistical}.
\begin{definition}
\label{def:GRDPG}
Let $d \geq 1$ be a positive integer and let
$\mathbf{I}_{a,b}=\operatorname{diag}(1, \ldots, 1,-1, \ldots,-1)$ be
a diagonal matrix with $a$ ``$1$'' followed by $b$ ``$-1$''; here $a \geq 1$ and $b \geq 0$ are two integers satisfying
$a+b=d$. Next
let $\mathcal{X}$ be a subset of $\mathbb{R}^{d}$ such that $x^{\top}
\mathbf{I}_{a, b} y \in[0,1]$ for all $x,y \in \mathcal{X}$. Let
$\bigl[X_{1} \mid \dots \mid X_{n} \bigr]^{\top} $ be a $n \times d$ matrix with 
rows $X_i \in \mathcal{X}$. Suppose $\mathbf{A}$ is a random, symmetric
matrix with entries in $\{0,1\}$ such that, conditional on $\mathbf{X}$, the entries $\mathbf{A}_{ij}$ for $i
< j$ are {\em independent} Bernoulli random variables with success
probabilities $\rho_n X_i^{\top} \mathbf{I}_{a,b} X_j$, i.e.,
$$\mathbb{P}[\mathbf{A} \mid \mathbf{X}]=\prod_{i<j}
\left(\rho_nX_{i}^{\top} \mathbf{I}_{a,b}X_{j}\right)^{\mathbf{A}_{i
j}}\left(1-\rho_nX_{i}^{\top}\mathbf{I}_{a,b}
X_{j}\right)^{1-\mathbf{A}_{i j}}.$$ We then say that $\mathbf{A} \sim
\operatorname{GRDPG}_{a,b}(\mathbf{X},\rho_n)$ is the adjacency matrix
of a generalised random dot product graph with latent positions
$\mathbf{X}$, signature $(a, b)$ and sparsity factor $\rho_n  \in
[0,1]$.
Here $\mathbf{A}_{i j}=1$ if there is
an edge between the $i$th and $j$th node and $\mathbf{A}_{i j}=0$
otherwise.
\end{definition}

Note that the graphs generated by our model are loops-free. The parameter $\rho_n$ is assumed to be either a
constant $\rho_{n}=1$ or, if not, that $\rho_{n} \rightarrow 0$; 
the case where $\rho_n \rightarrow c$ for some constant $c > 0$ can be
transformed to the case of $\rho_n \rightarrow 1$ by scaling the
domain of $\mathcal{X}$ accordingly. Since the
average degree of the graph grows as $n\rho_n$, the cases of
$\rho_{n}=1$ and $\rho_{n} \rightarrow 0$ correspond to the dense
and semi-sparse regime, respectively. Semi-sparse here means the sparsity factor $\rho_n$ satisfies $n \rho_n = \omega(\log n)$ which will be specified later in Condition 3 in Section~\ref{sec:ase}. We denote by
$\mathbf{P}=\rho_n\mathbf{X}\mathbf{I}_{a,b}\mathbf{X}^{T}$ the 
matrix of edge probabilities, i.e., $p_{ij}$ is the the probability of having an edge
between two vertices $i$ and $j$, and the (undirected) edges are assumed to be
mutually independent. Finally we note that Definition~\ref{def:GRDPG} is for undirected graphs. The setting for directed graphs is discussed later in Appendix~\ref{sec:directed}.

\begin{remark}
  \label{rem:SBM_MMSBM}
It is easy to verify that the standard stochastic block model and
mixed membership model are special cases of GRDPG
\citep{rubin2017statistical}. Indeed, a 
$K$-blocks mixed membership stochastic block model graph on $n$
vertices with blocks probabilities matrix $\mathbf{B}$ and 
sparsity factor $\rho_n$ is generated as follows. Let $\bm{\pi}_1,
\dots, \bm{\pi}_n$ be stochastic vectors in $\mathbb{R}^{K}$, i.e.,
$\bm{\pi}_i = (\pi_{i1}, \dots, \pi_{iK})^{\top}$ is a non-negative vector
with $\sum_{k=1}^{K} \pi_{ik} = 1$ for all $i$. 
Then given $\bm{\pi}_1, \dots, \bm{\pi}_n$, the edges between the
vertices are {\em independent} Bernoulli random variables with success
probabilities
$$\mathbf{P}_{ij} = \rho_n \bm{\pi}_i^{\top} \mathbf{B} \bm{\pi}_j
= \rho_n \sum_{k=1}^{K} \sum_{\ell=1}^{K} \pi_{ik} \pi_{j \ell} \mathbf{B}_{kl}$$
Note that if the $\bm{\pi}_i$'s are all elementary vectors
i.e., for each $i$, $\bm{\pi}_{i}$ contains a single entry equal to $1$,
then the mixed membership model reduces to that of the standard
stochastic block model. Writing $\bm{\Pi}=\left(\boldsymbol{\pi}_{1},
  \cdots, \boldsymbol{\pi}_{n}\right)^{\top}$ as the $n \times K$ matrix
whose rows are the $\bm{\pi}_i$, we have $\mathbf{P} = \rho_n
\bm{\Pi} \mathbf{B} \bm{\Pi}^{\top}$.
Now choose $\nu_{1}, \ldots, \nu_{K} \in \mathbb{R}^{d}$
for some $d = \operatorname{rank}(\mathbf{B}) \leq K$ such that $\nu_{k} ^{\top} \mathbf{I}_{a,b}
\nu_{l}=\mathbf{B}_{k l},$ for all $k, l \in\{1,\ldots, K\}$; 
here $a$ is the number of positive eigenvalues of $\mathbf{B}$ and
$b=d-a$. Then the collection $\{X_i=\sum_{k=1}^{K} \pi_{i k}
\nu_{k}, \colon i=1, \ldots, n\}$ defines the latent positions
for the GRDPG corresponding to the above mixed membership
SBM. Another special case of GRDPG is the Popularity Adjusted Block Model (PABM) proposed by \citep{sengupta2018block}. See Section~\ref{leeds_butterfly} for hypothesis testing of community memberships in PABM. 
We emphasize here that any {\em undirected}
independent edge random graphs on $n$ vertices where the $n \times n$
edge probabilities matrix $\mathbf{P}$ is of low-rank, that is
$d = \operatorname{rank}(\mathbf{P}) \ll n$, can be represented as a
GRDPG. Indeed, since $\mathbf{P}$ is symmetric it has eigendecomposition $\mathbf{P} =
\mathbf{U} \mathbf{S} \mathbf{U}^{\top}$ where $\mathbf{U}$ is a $n
\times d$ matrix of orthonormal eigenvectors and $\mathbf{S}$ is the
$d \times d$ diagonal matrix for the corresponding non-zero
eigenvalues of $\mathbf{P}$. We can then define the
latent positions as the rows of the $n \times d$ matrix $\mathbf{X} =
\mathbf{U} |\mathbf{S}|^{1/2}$ where the $|\cdot|$ operation is
applied elementwise.
\end{remark}

\subsection{Non-identifiability in generalized random dot product graphs}
  Non-identifiability is an intrinsic property of generalized random
  dot product graphs. In particular, if $\mathbf{Q}$ is a $(a+b)
  \times (a+b)$ matrix such that $\mathbf{Q} \mathbf{I}_{a,b}
  \mathbf{Q}^{\top} = \mathbf{I}_{a,b}$ then $\mathbf{X}$ and
  $\mathbf{X} \mathbf{Q}$ induce the same random graph model, i.e., 
   $\mathbf{A}_1 \sim \operatorname{GRDPG}_{a,b}(\mathbf{X},\rho_n)$
   and $\mathbf{A}_2 \sim \operatorname{GRDPG}_{a,b}(\mathbf{X}
   \mathbf{Q}, \rho_n)$ are identically distributed. The following result
   provides a converse statement to this observation. 
   \begin{proposition}
     \label{prop:non_identifiable1}
     Let $\mathbf{X}$ and $\mathbf{Y}$ be $n \times d$ matrices of
     full-column rank. Then $\mathbf{X}$ and $\mathbf{Y}$ induces the
     same GRDPG model if $\mathbf{X} \mathbf{I}_{a,b}
     \mathbf{X}^{\top} = \mathbf{Y} \mathbf{I}_{a,b}
     \mathbf{Y}^{\top}$. The condition $\mathbf{X} \mathbf{I}_{a,b}
     \mathbf{X}^{\top} = \mathbf{Y} \mathbf{I}_{a,b} \mathbf{Y}^{\top}$ is satisfied if
     and only if there exists an invertible matrix $\mathbf{Q}$ such
     that $\mathbf{X} = \mathbf{Y} \mathbf{Q}$ and
$\mathbf{Q} \mathbf{I}_{a,b} \mathbf{Q}^{\top} =
     \mathbf{I}_{a,b}$
   \end{proposition}
Any matrix $\mathbf{Q}$ satisfying $\mathbf{Q} \mathbf{I}_{a,b} \mathbf{Q}^{\top} =
     \mathbf{I}_{a,b}$ is
said to be an indefinite orthogonal matrix with signature
$(a,b)$. If $b = 0$ then $\mathbf{Q}$ is also an orthogonal matrix.  
See Chapter 7 of \citep{serre}
for further discussion of indefinite orthogonal matrices.

   \begin{remark}
     \label{rem:indefinite_properties}
   We now note several simple but useful facts regarding indefinite orthogonal
   matrices. Firstly, if $\mathbf{Q}$ is indefinite orthogonal with
   respect to $\mathbf{I}_{a,b}$
   then $\mathbf{Q}^{-1} = \mathbf{I}_{a,b} \mathbf{Q}^{\top}
   \mathbf{I}_{a,b}$. Secondly, $\mathbf{Q}$ is indefinite orthogonal if
   and only if $\mathbf{Q}^{\top}$ is indefinite orthogonal.
   Finally, an orthogonal matrix
   $\mathbf{Q}$ is also an indefinite orthogonal matrix with respect
   to $\mathbf{I}_{a,b}$ if and only if
   $\mathbf{Q}$ is $(a,b)$ block-diagonal, i.e., $\mathbf{Q} =
   \Bigl[\begin{smallmatrix} \mathbf{Q}_1 & 0 \\ 0 &
     \mathbf{Q}_{2} \end{smallmatrix} \Bigr]$ where $\mathbf{Q}_1$ and
   $\mathbf{Q}_2$ are $a \times a$ and $b \times b$ orthogonal
   matrices, respectively.
\end{remark}
   
   The following result shows that, given any edge probabilities
   matrix $\mathbf{P}$ of a generalized random dot product graph, the eigendecomposition $\mathbf{P} = \mathbf{U}
   \mathbf{S} \mathbf{U}^{\top}$ provides a latent positions
   representation of $\mathbf{P}$ that has minimum spectral norm and
   minimum Frobenius norm among all possible latent positions representations of $\mathbf{P}$. 
   \begin{proposition}
     \label{prop:non_identifiable2}
     Let $\mathbf{P}$ be a $n \times n$ symmetric matrix of rank
     $d$. Suppose $\mathbf{P}$ has $a$
     positive and $b$ negative eigenvalues, with $a + b = d$. Let $\mathbf{U} \mathbf{S} \mathbf{U}^{\top}$ be the
     eigendecomposition of $\mathbf{P}$ where $\mathbf{S}$ is a $d
     \times d$ diagonal matrix containing the non-zero eigenvalues and 
     $\mathbf{U}$ is the $n \times d$ matrix whose columns are the
     corresponding orthonormal eigenvectors. Then $\mathbf{Z} =
     \mathbf{U} |\mathbf{S}|^{1/2}$, where the $|\cdot|$ operation is
applied elementwise, is a latent positions
     representation for $\mathbf{P}$, i.e., $\mathbf{P} = \mathbf{Z}
     \mathbf{I}_{a,b} \mathbf{Z}^{\top}$. Furthermore, for any matrix
     $\mathbf{X}$ such that $\mathbf{P} = \mathbf{X} \mathbf{I}_{a,b}
     \mathbf{X}^{\top}$, we have $\|\mathbf{Z}\|_{F} \leq
     \|\mathbf{X}\|_{F}$ and $\|\mathbf{Z}\| \leq \|\mathbf{X}\|$. 
     Here $\|\cdot\|$ and $\|\cdot\|_{F}$ denote the spectral and
     Frobenius norms.
     Finally, $\|\mathbf{Z}\|_{F} = \|\mathbf{X}\|_{F}$ if and only if $\mathbf{Z} = \mathbf{X} \mathbf{W}$
     for some block orthogonal $\mathbf{W}$, i.e., $\mathbf{W}
     = \Bigl[\begin{smallmatrix} \mathbf{W}_1 & 0 \\ 0 &
       \mathbf{W}_{2} \end{smallmatrix}\Bigr]$ where $\mathbf{W}_1$
     and $\mathbf{W}_2$ are $a \times a$ and $b \times b$ orthogonal
     matrices.
   \end{proposition}
   The representation $\mathbf{Z} = \mathbf{U}
   |\mathbf{S}|^{1/2}$ has the desirable property that it minimizes
   the Frobenius norm among all representations of $\mathbf{P}$ and is
   unique up to orthogonal transformation. This property
   allows $\mathbf{Z}$ to serve as the {\em canonical} representation for
   mathematical analysis. Nevertheless, $\mathbf{Z}$ suffers from one important
   conceptual limitation in that $\mathbf{Z}$ is a function of $\mathbf{P}$ and
   hence cannot be determined prior to specifying $\mathbf{P}$. In
   other words, $\mathbf{Z}$ is generally not suitable as a {\em
     generative} representation for $\mathbf{P}$.

\subsection{Hypothesis test for equality of latent positions}
\label{sec:hypothesis_test}
Suppose we are given a generalised random dot product graph. Our
first hypothesis testing problem is to determine whether or not two given nodes
$i$ and $j$ in the graph have the same latent position, i.e., given two nodes $i$ and $j$ with $i \neq j$, we are interested in
testing the hypothesis
\begin{equation}
  \label{eq:test_grdpg}
  \mathbb{H}_{0} \colon X_{i}=X_{j} \quad \text{versus} \quad \mathbb{H}_{A} \colon X_{i} \neq X_{j}
\end{equation}
Our second hypothesis testing problem concerns the hypothesis of equality up to scaling
between the latent positions $X_i$ and $X_j$. The motivation behind
this hypothesis test is as follows. Recall that for SBM graphs, any two vertices $i$ and $j$ that are
assigned to the same block will have the same {\em expected}
degree. The degree-corrected SBM \citep{karrer2011stochastic} relaxes
this restriction on the SBM by incorporating degree heterogeneity
through a vector of degree parameters $(\theta_1, \dots, \theta_n)$,
i.e, the probability of connection between two vertices $i$ and $j$ is
given by $\theta_i \theta_j \mathbf{B}_{\tau_i, \tau_j}$ where
$\tau_i$ and $\tau_j$ are the community assignments of vertices $i$
and $j$, respectively. Writing the edge probabilities matrix for a
degree-corrected SBM as
$\mathbf{P}=\rho_n\mathbf{\Theta}\mathbf{\Pi}\mathbf{B}\mathbf{\Pi}^{\top}\mathbf{\Theta}$,
where $\mathbf{\Theta}=\operatorname{diag}(\theta_1,\ldots,\theta_n)$,
we see that a degree-corrected SBM is a special case of a GRDPG
with latent positions of the form $X_i = \theta_i \sum_{k}
\pi_{ik} \delta_{\nu_k}$ where $\delta_{\nu_k}$ is the Dirac measure
at the point masses $\nu_1, \nu_2, \dots, \nu_K$ that generates the
block probabilities matrix $\mathbf{B}$. Now consider testing 
equality of membership $\mathbb{H}_0 \colon \tau_i = \tau_j$. This is equivalent to testing 
$\mathbb{H}_0 \colon X_i = \tfrac{\theta_i}{\theta_j} X_j$. Since the
degree-correction factors $(\theta_1, \dots, \theta_n)$ are generally
unknown, we will consider the more general 
test \begin{equation}
  \label{eq:projection_test1}
  \mathbb{H}_0 \colon \frac{X_i}{\|X_i\|} = \frac{X_j}{\|X_j\|} \quad \text{versus} \quad  \mathbb{H}_A \colon
  \frac{X_i}{\|X_i\|} \not = \frac{X_j}{\|X_j\|}
\end{equation}
where $\|X_i\|$ denotes the $\ell_2$ norm of $X_i$.

\subsection{Adjacency and Laplacian spectral embedding}
\label{sec:ase}
It was shown in \citep{rubin2017statistical} that the spectral
decomposition of the adjacency and Laplacian matrices provide consistent
estimates of the latent positions of a GRDPG. In this article, we will
use these spectral embedding representations to construct appropriate
test statistics for testing the hypothesis in
Eq.~\eqref{eq:test_grdpg} and Eq.~\eqref{eq:projection_test1}. We define these spectral embeddings below.

\begin{definition}
  \label{def:ASE}
Let $\mathbf{A} \in\{0,1\}^{n \times n}$ be the adjacency matrix for
an undirected graph and let $d$ be a positive integer specifying the embedding dimension. Consider the
eigendecomposition
$$\mathbf{A} = \sum_{i=1}^{n} \hat{\bm{\lambda}}_i \hat{\bm{u}}_i
\hat{\bm{u}}_i^{\top}, \quad
|\hat{\lambda}_1| \geq |\hat{\lambda}_2| \geq \dots \geq
|\hat{\lambda}_n|.$$
Now let $\hat{\mathbf{S}} \in \mathbb{R}^{d \times d}$ be the diagonal matrix
with diagonal entries $(\hat{\lambda}_{\sigma(1)}, \dots,
\hat{\lambda}_{\sigma(d)})$ where $\sigma$ is a permutation of
$\{1,2,\dots,d\}$ such that $\hat{\lambda}_{\sigma(1)} \geq
\hat{\lambda}_{\sigma(2)} \geq \dots \geq \hat{\lambda}_{\sigma(d)}$ 
and
let $\hat{\mathbf{U}}$ be the $n \times d$ matrix whose columns are
the corresponding orthonormal eigenvectors
$\hat{\bm{u}}_{\sigma(1)}, \dots, \hat{\bm{u}}_{\sigma(d)}$.
We introduce $\sigma$ so that, for the diagonal entries of $\hat{\mathbf{S}}$, the positive eigenvalues
of $\mathbf{A}$ appear before the negative eigenvalues. The adjacency spectral embedding of $\mathbf{A}$ into
$\mathbb{R}^{d}$ is the $n \times d$ matrix
$$\hat{\mathbf{X}}=\Bigl[ |\hat{\lambda}_{\sigma(1)}|^{1/2} \hat{\bm{u}}_{\sigma(1)},
|\hat{\lambda}_{\sigma(2)}|^{1/2} \hat{\bm{u}}_{\sigma(2)} , \dots, |\hat{\lambda}_{\sigma(d)}|^{1/2} \hat{\bm{u}}_{\sigma(d)}
 \Bigr] = \hat{\mathbf{U}}|\hat{\mathbf{S}}|^{1 /
  2}$$
where $|\hat{\mathbf{S}}|$ denote the element-wise absolute value of
$\hat{\mathbf{S}}$. We also denote by $\hat{X}_i$ the $i$th row of
$\hat{\mathbf{X}}$. 
\end{definition}

\begin{definition}
  \label{def:LSE}
  Let $\mathbf{A} \in\{0,1\}^{n \times n}$ be the
adjacency matrix for an undirected graph and let $d$ be a positive
integer specifying the embedding dimension. Define $\mathbf{L}=
\mathbf{L}(\mathbf{A}) = \mathbf{D}^{-1 / 2} \mathbf{A D}^{-1 / 2}$ as
the normalized Laplacian of $\mathbf{A}$ where $\mathbf{D} \in
\mathbb{R}^{n \times n}$ is a diagonal matrix whose diagonal entry
$\mathbf{D}_{i i}=\sum_{j} \mathbf{A}_{i j}$ is the degree of the
$i$th node.
Now consider the eigendecomposition
$\mathbf{L}=\breve{\mathbf{U}} \breve{\mathbf{S}}
\breve{\mathbf{U}}^{\top}+\breve{\mathbf{U}}_{\perp}
\breve{\mathbf{S}}_{\perp} \breve{\mathbf{U}}_{\perp}^{\top}$, where
$\breve{\mathbf{S}} \in \mathbb{R}^{d \times d}$ is the diagonal
matrix with entries given by the top $d$ eigenvalues of $\mathbf{L}$
in magnitude arranged in decreasing order,
and $\breve{\mathbf{S}}_{\perp}$ is 
the $(n - d) \times (n - d)$ diagonal
matrix whose diagonal entries are the remaining $n-d$ eigenvalues of $\mathbf{L}$,
$\breve{\mathbf{U}}$ is the
$n \times d$ matrix whose columns 
are the orthonormal
eigenvectors corresponding to the eigenvalues in $\breve{\mathbf{S}}$
and $\breve{\mathbf{U}}_{\perp}$ is the $n \times (n-d)$ matrix whose
columns are the remaining orthonormal eigenvectors. The Laplacian spectral embedding of $\mathbf{A}$ into
$\mathbb{R}^{d}$ is the $n \times d$ matrix defined by
$\breve{\mathbf{X}}=\breve{\mathbf{U}}|\breve{\mathbf{S}}|^{1
/ 2}$. We denote by $\breve{X}_i$ the $i$th row of $\breve{\mathbf{X}}$. 
\end{definition}

We note that the spectral embedding of
the {\em normalized} Laplacian matrix
appeared prominently in the context of manifold learning algorithms such
as Laplacian eigenmaps and diffusion maps
\citep{belkin03:_laplac,coifman06:_diffus_maps} and community
detection via spectral clustering \citep{chaudhuri,ng_spectral,rohe2011spectral,luxburg08:_consis}.

In the remainder of this paper we will assume that as $n \rightarrow
\infty$, the $n \times d$ matrix of latent positions $\mathbf{X} = [X_1, X_2, \dots, X_n]^{\top}$ and the
sparsity factor $\rho_n$ satisfies the following three conditions.
\begin{condenum}
  \label{cond:regularity}
\item The matrix $\mathbf{X}$ is a $n \times d$ matrix with $d$ not
  depending on $n$ and $\sigma_1(\mathbf{X}) \asymp \sigma_d(\mathbf{X})=\Theta(n)$
  where $\sigma_1(\mathbf{X}) \geq \sigma_2(\mathbf{X}) \geq \cdots \geq
  \sigma_d(\mathbf{X})$ are the singular
  values of $\mathbf{X}$.
\item The latent positions $X_i$ belong to a {\em fixed} compact set
  $\mathcal{K}$ not depending on $n$ and there exists a {\em fixed} constant $c > 0$ such that
  $X_i^{\top}\mathbf{I}_{a,b}X_j \geq c$ for all $i,j$.
\item The sparsity factor $\rho_n$ satisfies $n \rho_n = \omega(\log
  n)$. Here we write that sequences $a_{n}=\omega\left(b_{n}\right)$ if for any positive constant $C$, there exists an integer $n_{0} \geq 1$ such that $a_{n} > C b_{n}$ for all $n \geq n_{0}$.
\end{condenum}
Condition~1 assumes that the singular values of $\mathbf{X}$ are all of
the same order and grow linearly with $n$. Condition~2 assumes that
the latent positions $\{X_i\}$ are all bounded in $\ell_2$ norm and
that the minimum edge probability between any two vertices, before scaling by the sparsity
parameter $\rho_n$, does not converge to $0$. This assumption prevents
the setting wherein, as $n \rightarrow \infty$, some vertices $v_i$
have latent positions $X_i$ for which $\|X_i\| \rightarrow 0$ and hence
$v_i$ became isolated. Condition~3 assumes that the average degree of
the graph grows faster than some poly-logarithmic function of
$n$. Note that the poly-logarithmic regime in $n$ is necessary for
spectral methods to work, e.g., if $n \rho_n = o(\log n)$ then the
eigenvalues and eigenvectors of $\mathbf{A}$ are no longer consistent
estimate of the corresponding eigenvalues and eigenvectors of the edge
probabilities matrix $\mathbf{P} = \rho_n\mathbf {X} \mathbf{I}_{a,b}
\mathbf{X}^{\top}$. Without a consistent estimate of the latent
positions we can not obtain an asymptotically valid and consistent
test procedure for the hypothesis that two arbitrary vertices have the
same latent positions.

Given the above conditions, the following result provides a central limit
theorem for the rows of the adjacency spectral embedding
$\hat{\mathbf{X}}$ around the latent positions representation $\mathbf{Z}$ obtained from the
eigendecomposition of $\mathbf{P}$ (see
Proposition~\ref{prop:non_identifiable2}). Analogous results for the Laplacian
spectral embeddings are mentioned in the proof of Theorem~\ref{thm2}  given in Appendix~\ref{appb.7}. This is done purely
for ease of exposition as (1) the limit results for the adjacency
spectral embedding are simpler to present compared to its Laplacian
counterpart and (2) a detailed discussion of adjacency spectral embedding
is sufficient to demonstrate the main technical contributions of this
paper. More specifically, to convert these limit results into
appropriate test statistics we have to first obtain consistent estimates of the limiting
covariance matrices, then show the limiting distribution of the test
statistics under the null hypothesis and local alternative hypothesis
and finally derive explicit expressions for the non-centrality
parameters under the local alternative.
\begin{theorem}
  \label{thm0}
  Let $\mathbf{A}^{(n)} \sim \mathrm{GRDPG}(\mathbf{X}^{(n)}, \rho_n)$
  be a {\em sequence} of generalized random dot product graphs on $n$ vertices with signature
  $(a,b)$. Suppose that, as $n \rightarrow \infty$, the 
  $\mathbf{X}^{(n)} = [X^{(n)}_1, X^{(n)}_2, \dots, X^{(n)}_n]$ satisfies Condition~1
  through Condition~3. Let $\mathbf{Z}^{(n)} = \mathbf{U}^{(n)} |\mathbf{S}^{(n)}|^{1/2}$
be the $n \times d$ matrix where $\mathbf{U}^{(n)} \mathbf{S}^{(n)}
(\mathbf{U}^{(n)})^{\top} = \rho_n \mathbf{X}^{(n)} \mathbf{I}_{a,b} (\mathbf{X}^{(n)})^{\top}$ is the eigendecomposition of the
edge probabilities matrix and $Z_i^{(n)}$ be the $i$th row of $\mathbf{Z}^{(n)}$. Note that $\rho_n^{1/2} \mathbf{X}^{(n)} =
\mathbf{Z}^{(n)} \mathbf{Q}_{\mathbf{X}^{(n)}}$ for some indefinite orthogonal
transformation $\mathbf{Q}_{\mathbf{X}^{(n)}}$. Define $\bm{\Sigma}(Z_i^{(n)})$ as the $d \times
d$ matrix of the form $$\bm{\Sigma}(Z_i^{(n)}) = n (\mathbf{S}^{(n)})^{-1}
\sum_{k=1}^{n} Z_k^{(n)} (Z_k^{(n)})^{\top} (Z_k^{(n)})^{\top} \mathbf{I}_{a,b} Z_i^{(n)} (1 -
(Z_k^{(n)})^{\top} \mathbf{I}_{a,b} Z_i^{(n)}) (\mathbf{S}^{(n)})^{-1}$$
Then there exists a sequence of $(a,b)$ {\em block-orthogonal} matrices
$\mathbf{W}_n$ (see Remark~\ref{rem:indefinite_properties}) such that for any index $i$
  \begin{equation}
    \label{eq:convergence_Z}
  \sqrt{n} \bm{\Sigma}(Z_i^{(n)})^{-1/2} \bigl( \mathbf{W}_n \hat{X}_i^{(n)} - Z_i^{(n)}\bigr) \rightsquigarrow \mathcal{N}\bigl(0,
  \mathbf{I})
  \end{equation}
  Note that $\mathbf{W}_n$ is unknown here but we do not need its value to construct the test statistics. Furthermore, for any pair of indices $i \not = j$, the vectors $\sqrt{n}\bigl( \mathbf{W}_n \hat{X}_{i}^{(n)} -
  Z_i^{(n)} \bigr)$ and $\sqrt{n}\bigl(\mathbf{W}_n \hat{X}_{j}^{(n)} - Z_j^{(n)} \bigr)$ are asymptotically independent. 
\end{theorem}
Theorem~\ref{thm0} comes from a restatement of
Theorem~4 of \citep{rubin2017statistical} to the setting of the
current paper: Under the same setting as Theorem~\ref{thm0}, for any $n$ and any $X^{(n)}_i$, let
  $\bm{\Sigma}(X^{(n)}_i; \mathbf{X}^{(n)})$ be a $d \times d$ matrix of the form
  \begin{equation}
    \label{eq:cov_X_theoretical}
    \begin{split}
  \bm{\Sigma}(X_i^{(n)}; \mathbf{X}^{(n)}) &= n \rho_n^{-1} \mathbf{I}_{a,b}
  \mathbf{M}_n^{-1} \sum_{k=1}^{n} X_k^{(n)} (X_k^{(n)})^{\top}
  p_{ik}^{(n)} (1 - p_{ik}^{(n)}) \mathbf{M}_n^{-1} \mathbf{I}_{a,b}
\end{split}
\end{equation}
where $\mathbf{M}_n = (\mathbf{X}^{(n)})^{\top}
  \mathbf{X}^{(n)}$ and $p^{(n)}_{ik} = \rho_n (X_i^{(n)})^{\top} \mathbf{I}_{a,b} X_k^{(n)}$ is the edge
  probability between the $i$th and $k$th vertices in $\mathbf{A}^{(n)}$. Then there exists a sequence of
  $(a,b)$ {\em block-orthogonal} matrices $\mathbf{W}_n$ (see Remark~\ref{rem:indefinite_properties}) and a
  sequence of indefinite orthogonal matrices $\mathbf{Q}_{\mathbf{X}^{(n)}}$
  such that for any index $i$
  \begin{equation}
    \label{eq:convergence}
  \sqrt{n} \bm{\Sigma}(X_i^{(n)}; \mathbf{X}^{(n)})^{-1/2} \bigl(\mathbf{Q}_{\mathbf{X}^{(n)}}^{\top} \mathbf{W}_n \hat{X}_i^{(n)} -
  \rho_n^{1/2} X_i^{(n)}\bigr) \rightsquigarrow \mathcal{N}\bigl(0,
  \mathbf{I})
  \end{equation}
  and for any {\em fixed} $m$ not depending on $n$ and any
  finite set of {\em distinct} indices $\{i_1, i_2, \dots, i_{m}\}$, 
  the vectors $r_{i_k}^{(n)} = \sqrt{n}\bigl(\mathbf{Q}_{\mathbf{X}^{(n)}}^{\top} \mathbf{W}_n \hat{X}_{i_k}^{(n)} -
  \rho_n^{1/2} X_{i_k}^{(n)}\bigr)$ for $k \leq m$ are {\em
    asymptotically}, mutually independent.

Theorem~\ref{thm0} is then established by reformulating the above result so that $\hat{\mathbf{X}}^{(n)}$ is centered around $\mathbf{Z}^{(n)}$.  See the discussion that begins Section~\ref{sec:proof_main} of the Appendix
for more details. This makes it more convenient for the subsequent technical
derivations as the orthogonal transformation mapping
$\hat{X}_i^{(n)}$ to $Z_i^{(n)}$ is much simpler than the
indefinite orthogonal transformation mapping $\hat{X}_i^{(n)}$ to
$\rho_n^{1/2} X_i^{(n)}$. For example the $\ell_2$ norm is invariant with respect to
orthogonal transformations but not invariant with respect to indefinite orthogonal transformations. 

\begin{remark}
  \label{rem:adaptation}
The statement of Theorem~\ref{thm0} assumes that we are given a {\em sequence} of matrices
$\{\mathbf{X}^{(n)}\}$ where for each $n$, $\mathbf{X}^{(n)}$ is a $n
\times d$ matrix of latent positions for a GRDPG graph on $n$
vertices and furthermore, the latent positions in $\mathbf{X}^{(n)}$ need
not be related to those in $\mathbf{X}^{(n')}$ for $n' \not =
n$; rather we only assume that the sequence $\{\mathbf{X}^{(n)}\}$ satisfies
Condition~1 and Condition~2 above. As the notations in Theorem~\ref{thm0}
are quite cumbersome, for ease of exposition, we will henceforth drop
the index $n$ from most of the notations in Theorem~\ref{thm0}. 
For example we will only write $\mathbf{A}$, $\mathbf{X}$, 
$\mathbf{Q}_{\mathbf{X}}, X_i, \mathbf{Z}, Z_i$ in place of $\mathbf{A}^{(n)}$,
$\mathbf{X}^{(n)}$, $\mathbf{Q}_{\mathbf{X}}^{(n)}$, $X_i^{(n)}, \mathbf{Z}^{(n)}$ and $Z_i^{(n)}$; the
covariance matrices in Theorem~\ref{thm0} and  Eq.~\eqref{eq:cov_X_theoretical} are denoted by $\bm{\Sigma}(Z_i)$ and 
$\bm{\Sigma}(X_i)$, respectively, and $p_{ik}^{(n)}$ is replaced by $p_{ik}$. In
this form we can view Theorem~\ref{thm0} as providing a multivariate normal approximation
for $\sqrt{n}\bigl( \mathbf{W}_n \hat{X}_{i} - Z_i \bigr)$ as $n$ increases.
\end{remark}

\section{Test Statistics Using Adjacency Spectral Embedding}
\label{sec3}
We now discuss how the limit results in Theorem~\ref{thm0} can be adapted to construct test statistics for
testing the hypothesis of equality and equality up to scaling as
described in Section~\ref{sec:hypothesis_test}. 
See Table~\ref{Notation} for a summary of several notations that are
frequently used throughout this paper.

\begin{table*}
\caption{Frequently used notations in this paper.}
\label{Notation}
\begin{tabular}{@{}cc@{}}
\hline
Notation & Definition\\
\hline
$U_i$& the $i$th row of matrix $\mathbf{U}$ where
$\mathbf{U} \mathbf{S} \mathbf{U}^{\top}$ is the eigendecomposition of
$\rho_n \mathbf{X} \mathbf{I}_{a,b} \mathbf{X}^{\top}$\\
$\hat{U}_i$& the $i$th row of matrix $\hat{\mathbf{U}}$, where $\hat{\mathbf{X}} = \hat{\mathbf{U}}|\hat{\mathbf{S}}|^{1 /
  2}$(see Definition~\ref{def:ASE})
\\
$X_i$& the $i$th row of the latent position matrix $\mathbf{X}$ \\
$\hat{X}_i$& the $i$th row of the
adjacency spectral embedding $\hat{\mathbf{X}}$ obtained from
             $\mathbf{A}$ (see Definition~\ref{def:ASE}) \\ 
$\check{X}_i$& the $i$th row of the Laplacian spectral embedding
               $\check{\mathbf{X}}$ obtained from $\mathbf{A}$ (see Definition~\ref{def:LSE}) \\
$Z_i$ & the $i$th row of the matrix $\mathbf{Z} = \mathbf{U}
               |\mathbf{S}|^{1/2}$ (see
        Proposition~\ref{prop:non_identifiable2}), i.e., $Z_i = |\mathbf{S}|^{1/2} U_i$ \\
$\mathbf{Q}_{\mathbf{X}}$ & the {\em
  indefinite} orthogonal transformation such that $\rho_n^{1/2} \mathbf{X} =
\mathbf{Z} \mathbf{Q}_{\mathbf{X}}$\\
\hline
\end{tabular}
\end{table*}

\subsection{Testing $\mathbb{H}_0 \colon X_i = X_j$}
\label{sec:ase_test1}
Let $\mathbf{A}$ be a graph on $n$ vertices generated from the model
$\operatorname{GRDPG}_{a,b}(\mathbf{X},\rho_n)$ with
signature $(a,b)$ and sparsity factor $\rho_n$, where $a+b=d$ and $n
\rho_{n}=\omega(\log n)$. Given two vertices $i$ and $j$ in $\mathbf{A}$, we wish
to test the null hypothesis $\mathbb{H}_0 \colon X_i = X_j$
against the alternative hypothesis $\mathbb{H}_A \colon X_i \not =
X_j$.

Recall Theorem~\ref{thm0}. Then for $X_i = X_j$, we have
\begin{equation}
  \label{eq:t_ase_pre0}
n\bigl(\hat{X}_i-\hat{X}_j\bigr)^{\top} \mathbf{W}_n^{\top}
\bigl(\bm{\Sigma}(Z_i) + \bm{\Sigma}(Z_j)\bigr)^{-1}
\mathbf{W}_n \bigl(\hat{X}_i-\hat{X}_j\bigr) \rightsquigarrow
\chi^2_{d}
\end{equation}
Our objective is to convert Eq.~\eqref{eq:t_ase_pre0} into
an appropriate test
statistic that depends only on the $\{\hat{X}_i\}$. It is thus sufficient to find a
consistent estimate for $\mathbf{W}_n^{\top} \bigl(\bm{\Sigma}(Z_i) +
\bm{\Sigma}(Z_j)\bigr)^{-1} \mathbf{W}_n$ in terms of the
$\{\hat{X}_i\}$. The following lemma provides one such estimate.
 \begin{lem}
  \label{lem:conv1}
  Assume the setting in Theorem~\ref{thm0}. Define
  $\hat{\bm{\Sigma}}(\hat{X}_i)$ as the $d \times d$ matrix of the form
  \begin{equation}
    \label{th1_2}
    \begin{split}
  \hat{\boldsymbol{\Sigma}}(\hat{X}_i) &=  n\mathbf{I}_{a,b} \bigl(\hat{\mathbf{X}}^{\top}\hat{\mathbf{X}}\bigr)^{-1}
  \Bigl[\sum_{k=1}^n \hat{X}_k \hat{X}_k^{\top}
   \hat{X}_i^{\top} \mathbf{I}_{a,b} \hat{X}_k (1 - \hat{X}_i^{\top} \mathbf{I}_{a,b} \hat{X}_k)\Bigr]
   \bigl(\hat{\mathbf{\mathbf{X}}}^{\top}\hat{\mathbf{X}}\bigr)^{-1}\mathbf{I}_{a,b} \\
   &= n \hat{\mathbf{S}}^{-1} \Bigl[ \sum_{k=1}^n  \hat{X}_k \hat{X}_k^{\top}
   \hat{X}_i^{\top} \mathbf{I}_{a,b} \hat{X}_k (1 - \hat{X}_i^{\top}
   \mathbf{I}_{a,b} \hat{X}_k)\Bigr] \hat{\mathbf{S}}^{-1}.
\end{split}
\end{equation}
  If $\mathbf{X}$ satisfies Condition~1 through Condition~3 as $n
  \rightarrow \infty$, then
  \begin{equation}
    \label{eq:conv1}
\hat{\boldsymbol{\Sigma}}(\hat{X}_i) - \mathbf{W}_n^{\top} \bm{\Sigma}(Z_i) \mathbf{W}_n
\overset{\mathrm{a.s}}{\longrightarrow} 0.
\end{equation}
\end{lem}
Theorem~\ref{thm0} together with Lemma~\ref{lem:conv1} implies
the following large sample limiting behavior of the test statistic based
on the Mahalanobis distance between $\hat{X}_i$ and $\hat{X}_j$ under
both the null hypothesis and {\em local} alternative hypothesis.
\begin{theorem}\label{thm1}
  Let $\mathbf{X}$ be a $n \times d$ matrix and let $\mathbf{A} \sim
  \mathrm{GRDPG}_{a,b}(\mathbf{X}, \rho_n)$.
  Let $\hat{\mathbf{X}}$ be the adjacency spectral embedding of
  $\mathbf{A}$ into $\mathbb{R}^{d}$. Define the test statistic 
  \begin{equation}
  \label{th1_1}
  T_{\mathrm{ASE}}(\hat{X}_i, \hat{X}_j)
  =n(\hat{X}_i - \hat{X}_j)^{\top}\Bigl(\hat{\boldsymbol{\Sigma}}(\hat{X}_i)
  + \hat{\boldsymbol{\Sigma}}(\hat{X}_j)\Bigr)^{-1} (\hat{X}_i - \hat{X}_j)
  \end{equation}
where the $d \times d$ matrices $\hat{\boldsymbol{\Sigma}}(\hat{X}_i)$ and
$\hat{\boldsymbol{\Sigma}}(\hat{X}_j)$ are as given in
Lemma~\ref{lem:conv1}. Then under the null hypothesis $\mathbb{H}_0 \colon X_i=X_{j}$ and for
$n \rightarrow \infty$ with $n \rho_n = \omega(\log n)$, we have
$$T_{\mathrm{ASE}}(\hat{X}_i, \hat{X}_j) \rightsquigarrow \chi_d^2.$$
Let $\boldsymbol{\Sigma}(X_i)$ be as defined
in Eq.~\eqref{eq:cov_X_theoretical} and let $\mu > 0$ be a finite constant such that 
\begin{equation}
  \label{eq:noncentral1}
n\rho_n(X_i-X_j)^{\top}\bigl(\boldsymbol{\Sigma}(X_i) + \boldsymbol{\Sigma}(X_j)\bigr)^{-1}
(X_i-X_j) \rightarrow \mu
\end{equation}
Then, under a local alternative $X_i \not = X_j$, we have $T_{\mathrm{ASE}}(\hat{X}_i, \hat{X}_j) \rightsquigarrow \chi_d^2\left(\mu\right)$
where $\chi_d^2\left(\mu\right)$ is the
noncentral chi-square distribution with $d$ degrees of freedom and
noncentrality parameter $\mu$.
\end{theorem} 
Theorem \ref{thm1} indicates that for a chosen significance level
$\alpha$, we will reject $\mathbb{H}_0$ if $T_{\mathrm{ASE}}(\hat{X}_i, \hat{X}_j)> c_{1-\alpha}$, where
$c_{1-\alpha}$ is the $100 \times (1-\alpha)$th percentile of the $\chi^2_{d}$ distribution.

\begin{remark}
  \label{rem:thm1}
  The sparsity factor $\rho_n$ does not
appear in the test statistic of Eq.~\eqref{th1_1}. This might
seems surprising since, if $\rho_n \rightarrow 0$ then the graphs
become sparser and we have less signal. The main reason why this
does not affect the limiting behavior of $T_{\mathrm{ASE}}$ is that while the error rate for $\|\mathbf{W}_n
\hat{X}_i - Z_i\|$ becomes larger relative to $\|Z_i\|$ and
$\|\hat{X}_i\|$, both of which are also converging to $0$ as $\rho_n
\rightarrow 0$, it does not increase in absolute terms. See the
statement of Theorem~\ref{thm0}. In contrast, the sparsity
parameter $\rho_n$ appears in the condition for the local alternative
in Eq.~\eqref{eq:noncentral1}; our interpretation of this condition is 
that as $\rho_n \rightarrow 0$ then we need a
larger distance between $X_i - X_j$ to compensate for the decrease in
magnitude of the edge probabilities, i.e., if
Eq.~\eqref{eq:noncentral1} holds then $X_i$ is sufficiently close to
$X_j$ so that $\|\bm{\Sigma}(X_i) - \bm{\Sigma}(X_j)\| \rightarrow 0$ and Eq.~\eqref{eq:noncentral1} is equivalent to the condition
$X_j = X_i + v$ where
$$\frac{1}{2} n \rho_n v^{\top} \bigl(\bm{\Sigma}(X_i)\bigr)^{-1} v
\longrightarrow \mu.$$
Finally we emphasize that the condition in Eq.~\eqref{eq:noncentral1}
is invariant with respect to the choice of the $\{X_i\}$, i.e., the
value of $\mu$ is not affected by the non-identifiability of the
latent positions $\{X_i\}$.
\end{remark}

\subsection{Testing with degree-correction $\mathbb{H}_0 \colon
  X_i/\|X_i\| = X_j/\|X_j\|$ }
\label{sec:degree_corrected}
We now discuss a test statistic for testing $\mathbb{H}_0 \colon
X_i/\|X_i\| = X_j/\|X_j\|$. We start by considering an empirical
Mahalanobis distance between $\hat{X}_i/\|\hat{X}_i\|$ and
$\hat{X}_j/\|\hat{X}_j\|$.  
\begin{theorem}\label{thm3}
Consider the setting in Theorem~\ref{thm1}. Let $s(\xi) = \xi/\|\xi\|$ be the transformation that projects
any vector $\xi \in \mathbb{R}^{d}$ onto the unit sphere in
$\mathbb{R}^{d}$ and denote by $\mathbf{J}(\xi)$ the Jacobian of $s(\xi)$, i.e.,
$$\mathbf{J}(\xi)= \frac{1}{\|\xi\|}\Bigl(\mathbf{I} -
\frac{\xi\xi^{\top}}{\|\xi\|^2} \Bigr).$$
Next recall the definition of
$\hat{\bm{\Sigma}}(\hat{X}_i)$ given in Lemma~\ref{lem:conv1} and define the test statistic
\begin{equation}
  \label{eq:thm3_1}
  G_{\mathrm{ASE}}(\hat{X}_i,\hat{X}_j)=n\bigl(s(\hat{X}_i)-s(\hat{X}_j)\bigr)^{\top}\Bigl(\mathbf{J}(\hat{X}_i)\Bigl[\hat{\mathbf{\Sigma}}(\hat{X}_i)+\tfrac{\|\hat{X}_i\|^2}{\|\hat{X}_j\|^2}\hat{\mathbf{\Sigma}}(\hat{X}_j)\Bigr]\mathbf{J}(\hat{X}_i)\Bigr)^{\dagger}\bigl(s(\hat{X}_i)-s(\hat{X}_j)\bigr).
\end{equation}
Here $(\cdot)^{\dagger}$ denotes the Moore-Penrose pseudoinverse of a matrix. 
Then under the null hypothesis $\mathbb{H}_0 \colon X_i/\|X_i\| = X_j/\|X_j\|$ and for
$n \rightarrow \infty$ with $n \rho_n = \omega(\log n)$, we have
$$G_{\mathrm{ASE}}(\hat{X}_i, \hat{X}_j) \rightsquigarrow \chi_{d-1}^2.$$
Now recall the definition of $\bm{\Sigma}(Z_i)$ in
Theorem~\ref{thm0}.
Let $\mu > 0$ be a finite constant such that
\begin{equation}
  \label{eq:local_alt_ase2}
  n \bigl(s(Z_i)-s(Z_j)\bigr)^{\top}\Bigl(\mathbf{J}(Z_i) \Bigl(
  \boldsymbol{\Sigma}(Z_i) + \tfrac{\|Z_i\|^2}{\|Z_j\|^2}
  \bm{\Sigma}(Z_j) \Bigr) \mathbf{J}(Z_i)\Bigr)^{\dagger} \bigl(s(Z_i)-s(Z_j)\bigr) \rightarrow \mu.
\end{equation}
Then under a local alternative $X_i/\|X_i\| \not = X_j/\|X_j\|$, we have $G_{\mathrm{ASE}}(\hat{X}_i, \hat{X}_j)
\rightsquigarrow \chi_{d-1}^2\left(\mu\right)$ where $\chi_{d-1}^2\left(\mu\right)$ is the
noncentral chi-square with $d-1$ degrees of freedom and
noncentrality parameter $\mu$. 
\end{theorem} 

There is a difference in the scaling of $n \rho_n$ for
the non-centrality parameter in Eq.~\eqref{eq:noncentral1} versus the
scaling of $n$ for Eq.~\eqref{eq:local_alt_ase2}. This difference is
due to the difference in the scale of $Z_i$ versus $X_i$, i.e.,
$\|Z_i\| = \Theta(\rho_n^{1/2})$ while $\|X_i\| = \Theta(1)$. This
difference does not manifest itself in the scale of $s(Z_i)$ versus
$s(X_i)$ but rather in the scale of the Jacobian $\mathbf{J}(Z_i)$
versus $\mathbf{J}(X_i)$. Indeed, we have $s(c \xi) = s(\xi)$ but
$\mathbf{J}(c \xi) = c^{-1} \mathbf{J}(\xi)$ for any vector $\xi \in
\mathbb{R}^{d}$ and any constant $c > 0$. We also note that the condition in
Eq.~\eqref{eq:local_alt_ase2} is specified in terms of the $\{Z_i\}$
instead of the $\{X_i\}$. That is to say, the non-centrality
parameter for Theorem~\ref{thm3} might not be invariant with respect
to the choice of non-identifiability transformations of the latent
positions $\{X_i\}$. The following result provides a different representation of $\mu$ that is invariant to the non-identifiability in the latent positions $\{X_i\}$.
\begin{proposition}
  \label{prop:equality_ncp}
  Consider the test statistic $G_{\mathrm{ASE}}(\hat{X}_i, \hat{X}_j)$
  in Theorem~\ref{thm3}. Let $s'(\xi)$ be the transformation $s'(\xi) = \xi/\|\xi\|_{\mathbf{I}_{a,b}}$ where
  $\|\xi\|^2_{\mathbf{I}_{a,b}}=\xi^{\top}\mathbf{I}_{a,b}\xi$. Denote
  by $\mathbf{J}'(\xi)$ the Jacobian of $s'$, i.e., 
 $$\mathbf{J}^{'}(\xi)= \frac{1}{\|\xi\|_{\mathbf{I}_{a,b}}}\Bigl(\mathbf{I} -
   \frac{ \xi\xi^{\top}\mathbf{I}_{a,b}}{\|\xi\|_{\mathbf{I}_{a,b}}^2}\Bigr).$$
   Note that $s'(\xi) = s(\xi)$ and $\mathbf{J}'(\xi) =
   \mathbf{J}(\xi)$ whenever $b = 0$. However, for $b > 0$, $s'(\xi)
   \not = s(\xi)$ and $\mathbf{J}'(\xi)$ is not necessarily  symmetric
   matrix. The condition in Eq.~\eqref{eq:local_alt_ase2} is then
   equivalent to the condition 
    \begin{gather}
   \label{eq:indefinite_alt_cond2}
 n \rho_n \bigl(s^{'}(X_i)-s^{'}(X_j)\bigr)^{\top}\Bigl(\mathbf{J}^{'}(X_i)
 \Bigl( \boldsymbol{\Sigma}(X_i) +  \tfrac{\|X_i\|^2_{\mathbf{I}_{a,b}}}{\|X_j\|_{\mathbf{I}_{a,b}}^2}
  \bm{\Sigma}(X_j) \Bigr) \mathbf{J}^{'}(X_i)^{\top}\Bigr)^{\dagger}
 \bigl(s^{'}(X_i)-s^{'}(X_j)\bigr) \rightarrow \mu.
 \end{gather}
\end{proposition}  

\subsection{Relationship with previous work}
\label{sec:related_work}
The problem of membership testing in degree-corrected and mixed
membership stochastic block model graphs had previously been considered
in \citep{fan2019simple}. The test statistics in \citep{fan2019simple}
are closely related to that of the current paper.  In particular
their test statistics are based on the Mahalanobis distance between
$\hat{U}_i$ and $\hat{U}_j$; here $\hat{U}_i$ denote the $i$th row of
the $n \times d$ matrix $\hat{\mathbf{U}}$ whose columns are the
eigenvectors corresponding to the $d$ largest eigenvalues in magnitude
of the adjacency matrix $\mathbf{A}$.  Recall that our embedding
$\hat{X}_i$ in Definition~\ref{def:ASE} are obtained by scaling the
eigenvectors $\hat{\mathbf{U}}$ by the square-root of the eigenvalues
$|\hat{\mathbf{S}}|^{1/2} = \mathrm{diag}(|\hat{\lambda}_1|^{1/2},
\dots, |\hat{\lambda}_d|^{1/2})$. The motivation for considering
$\hat{U}_i - \hat{U}_j$ is that $\hat{\mathbf{U}}$ is an estimate, up
to orthogonal transformation, for the $n \times d$ matrix $\mathbf{U}$
whose columns are the eigenvectors corresponding to the non-zero
eigenvalues of $\rho_n \mathbf{X} \mathbf{I}_{a,b}
\mathbf{X}^{\top}$. Furthermore, $X_i = X_j$ if and only if $U_i =
U_j$. Thus both $\hat{U}_i - \hat{U}_j$ and $\hat{X}_i - \hat{X}_j$
can be used to construct test statistics for $\mathbb{H}_0 \colon X_i
= X_j$. As $\hat{X}_i$ and $\hat{U}_i$ are invertible
linear transformations of one another, and Mahalanobis distance is
invariant to invertible linear transformations, the test
statistics based on $\hat{U}_i - \hat{U}_j$ and $\hat{X}_i -
\hat{X}_j$ are identical, i.e.,
$$(\hat{X}_i-\hat{X}_j)^{\top}\bigl(\hat{\bm{\Sigma}}(\hat{X}_i) +
\hat{\bm{\Sigma}}(\hat{X}_j)\bigr)^{-1}(\hat{X}_i-\hat{X}_j) =
(\hat{U}_i - \hat{U}_j)^{\top} \Bigl(|\hat{\mathbf{S}}|^{-1/2} \bigl(\hat{\bm{\Sigma}}(\hat{X}_i) +
\hat{\bm{\Sigma}}(\hat{X}_j)\bigr) |\hat{\mathbf{S}}|^{-1/2}\Bigr)^{-1}
(\hat{U}_i - \hat{U}_j)^{\top}.$$
The following result is
thus a reformulation of Theorem~\ref{thm1} in the current paper to the
Mahalanobis distance for $\hat{U}_i - \hat{U}_j$, and is a generalization of
Theorem~1 and Theorem~2 in \citep{fan2019simple}.

\begin{cor}\label{cor1}
Consider the setting in Theorem~\ref{thm1}. Now define the test statistic
\begin{equation}
  \label{eq:T_tilde_ASE}
  \begin{split}
  \tilde{T}_{\mathrm{ASE}}(\hat{U}_i,\hat{U}_j)&=n^2 \rho_n (\hat{U}_i-\hat{U}_j)^{\top}\bigl(\hat{\boldsymbol{\Sigma}}(\hat{U}_i)
  +
  \hat{\boldsymbol{\Sigma}}(\hat{U}_j)\bigr)^{-1}(\hat{U}_i-\hat{U}_j)
  \\
  &= n (\hat{X}_i-\hat{X}_j)^{\top}\bigl(\hat{\bm{\Sigma}}(\hat{X}_i) + \hat{\bm{\Sigma}}(\hat{X}_j)\bigr)^{-1}(\hat{X}_i-\hat{X}_j).
  \end{split}
  \end{equation}
where the $d \times d$ matrices $\hat{\boldsymbol{\Sigma}}(\hat{U}_i)$ and
$\hat{\boldsymbol{\Sigma}}(\hat{U}_j)$ are given by
\begin{equation}
  \label{eq:hat_Sigma_hat_U}
  \begin{split}
  \hat{\mathbf{\Sigma}}(\hat{U}_i) &= n \rho_n
|\hat{\mathbf{S}}|^{-1/2} \hat{\bm{\Sigma}}(\hat{X}_i)
|\hat{\mathbf{S}}|^{-1/2} \\ &= n^2 \rho_n \mathbf{I}_{a,b} |\hat{\mathbf{S}}|^{-3/2} \Bigl[\sum_{k=1}^n
\hat{X}_k\hat{X}_k^{\top}\hat{X}_i^{\top}\mathbf{I}_{a,b}
\hat{X}_{k}\bigl(1-\hat{X}_i^{\top} \mathbf{I}_{a,b}\hat{X}_{k}
\bigr)\Bigr]|\hat{\mathbf{S}}|^{-3/2} \mathbf{I}_{a,b}.
\end{split}
\end{equation}
Then under the null hypothesis $\mathbb{H}_0 \colon X_i=X_{j}$ and for
$n \rightarrow \infty$ with $n \rho_n = \omega(\log n)$, we have
$$\tilde{T}_{\mathrm{ASE}}(\hat{U}_i, \hat{U}_j) \rightsquigarrow \chi_d^2.$$
Now recall the expression for $\bm{\Sigma}(Z_i)$ in
Theorem~\ref{thm0} and let 
\begin{equation}
  \label{eq:Sigma_U}
  \begin{split}
  \boldsymbol{\Sigma}(U_i) &= n \rho_n |\mathbf{S}|^{-1/2}
  \bm{\Sigma}(Z_i) |\mathbf{S}|^{-1/2} = n^2 \rho_n \mathbf{S}^{-1} \sum_{k} U_{k}
  U_k^{\top} p_{ik} (1 - p_{ik}) \mathbf{S}^{-1}
 \end{split}
 \end{equation}
where $p_{ik} = \rho_n X_i^{\top} \mathbf{I}_{a,b} X_k = Z_i^{\top}
\mathbf{I}_{a,b} Z_k$ is the probability of the $ik$ edge. Let $\mu >
0$ be a finite constant such that $X_i \not = X_j$ satisfies a
local alternative where
$$n^2 \rho_n (U_i - U_j)^{\top}
\bigl(\boldsymbol{\Sigma}(U_i) +
\boldsymbol{\Sigma}(U_j)\bigr)^{-1}(U_i - U_j) = n \rho_n (X_i -
X_j)^{\top}\bigl(\boldsymbol{\Sigma}(X_i) +
\boldsymbol{\Sigma}(X_j)\bigr)^{-1} (X_i-X_j) \rightarrow \mu.$$
Then $\tilde{T}_{\mathrm{ASE}}(\hat{U}_i, \hat{U}_j) \rightsquigarrow
\chi_d^2\left(\mu\right)$ where $\chi_d^2\left(\mu\right)$ is the
noncentral chi-square distribution with $d$ degrees of freedom and
noncentrality parameter $\mu$.
\end{cor} 
For the case of testing the degree-corrected hypothesis $\mathbb{H}_0
\colon X_i/\|X_i\| = X_j/\|X_j\|$, \citep{fan2019simple} construct a
test statistic using the Mahalanobis distance between
$\hat{U}_{i,2:d}/\hat{U}_{i1}$ and $\hat{U}_{j,2:d}/\hat{U}_{j1}$
Here $\hat{U}_{i,2:d} = (\hat{U}_{i2},\hat{U}_{i3}, \dots,
\hat{U}_{id})$ and $\hat{U}_{j,2:d} =  (\hat{U}_{j2},\hat{U}_{j3}, \dots,
\hat{U}_{jd})$ are vectors with the the first coordinate
of $\hat{U}_{i}$ and $\hat{U}_j$ removed, respectively; recall that the columns of $\hat{\mathbf{U}}$ are
ordered such that the first column is the eigenvector corresponding to
the largest eigenvalue of $\mathbf{A}$. 
The transformation $\hat{U}_{i} \mapsto
\hat{U}_{i,2:d}/\hat{U}_{i1}$ was motivated by the spectral clustering
using ratio of eigenvectors (SCORE) procedure described in
\cite{jin2015fast}. The following result shows that the Mahalanobis distance
between $\hat{U}_{i,2:d}/\hat{U}_{i1}$ and
$\hat{U}_{j,2:d}/\hat{U}_{j1}$ is the same as the Mahalanobis distance between
$\hat{X}_{i,2:d}/\hat{X}_{i1}$ and $\hat{X}_{j,2:d}/\hat{X}_{j1}$.
\begin{proposition}
  \label{prop:equivalence}
  Consider the setting in Theorem~\ref{thm3}. Let $\tilde{s} \colon
  \mathbb{R}^{d} \mapsto \mathbb{R}^{d-1}$ for $d \geq 2$ be defined
  by $\tilde{s}(\xi) = \xi_{2:d}/\xi_{1}$ for $\xi = (\xi_1, \xi_2,
  \dots, \xi_d)$. Denote by $\tilde{\mathbf{J}}$ the Jacobian
  transformation for $\tilde{s}$, i.e., $\tilde{\mathbf{J}}$ is the
  $(d-1) \times d$ matrix of the form
  $\tilde{\mathbf{J}}(\xi) = \tfrac{1}{\xi_1}\bigl[-\tilde{s}(\xi) \mid
  \mathbf{I} \bigr]$. 
  Now define the matrices
  \begin{gather}
    \label{eq:G_tilde_U_cov1}
    \widehat{\mathrm{var}}[\tilde{s}(\hat{U}_i) -
    \tilde{s}(\hat{U}_j)] = \tilde{\mathbf{J}}(\hat{U}_i) \hat{\boldsymbol{\Sigma}}(\hat{U}_i) \tilde{\mathbf{J}}(\hat{U}_i)^{\top} +
\tilde{\mathbf{J}}(\hat{U}_j) \hat{\boldsymbol{\Sigma}}(\hat{U}_j)
\tilde{\mathbf{J}}(\hat{U_j})^{\top}, \\
    \label{eq:G_tilde_U_cov2}
    \widehat{\mathrm{var}}[\tilde{s}(\hat{X}_i) -
    \tilde{s}(\hat{X}_j)]  =  \tilde{\mathbf{J}}(\hat{X}_i) \hat{\bm{\Sigma}}(\hat{X}_i)
\tilde{\mathbf{J}}(\hat{X}_i)^{\top} + \tilde{\mathbf{J}}(\hat{X}_j) \hat{\bm{\Sigma}}(\hat{X}_j)
\tilde{\mathbf{J}}(\hat{X}_j)^{\top},
\end{gather}
where the $d \times d$ matrices $\hat{\boldsymbol{\Sigma}}(\hat{U}_i)$ and
$\hat{\boldsymbol{\Sigma}}(\hat{U}_j)$ are as given in
Corollary~\ref{cor1} and the $d \times d$ matrices
$\hat{\boldsymbol{\Sigma}}(\hat{X}_i)$ and
$\hat{\bm{\Sigma}}(\hat{X}_j)$ are as given in Theorem~\ref{thm1}.
Then the test statistic based on the Mahalanobis distance for
$\tilde{s}(\hat{U}_i) - \tilde{s}(\hat{U}_j)$ is identical to 
that based on the Mahalanobis distance for
$\tilde{s}(\hat{X}_i) - \tilde{s}(\hat{X}_j)$, i.e.,
  \begin{equation}
    \label{eq:G_def_U}
    \begin{split}
    \tilde{G}_{\mathrm{ASE}}(\hat{U}_i,\hat{U}_j) & =
    n^2 \rho_n
    \bigl(\tilde{s}(\hat{U}_i) - \tilde{s}(\hat{U}_j)\bigr)^{\top}
    \Bigl(\widehat{\mathrm{var}}[\tilde{s}(\hat{U}_i) -
    \tilde{s}(\hat{U}_j)]\Bigr)^{-1}     \bigl(\tilde{s}(\hat{U}_i) -
    \tilde{s}(\hat{U}_j)\bigr) \\
    &=
n \bigl(\tilde{s}(\hat{X}_i) -
\tilde{s}(\hat{X}_j)\bigr)^{\top} 
\Bigl(\widehat{\mathrm{var}}[\tilde{s}(\hat{X}_i) - \tilde{s}(\hat{X}_j)]\Bigr)^{-1} 
  \bigl(\tilde{s}(\hat{X}_i) - \tilde{s}(\hat{X}_j)\bigr).
  \end{split}
\end{equation}
\end{proposition}
Note that the sparsity factor $\rho_n$ appeared in the scaling 
of both the Mahalanobis distance between $\hat{U}_i - \hat{U}_j$
(Eq.~\eqref{eq:T_tilde_ASE}) and the Mahalanobis distance
between $\tilde{s}(\hat{U}_i) - \tilde{s}(\hat{U}_j)$
(Eq.~\eqref{eq:G_def_U}). However, since $\rho_n$ also appeared in
the definition of $\hat{\bm{\Sigma}}(\hat{U}_i)$
(Eq.~\eqref{eq:hat_Sigma_hat_U}), these $\rho_n$ factors cancel out and
the calculation of the test statistic does not depend on $\rho_n$
(which is generally assumed to be unknown). 
The main reason for including $\rho_n$ in the statement of
Corollary~\ref{cor1} and Proposition~\ref{prop:equivalence} is that if
$\rho_n \rightarrow 0$ then the covariance matrix $\bm{\Sigma}(U_i)$
in Eq.~\eqref{eq:Sigma_U} remains bounded and this simplifies the
proof of these results; see Remark~\ref{4.2} for further discussions.

Given the equivalence of the test statistics in
Proposition~\ref{prop:equivalence}, the following result is analogous
to Theorem~\ref{thm3} and provide a limiting distribution for
the Mahalanobis distance of $\tilde{s}(\hat{X}_i) -
\tilde{s}(\hat{X}_j)$.
\begin{cor}
  \label{cor2}
  Consider the setting in Theorem~\ref{thm3} and let
  $\tilde{G}_{\mathrm{ASE}}(\hat{U}_i, \hat{U}_j)$ be the test
  statistic as defined in Eq.~\eqref{eq:G_def_U} of Proposition~\ref{prop:equivalence}.
Then under $\mathbb{H}_0 \colon X_i/\|X_i\| = X_j/\|X_j\|$ and for
$n \rightarrow \infty$ with $n \rho_n = \omega(\log n)$, we have
$$\tilde{G}_{\mathrm{ASE}}(\hat{U}_i, \hat{U}_j) \rightsquigarrow \chi_{d-1}^2.$$
Furthermore, let $\mu$ be a constant such that
$X_i/\|X_i\| \not = X_j/\|X_j\|$ satisfies a local alternative where
\begin{gather}
  \label{eq:local_alt_ase_cor2b}
n \bigl(\tilde{s}(Z_i) -
\tilde{s}(Z_j))^{\top}(\tilde{\mathbf{J}}(Z_i) \bm{\Sigma}(Z_i)
\tilde{\mathbf{J}}(Z_i)^{\top}\!  + \tilde{\mathbf{J}}(Z_j)
\bm{\Sigma}(Z_j) \tilde{\mathbf{J}}(Z_j)^{\top})^{-1}(\tilde{s}(Z_i) -
\tilde{s}(Z_j)) \rightarrow \mu.
\end{gather}
Then $\tilde{G}_{\mathrm{ASE}}(\hat{U}_i, \hat{U}_j)
\rightsquigarrow \chi_{d-1}^2\left(\mu\right)$.
Finally, the condition in
Eq.~\eqref{eq:local_alt_ase_cor2b} is equivalent to the condition
in Eq.~\eqref{eq:local_alt_ase2} and hence, by
Proposition~\ref{prop:equality_ncp}, also equivalent to the
condition in Eq.~\eqref{eq:indefinite_alt_cond2}.
\end{cor}
We now compare Corollary~\ref{cor1} and Corollary~\ref{cor2} to the corresponding results in
Theorem~1 through Theorem~4 of \citep{fan2019simple}.
\begin{enumerate}
\item The theoretical results in \citep{fan2019simple} assume that (1) the
  eigenvalues of the edge probabilities matrix $\mathbf{P}$ has
  distinct eigenvalues, (2) the block probabilities matrix
  $\mathbf{B}$ is of full-rank and (3) in the setting of the degree-corrected
  SBMs, the matrix $\mathbf{B}$ is assumed to be positive definite. These assumptions are not needed for the theoretical results in this
  paper, and by removing these assumptions, our results are applicable
  to all random graphs models whose edge probabilities matrix
  $\mathbf{P}$ exhibits a low-rank structure.   Note that the notations in our
  paper differ slightly from that in \citep{fan2019simple}. In particular
  the matrices $\mathbf{B}$
  and $\mathbf{P}$ in our paper correspond to the $\mathbf{P}$
  and $\mathbf{H}$ in \citep{fan2019simple}, respectively. 

\item The estimated covariance matrices $\hat{\bm{\Sigma}}(\hat{X}_i)$
  and $\tilde{\mathbf{J}}(\hat{X}_i) \hat{\bm{\Sigma}}(\hat{X}_i)
  \tilde{\mathbf{J}}(\hat{X}_i)^{\top}$ used in Corollary~\ref{cor1} and Corollary~\ref{cor2} can be
  written explicitly in terms of the adjacency spectral embedding
$\{\hat{X}_i\}$. The covariance matrices in \citep{fan2019simple}
are more complicated and requires debiasing of the eigenvalues. The
main reason behind this difference is because the
$\{\hat{X}_i\}$ already capture the joint dependence
between the eigenvalues and eigenvectors of $\mathbf{A}$ and thus
leads to a more direct estimator for the covariance. A more detailed
explanation of this difference is given below. 
\\
\\
Consider for example the expression for the covariance $\bm{\Sigma}(U_i)$ as given
in Eq.~\eqref{eq:Sigma_U} of Corollary~\ref{cor1}; the second part of
this expression using the rows of the eigenvectors $\mathbf{U}$ is
identical to that in Lemma~2 of
\citep{fan2019simple}.
Since the adjacency spectral embedding
$\{\hat{X}_i\}$ are consistent estimates for $X_i$, our estimate
$\hat{\bm{\Sigma}}(\hat{U}_i)$ given in Corollary~\ref{cor1} can be
constructed directly using the $\hat{X}_i$, e.g., $\hat{p}_{ik} = \hat{X}_i^{\top}
\mathbf{I}_{a,b} \hat{X}_k$ is a consistent estimate for $p_{ik}$, and
$\hat{X}_k \hat{X}_k^{\top}$ is a consistent estimate for
$Z_k Z_k^{\top}$. The challenges in working with
$\hat{\mathbf{X}}$ is in understanding how the non-identifiability in $\mathbf{X}$ affects
the estimation of the covariance matrices in Theorem~\ref{thm3} and
Corollary~\ref{cor2} and the resulting non-centrality parameters.
If we only use the $\hat{U}_k$ to construct our test
statistic as done in \citep{fan2019simple}, then we still need to estimate the variance
terms $p_{ik} ( 1 - p_{ik})$ in Eq.~\eqref{eq:Sigma_U};
\citep{fan2019simple} estimated these variances by using the squared
residuals $(a_{ik} - \hat{p}_{ik})^2$. These squared residuals are, however, biased
and therefore \citep{fan2019simple} had to first debiased the
residuals $a_{ik} - \hat{p}_{ik}$ through a one-step update for the
eigenvalues in $\hat{\mathbf{S}}_k$. In effect, by decoupling the analysis 
of $\hat{\mathbf{U}}$ from that of $\hat{\mathbf{S}}$, it became
quite harder to directly estimate the variance terms $\{p_{ik}(1 -
p_{ik})\}$.
\\
\par
In summary, the perspective of latent
positions estimation as is done in this paper brings some added
conceptual complexity in the analysis due to the non-identifiablity of
the latent positions but it pays dividend in simplifying the
theoretical results since the random graphs distribution are defined using the latent
positions $\mathbf{X}$ and thus almost all quantities associated with this distribution can be directly estimated using the
$\hat{\mathbf{X}}$. 

\item For testing the hypothesis $\mathbb{H}_0 \colon
  X_i/\|X_i\| = X_j/\|X_j\|$, we can use either the test statistic in
  Theorem~\ref{thm3} or the test statistic in
  Corollary~\ref{cor2}. The last statement in Corollary~\ref{cor2}
  indicates that these two test statistics have the same non-centrality
  parameters and hence they have the same limiting
  distributions under both the null and local alternative hypothesis. Nevertheless their finite-sample performance can be somewhat different; see Section~\ref{sec:comparison} for simulation results illustrating this claim. We also note that while there are numerous equivalent
  conditions for the non-centrality parameters, the condition 
  in Eq.~\eqref{eq:indefinite_alt_cond2} is much simpler than that
  of Eq.~\eqref{eq:local_alt_ase_cor2b}. Indeed, Eq.~\eqref{eq:local_alt_ase_cor2b} depends on
  $\tilde{\mathbf{J}}(Z_i)$ where $Z_i$ is
  only defined implicitly through the eigendecomposition of
  $\mathbf{X} \mathbf{I}_{a,b} \mathbf{X}^{\top}$. For an
  illustrative, albeit contrived example, suppose we know the
  functions $\bm{\Sigma}(X_i)$ and $\bm{\Sigma}(Z_i)$. It is then easy
  to see how the change in $X_i - X_j$ affects the 
  resulting non-centrality parameter $\mu$. In contrast it is not clear how 
  a change in $Z_i - Z_j$ impacts $\mu$ in Eq.~\eqref{eq:local_alt_ase_cor2b} since not all values of  
  $Z_i - Z_j$ are valid due to the constraint that $\mathbf{U}$
  have orthonormal columns. 
\item In \citep{fan2019simple}, the authors did not derive an expression
  for the non-centrality parameter for the test statistic
  $\tilde{G}_{\mathrm{ASE}}$ as given in
  Corollary~\ref{cor2} of the 
  current paper; rather, in the context of degree-corrected mixed-membership SBM,
  Theorem~3 of \citep{fan2019simple} shows a slightly weaker result in that
  the power of $\tilde{G}_{\mathrm{ASE}}$
  converges to $1$ whenever $\lambda_2(\pi_i \pi_i^{\top} + \pi_j
  \pi_j^{\top}) \gg (n \rho_n)^{-1}$; here $\lambda_2(\cdot)$ denotes
  the second largest eigenvalue of $(\cdot)$.
\end{enumerate}

\subsection{Model selection for block models}
\label{sec:model_selection}
An important class of inference problems in networks analysis is that of model
selection wherein, given an observed graph and a set of candidate
models, choose the most parsimonious model from which the graph
might have been generated. A popular example of model selection is
in determining the number of communities in a stochastic block model
graph, and there are numerous procedures developed for this
problem, including those based on spectral information, BIC,
cross-validation and likelihood ratio statistics. See for
example  
\citep{bickel2016hypothesis,hu2020corrected,le2015estimating,lei2016goodness,ma2021determining,wang2017likelihood}
and the references therein.

Another simple yet non-trivial model selection problem
is to decide whether an observed graph is generated from
a SBM versus a degree-corrected SBM. For this problem,
\citep{yan2014model} propose a log-likelihood ratio test in
the setting of Poisson stochastic block model and showed that when the
graph is sparse, the distribution of the log-likelihood ratio does
{\em not} converge to a chi-square random variable as commonly seen in classical
statistics due to the high-dimensionality of the parameters.
Nevertheless, they derive the unbiased estimations
of log-likelihood ratio's mean and variance in the limit of large
graphs to determine the appropriate threshold. Meanwhile \citep{chen2018network}
and \citep{li2020network} develop efficient cross-validation
approaches to do model selection. For example, \citep{chen2018network} propose 
a network cross-validation approach based on a block-wise
node-pair splitting technique together with community recovery
using spectral decomposition followed by {\em k}-means
clustering. By choosing the best model as the one that minimizes the
cross-validation loss, this method is not only able to choose between
the SBM and degree-corrected SBM, but also determine the number of
blocks; see also Algorithm~3 in \citep{li2020network}. 
Nevertheless neither \citep{chen2018network}
nor \citep{li2020network} provide test statistics with known limiting distributions for deciding between a SBM and a degree-corrected SBM.

In this paper, leveraging the limit results in Theorem~\ref{thm1} and
Theorem~\ref{thm3}, we propose another test statistic for selecting between the standard stochastic
block model (SBM) and a degree-corrected stochastic block model
(DCSBM) as follows. Given a graph generated from either a SBM or a
DCSBM, let $\mathbf{\Theta}=\operatorname{diag}(\theta_1,\ldots,\theta_n)$ be the
degree heterogeneity matrix; note that $\bm{\Theta} = c\mathbf{I}$ for some
constant $c > 0$ whenever the graph is a SBM.
We are thus interested in testing the hypothesis
\begin{equation}
  \label{eq:SBM_vs_DCSBM}
\mathbb{H}_{0}: \theta_1= \theta_2 = \dots = \theta_n, \quad
\text{versus} \quad \mathbb{H}_{A}:\theta_i \neq \theta_j \; \text{for
  at least one pair of}\; i,j
\end{equation}
If the true generative model is a SBM then
the $p$-values of the test statistic in Theorem \ref{thm1} for any
randomly selected pairs of nodes are (asymptotically) uniformly distributed
on $[0,1]$. We therefore have the following result.
\begin{theorem}\label{thm6}
Let $\mathbf{A}$ be either a $K$-blocks SBM graph or a $K$-blocks
degree-corrected SBM graph on $n$ vertices, where $K$ is either
assumed known or is consistently estimated.
Cluster the vertices of $\mathbf{A}$ into $K$ clusters using any clustering
algorithm that guarantees strong or perfect recovery.  Now for $m \geq 1$, randomly select $m$ different
pairs of nodes in each of the estimated $K$ communities and apply the test statistics in Theorem \ref{thm1}. Let $\xi_{ki}$ for $k=1,...,K$ and $i=1,...,m$ be the 
resulting $p$-values, i.e., $p_{ki}$ is the p-value of the test
statistic for the $i$th selected pair in the $k$th block.
Next let $\zeta_k=\max_{i} \xi_{ki}$ be the maximum of the $p$-values for
the $k$th block and define the test statistic
$$S_1=-2m\sum_{k=1}^K\log(\zeta_k).$$
Then under $\mathbb{H}_{0}: \theta_1 = \theta_2 = \dots = \theta_n$ and for
$n \rightarrow \infty$ with $n \rho_n = \omega(\log n)$, we have
$S_1 \rightsquigarrow \chi_{2K}^2.$
\end{theorem}
Theorem~\ref{thm6} depends on the fact that we can consistently estimate the number of communities $K$ 
as well as consistently recovery the community assignments. Examples
of procedures for estimating $K$ are discussed above and 
clustering algorithms that guarantee strong recovery include those
based on semidefinite programming, spectral clustering, and two-step
likelihood based approaches \citep{abbe2017community, gao_optimal}.

\begin{remark}
  \label{rem:model selection}
The test statistic in Theorem~\ref{thm6} can also be used to test the hypothesis that the number of communities in a SBM graph is $K$ against the number of communities is $K^{\prime}$ where $K^{\prime} > K$. In the special case where $K=1$, we are testing if a graph has one or more communities
(Erdős–Rényi graph vs SBM). A more powerful test can be defined via another form of Fisher combination of $p$-values. For $m \geq 1$, randomly select $m$ different pairs of nodes and apply the test statistics in Theorem~\ref{thm1} with $d=1$. Let $\xi_{k}$ for $k=1,...,m$ be the 
resulting $p$-values.
Now define the test statistic
$$S_2=-2\sum_{k=1}^m\log(\xi_k).$$
Then under $\mathbb{H}_{0}: K = 1$ and for
$n \rightarrow \infty$ with $n \rho_n = \omega(\log n)$, we have
$S_2 \rightsquigarrow \chi_{2m}^2$. This limiting property is established in the same way as Theorem~\ref{thm6} and numerical simulations indicated that the proposed test is able to determine the existence of the community structure correctly most of the time. 
\end{remark}

\section{Test Statistics Using Laplacian Spectral Embedding}
\label{sec:LSE}
In this section we study test statistics based on the Mahalanobis
distances between the Laplacian spectral embedding $\breve{X}_i$ and
$\breve{X}_j$. We obtain analogous results to those in
Section~\ref{sec:ase_test1} and Section~\ref{sec:degree_corrected} for
the adjacency spectral embedding. For conciseness we will only present result for the hypothesis test $\mathbb{H}_0 \colon X_i =
X_j$ here; test statistic for the degree-corrected hypothesis is discussed in Appendix~\ref{sec:app_lse}. 
\begin{theorem}\label{thm2}
 Recall the setting of Theorem~\ref{thm1}. Let $d_i$ denote the degree
 of the $i$th node and define
 $$\breve{Z}_{ik} =\frac{(\breve{\mathbf{X}}^{\top} \breve{\mathbf{X}})^{-1}
  \breve{X}_k}{\sqrt{d_k}}-\frac{\mathbf{I}_{a,b}\breve{X}_i }{2\sqrt{d_i}}.$$
  Now consider the test statistic
\begin{equation}
\label{th2_1}
T_{\mathrm{LSE}}(\breve{X}_i, \breve{X}_j)
=n^2 \rho_n\bigl(\breve{X}_i-\breve{X}_j\bigr)^{\top}\Bigl(\breve{\boldsymbol{\Sigma}}(\breve{X}_i)
+ \breve{\boldsymbol{\Sigma}}(\breve{X}_j)\Bigr)^{-1}\bigl(\breve{X}_i - \breve{X}_j \bigr).
\end{equation}
where $\breve{\bm{\Sigma}}(\breve{X}_i)$ and
$\breve{\bm{\Sigma}}(\breve{X}_j)$ are matrices of the form
\begin{equation}
\label{th2_2}
\breve{\mathbf{\Sigma}}(\breve{X}_i)=
  n^2 \rho_n \mathbf{I}_{a,b} \Bigl[\sum_{k=1}^n \breve{Z}_{ik}
  \breve{Z}_{ik}^{\top} \Bigl(\frac{\sqrt{d_k}}{\sqrt{d_i}}
  \breve{X}_i^{\top} \mathbf{I}_{a,b} \breve{X}_k - d_k
  (\breve{X}_i^{\top} \mathbf{I}_{a,b} \breve{X}_k)^2 \Bigr) \Bigr] \mathbf{I}_{a,b}. 
\end{equation}
Then under the null hypothesis $\mathbb{H}_0 \colon X_i=X_j$ and for
$n \rightarrow \infty$ with $n \rho_n = \omega(\log n)$, we have
$$T_{\mathrm{LSE}}(\breve{X}_i, \breve{X}_j) \rightsquigarrow \chi_d^2.$$
Next let $t_i = \sum_{j} \rho_n X_i^{\top} \mathbf{I}_{a,b} X_j$ be
the expected degree of the $i$th node and let $\mathbf{T} =
\mathrm{diag}(t_1, t_2, \dots, t_n)$ be the diagonal matrix of
expected degrees. Also let
$$\zeta_{ik} = \rho_n^{1/2}\Bigl( \frac{(\rho_n \mathbf{X}^{\top} \mathbf{T}^{-1}
  \mathbf{X})^{-1} X_{k}}{t_k} - \frac{\mathbf{I}_{a,b} X_i}{2t_i}\Bigr)$$ 
Now define
\begin{equation*}
 \tilde{\mathbf{\Sigma}}(X_i) = \frac{n^2 \rho^2_n}{t_i} 
    \mathbf{I}_{a,b} \Bigl[\sum_{k=1}^n
    \zeta_{ik} \zeta_{ik}^{\top} X_i^{\top} \mathbf{I}_{a,b} X_k
    (1 - \rho_n X_i^{\top} \mathbf{I}_{a,b} X_k) \Bigr] \mathbf{I}_{a,b}.
  \end{equation*}
Let $\mu > 0$ be a finite constant such that $X_i \not = X_j$ satisfies a local
alternative where
\begin{equation}
  \label{eq:alt_cond_lse}
  n^2 \rho_n^2\Bigl(\frac{X_i}{\sqrt{t_i}}-
\frac{X_j}{\sqrt{t_j}}\Bigr)^{\top}\bigl(\tilde{\boldsymbol{\Sigma}}(X_i)
+ \tilde{\boldsymbol{\Sigma}}(X_j)\bigr)^{-1}\Bigl(\frac{X_i}{\sqrt{t_i}}-
\frac{X_j}{\sqrt{t_j}}\Bigr)
\rightarrow \mu
\end{equation}
Then we have
$T_{\mathrm{LSE}}(\breve{X}_i, \breve{X}_j) \rightsquigarrow \chi_d^2(\mu)$ where $\chi_d^2\left(\mu\right)$ is the
noncentral chi-square distribution with $d$ degrees of freedom and noncentrality parameter $\mu$.
\end{theorem}
\begin{remark}
\label{4.2}
The sparsity factor $\rho_n$, which is generally unknown, appeared in both Eq.~\eqref{th2_1}
and Eq.~\eqref{th2_2} and thus cancels out. The
main reason for including $\rho_n$ in Eq.~\eqref{th2_2} is that if
$\rho_n \rightarrow 0$, then the covariance matrix as defined in
Eq.~\eqref{th2_2} remains bounded, i.e., the entries of
$\breve{\bm{\Sigma}}(\breve{X}_i)$ do not diverge to $\infty$. 
The boundedness of $\breve{\bm{\Sigma}}(\breve{X}_i)$ simplifies
the exposition in the proof of Theorem~\ref{thm2}. Similarly, the
factor $\rho_n^2$ appears in both the definition of
$\tilde{\bm{\Sigma}}(X_i)$ as well as the condition for $\mu$ in
Eq.~\eqref{eq:alt_cond_lse} and this is also done so that $\tilde{\bm{\Sigma}}(X_i)$ is bounded as $n
\rightarrow \infty$. Note that
$\breve{\bm{\Sigma}}(\breve{X}_i)$ is a consistent estimate of the
covariance matrix $\tilde{\bm{\Sigma}}(X_i)$; while Eq.~\eqref{th2_2}
appears somewhat complicated, we chose this representation because it
can be computed directly from the Laplacian spectral embeddings $\{\breve{X}_i\}$
together with the observed degrees $\{d_i\}$. Finally, similar to the
discussion in Remark~\ref{rem:thm1}, the condition
in Eq.~\eqref{eq:alt_cond_lse} is equivalent to $X_i = X_j + v$ where
$$\frac{n^2 \rho_n^2}{2 t_i} v^{\top}\bigl(\tilde{\bm{\Sigma}}(X_i)\bigr)^{-1}
v \rightarrow \mu.$$
Since $t_i$ is growing at order $\Theta(n \rho_n)$, the above
condition is analogous to that of Eq.~\eqref{eq:noncentral1}.
\end{remark} 

\begin{remark}
\label{rem:ncp_eq}
Theorem~\ref{thm1} and Theorem~\ref{thm2} show that 
the test statistics $T_{\mathrm{ASE}}$ and $T_{\mathrm{LSE}}$ both
converge to chi-squared random variables with the same degrees of
freedom under the null and local alternative
hypothesis. Since the power of the test
statistics under the alternative hypothesis is a monotone
increasing function of the non-centrality parameter, it is natural
to compare the non-centrality parameters for $T_{\mathrm{ASE}}$
against that of $T_{\mathrm{LSE}}$. We can see from the numerical
simulations in Section~\ref{sec4} that these non-centrality parameters
are almost identical. Derivations of the exact theoretical
relationships between these non-centrality parameters appear quite
challenging. At the current moment we are only able to show
that, for balanced SBMs where all diagonal elements of the block
probabilities matrices are equal to $p$ and all off-diagonal elements are
equal to $q$ with $p \not = q$ then, with $\rho_n \rightarrow 0$, the non-centrality
parameter for $T_{\mathrm{LSE}}$ is equal to the non-centrality
parameter for $T_{\mathrm{ASE}}$ plus a small but non-vanishing
constant. See Appendix~\ref{appb.8} for more details. 
 \end{remark}

\section{Simulation Studies}
\label{sec4}
We now conduct simulations to investigate the finite
sample performance of the proposed test statistics. For conciseness we
only focus on the undirected case here; simulation results for the
directed case are presented in Section~\ref{sec:directed_simu} of the
Appendix. 

We consider two mixed membership stochastic block model settings, with and without
degree correction factors. Both settings assume that the block probabilities matrix is a $3
\times 3$ matrix of the form 
$\mathbf{B}= 0.9 \bm{1} \bm{1}^{\top} - 0.6 \mathbf{I}$; note that $\mathbf{B}$ has one positive and two negative
eigenvalues. The first model setting (Model I) is
a mixed membership SBM setting {\em without} degree correction
factors where the nodes are assume to have 
one of seven possible membership vectors, namely
$$\pi_i \in \{(1,0,0),(0,1,0),(0,0,1),(0.5, 0.3, 0.2), (0.3, 0.2,
0.5), (0.2, 0.5, 0.3), (1 - 2\epsilon, \epsilon, \epsilon)\}$$
The value of $\epsilon \in (0,1/2)$ will be specified later.
Vertices $v_i$ with $\pi_i \in \{(1,0,0),(0,1,0),(0,0,1)\}$
represent pure nodes, i.e., nodes that belong to a single
community. The second model setting (Model II) is obtained by introducing degree
correction factors $\{\theta_i\}$ to Model I. Here the $\{\theta_i\}$ are independent samples from the uniform distribution
on the interval $\left[1,k\right]$ for some choice of $k \in \{1.1,
1.2,\dots,2\}$ that will be specified later. 


\subsection{Estimated powers for membership testing }
\label{s5.1}
\begin{table*}
\caption{Empirical estimates for the size and power for the test
statistics $T_{\mathrm{ASE}}$ and $T_{\mathrm{LSE}}$ for
mixed-membership stochastic block model graphs with various
choices of sparsity parameter $\rho$.
The null hypothesis corresponds to $\pi_i = \pi_j = (0.5, 0.3, 0.2)$ and the alternative hypothesis corresponds to $\pi_i =
(1,0,0)$ and $\pi_j = (1 - 10n^{-1/2}, 5n^{-1/2}, 5n^{-1/2})$. The
rows with labels $\mathrm{ncp}$ are the non-centrality parameters $\mu$ for the local alternative hypothesis. 
}
\label{T2}
\begin{tabular}{@{}ccccccccc@{}}
\hline
$\rho$& 0.3& 0.4& 0.5& 0.6& 0.7&0.8&0.9&1.0 \\
\hline
Size ($T_{\mathrm{ASE}}$)  &0.064& 0.068& 0.056& 0.040&0.064&0.048& 0.058&0.064 \\
Size ($T_{\mathrm{LSE}}$)  & 0.070&0.060& 0.070& 0.038& 0.056&0.044&0.048&0.048 \\ 
Power ($T_{\mathrm{ASE}}$) & 0.330& 0.460& 0.580& 0.730 &0.840&0.942 &0.986 &0.998 \\ 
Power ($T_{\mathrm{LSE}}$) & 0.336& 0.460& 0.586& 0.704 &0.818&0.916 &0.960 &0.998 \\ 
ncp ($T_{\mathrm{ASE}}$)& 3.7548& 5.3640 &7.2360&9.4608 &12.1879 &15.6971&20.6098&28.7589 \\
ncp ($T_{\mathrm{LSE}}$)& 3.7575& 5.3674 &7.2399&9.4648 &12.1915 &15.6990&20.6071&28.7427 \\
Theoretical Power ($T_{\mathrm{ASE}}$) & 0.3380& 0.4694& 0.6055& 0.7351 &0.8464&0.9293 &0.9786 &0.9976 \\ 
Theoretical Power ($T_{\mathrm{LSE}}$) & 0.3382& 0.4696& 0.6057& 0.7353 &0.8465&0.9293 &0.9786 &0.9976 \\    
\hline
\end{tabular}
\end{table*}

\label{sec4.1}
Given a graph generated from either Model I or Model II described above
we wish to test the hypothesis that two given nodes $i$ and $j$
have the same latent position, possibly up to scaling. We
consider test statistics based on both the adjacency and Laplacian
spectral embedding and we evaluate their size and power through
simulations. For Model I we set the null hypothesis as $\bm{\pi}_i = \bm{\pi}_j =
(0.5,0.3,0.2)$ and the alternative hypothesis as $\bm{\pi}_i = (1,0,0)$ and
$\bm{\pi}_j = (1 - 2cn^{-1/2}, cn^{-1/2}, cn^{-1/2})$ for some
constant $c$. Here $n$ is the number of vertices in the graph and
hence the alternative hypothesis represents a local alternative. 
Meanwhile, for Model II, once again, we set the null hypothesis as $\bm{\pi}_i = \bm{\pi}_j =
(0.5,0.3,0.2)$ and the alternative hypothesis as 
$\bm{\pi}_i = (1,0,0)$ and $\bm{\pi}_j = (1 - 2cn^{-1/2}, cn^{-1/2},
cn^{-1/2})$.
We set the significance level $\alpha$ to be $0.05$, i.e., the rejection
regions for our test statistics
are given by the $95$th percentile of the
chi-square distributions with appropriate degrees of freedom. Empirical estimates of
the size and power are based on $500$ Monte Carlo replicates. 

We first generate graphs on $n = 3100$ vertices according to the mixed
membership SBM in Model I. Among these $3100$ vertices, $100$ vertices
are assigned to have membership vector $\pi_i = (1 - 2cn^{-1/2},
cn^{-1/2}, cn^{-1/2})$ with $c = 5$, and the remaining $3000$ vertices
are equally assigned to the remaining membership vectors.
The empirical size and power of the test statistics $T_{\mathrm{ASE}}$
and $T_{\mathrm{LSE}}$ for testing $\mathbb{H}_0 \colon \pi_i = \pi_j$
against $\mathbb{H}_A \colon \pi_i \not = \pi_j$, under various choices
of sparsity factors $\rho$, are reported in
Table~\ref{T2}; the empirical size here refers to the null rejection
rate. Table~\ref{T2} also report the large-sample, {\em
theoretical} power computed according to the non-central chi-square
distribution with non-centrality parameters given in
Theorem~\ref{thm1} and Theorem~\ref{thm2}. We see that the empirical
estimates of the power are almost identical to the true theoretical
values. In addition, Figure~\ref{f2} in Appendix~\ref{App:plots} plots the empirical histograms
for $T_{\mathrm{ASE}}$ and $T_{\mathrm{LSE}}$ under the null
hypothesis for $\rho=1.0$ and show that the distributions of
$T_{\mathrm{ASE}}$ and $T_{\mathrm{LSE}}$ are well-approximated by the
$\chi^2_3$ distribution.
\begin{table*}[htbp]
\caption{Empirical estimates for the size and power for the test
statistics $G_{\mathrm{ASE}}$ and $G_{\mathrm{LSE}}$ for
the degree-corrected mixed-membership stochastic block model graphs with various
choices of degree heterogeneity parameter $k$. The alternative hypothesis corresponds to $\pi_i =
(1,0,0)$ and $\pi_j = (1 - 10n^{-1/2}, 5n^{-1/2}, 5n^{-1/2})$.
The rows with labels $\mathrm{ncp}$ are the non-centrality parameters $\mu$ for
the local alternative hypothesis as $k$ changes.
}
\label{T5}
\begin{tabular}{@{}ccccccccc@{}}
\hline
$k$& 1.3& 1.4& 1.5& 1.6& 1.7& 1.8& 1.9& 2.0\\
\hline
Size ($G_{\mathrm{ASE}}$) &0.082& 0.060& 0.062& 0.062& 0.068& 0.038& 0.058&0.062 \\
Size ($G_{\mathrm{LSE}}$) &0.088& 0.062& 0.058& 0.060& 0.072& 0.038& 0.060&0.056 \\
Power ($G_{\mathrm{ASE}})$ &0.472& 0.566& 0.602& 0.592& 0.672& 0.668 &0.768 &0.824 \\ 
Power ($G_{\mathrm{LSE}}$) &0.466& 0.566& 0.604& 0.582& 0.670& 0.662 &0.776 &0.832 \\ 
ncp ($G_{\mathrm{ASE}}$)& 4.5332& 5.3192 &6.4099&6.4703 &7.2737 &6.7527&9.2965&11.6185 \\
ncp ($G_{\mathrm{LSE}}$)& 4.5607& 5.3535 &6.4497&6.5202 &7.3236 &6.8013&9.3074&11.5597 \\
Theoretical Power ($G_{\mathrm{ASE}}$) & 0.4634& 0.5302& 0.6144& 0.6188 &0.6734&0.6386 &0.7848 &0.8722 \\ 
Theoretical Power ($G_{\mathrm{LSE}}$) & 0.4658& 0.5331& 0.6173& 0.6223 &0.6766&0.6420 &0.7853 &0.8705 \\  
\hline
\end{tabular}
\end{table*}

We next generate graphs on $n = 3100$ vertices according to the {\em degree-corrected} mixed membership SBM
in Model II. Once again there are $100$ vertices assigned to have
membership vectors $\pi_i = (1 - 2cn^{-1/2}, cn^{-1/2}, cn^{-1/2})$
with $c = 5$ and the remaining $3000$ vertices are equally assigned to the
remaining membership vectors. 
The empirical size and power of the test statistics $G_{\mathrm{ASE}}$ (Theorem~\ref{thm3}) and
$G_{\mathrm{LSE}}$ (Theorem~\ref{thm4} in Appendix), for $\rho = 0.25$ and various
choices of $k \in \{1.3,1.4,\dots,2\}$, are reported in
Table~\ref{T5}. Figure~\ref{f3} in Appendix~\ref{App:plots} plots the empirical histograms for
$G_{\mathrm{ASE}}$ and $G_{\mathrm{LSE}}$ under the  
null hypothesis when the degree heterogeneity parameters $\theta_i$ are
uniformly distributed in the interval $[1,2]$ and the sparsity factor
is $\rho = 0.25$. Once again we see that the empirical
estimates of the power are almost identical to the true theoretical
values and furthermore the distributions of
$G_{\mathrm{ASE}}$ and $G_{\mathrm{LSE}}$ are well-approximated by the
$\chi^2_2$ distribution. 

We also note that, in comparison to Table~\ref{T2},
Table~\ref{T5} shows more deviation between the theoretical and empirical
powers. We surmise that this is due to the additional variability in
the latent positions of a degree-corrected SBM compared to those for a SBM (indeed, 
two nodes in the same community of a DCSBM can have quite different
degrees profile). Estimation of the latent positions in a
DCSBM is therefore generally less accurate than those for a SBM and
thus we expect $G_{\mathrm{ASE}}$ (resp. $G_{\mathrm{LSE}}$) to
converge to the limiting $\chi^2_{d-1}$
somewhat slower than the convergence of $T_{\mathrm{ASE}}$ (resp. $T_{\mathrm{LSE}}$) to $\chi^2_{d}$.
\subsection{Model selection}
We now examine how the previous test statistics can be used to choose 
between the stochastic block model and the degree-corrected stochastic
block model. We perform $500$ Monte Carlo replicates where, in each
replicate, we do the following steps.
\begin{enumerate}
  \item Generate a $3$-blocks stochastic block model graph on $n = 1500$
vertices, equal block sizes, and block probabilities matrix $\mathbf{B} = 0.9 \bm{1}
\bm{1}^{\top} - 0.6 \mathbf{I}$.

\item Embed the graph into $\mathbb{R}^3$ using adjacency spectral embedding and then cluster
  these embedded vertices into $K = 3$ communities.
\item Select $m = 10$ pairs
of nodes from each community and compute $T_{\mathrm{ASE}}$ for each pair.

\item Convert these test statistic values into $p$-values based on the quantiles of the $\chi^2_3$
distributions. Compute the test statistic $S_1$ as defined in
Theorem~\ref{thm6} using these $p$-values.

\item Reject the null hypothesis that the graph is a $3$-blocks SBM
  graph if $S_1 > \chi^2_{6, 0.95}$, the $95$ percentile of the chi-square
  distribution with $6$ degrees of freedom.
  \end{enumerate}

We then perform another $500$ Monte Carlo replicates of the above
steps, except that we now allow for degree heterogeneity by sampling,
in addition to the above SBM parameters, 
a sequence of degree correction factors $\{\theta_1, \theta_2, \dots,
\theta_n\}$ which are iid uniform random variables in the interval $[1,k]$
for $k \in \{1.3, 1.4, \dots, 2\}$.
The number of times we reject the null hypothesis among the first and
second batch of these $500$ replicates is an estimate of the
significance level and power, respectively, for using
$S_1$ as a goodness of fit test for deciding between a SBM and a
degree-corrected SBM.
Table~\ref{T7} and Table~\ref{T8} reported the empirical size and power for various values of $\rho \in
\{0.3,0.4,\dots,1\}$ (under the null hypothesis) and $k \in \{1.3,
1.4,\dots,2\}$ (under the alternative hypothesis),
respectively. The results in Tables~\ref{T7} and \ref{T8} indicate
that the proposed model selection procedure frequently chooses the
correct generative model for the observed graphs.

\begin{table*}[htbp]
\caption{Empirical estimates of the significance level for using $S_1$
as a goodness of fit test statistic for deciding between a stochastic
block model and a degree-corrected stochastic block model.}
\label{T7}
\begin{tabular}{@{}ccccccccc@{}}
\hline
$\rho$&0.3& 0.4& 0.5& 0.6& 0.7& 0.8& 0.9& 1.0\\
\hline
Size& 0.072& 0.056& 0.044& 0.044& 0.068&0.056& 0.050&0.068 \\
\hline
\end{tabular}
\end{table*}

\begin{table*}[htbp]
\caption{Empirical estimates of the power for using $S_1$
as a goodness of fit test statistic for deciding between a stochastic
block model and a degree-corrected stochastic block model.}
\label{T8}
\begin{tabular}{@{}ccccccccc@{}}
\hline
$k$& 1.3& 1.4& 1.5& 1.6& 1.7& 1.8& 1.9& 2.0\\
\hline
Power& 0.400& 0.584& 0.674& 0.778 &0.846 &0.874 &0.904 &0.940\\ 
\hline
\end{tabular}
\end{table*}

\subsection{Power Comparison with Test Statistics in \citep{fan2019simple}}
\label{sec:comparison}
We discussed in Section~\ref{sec:related_work} the relationship
between our proposed test statistics and those studied in
\citep{fan2019simple}; in particular the test statistics
$T_{\mathrm{ASE}}$ and $\mathrm{G}_{\mathrm{ASE}}$ are asymptotically
equivalent to those studied in \citep{fan2019simple}. 
We now conduct numerical simulations to compare the finite sample
power of these test statistics under local alternatives. 
We used the same settings as those presented for Model I and Model II
in Section~\ref{s5.1}, except that the block probabilities matrix
$\mathbf{B}$ is now set to $\mathbf{B}= 0.3 \bm{1} \bm{1}^{\top} + 0.6
\mathbf{I}$. We chose this $\mathbf{B}$ because the theoretical results in
\citep{fan2019simple} require $\mathbf{B}$ to be
positive-semidefinite. The results are presented in Table~\ref{T5.31}
and Table~\ref{T5.32}. Table~\ref{T5.31} indicates that, for the Model I
setting, the (empirical) powers for all test statistics are almost
identical. In contrast, Table~\ref{T5.32} shows discernible
differences between these test statistics
for Model II. In particular our test statistics have higher
(finite-sample) power compared to those of \citep{fan2019simple};
these difference are statistically significant (confirmed via
McNemar's test \citep{mcnemar}). 
\begin{table}[htbp]
\begin{center}
\caption{Empirical estimates for the power of the test
statistics $T_{\mathrm{FAN}}$, $T_{\mathrm{ASE}}$ and $T_{\mathrm{LSE}}$ for mixed membership stochastic
block model graphs under local alternative hypothesis corresponding to $\pi_i =
(1,0,0)$ and $\pi_j = (1 - 10n^{-1/2}, 5n^{-1/2}, 5n^{-1/2})$ with various
choices of sparsity parameter $\rho$. The estimates are based on $500$
Monte Carlo replicates.}
\begin{tabular}{ccccccccccc}
\hline 
$\rho$& 0.1& 0.2& 0.3& 0.4& 0.5&0.6& 0.7&0.8&0.9&1.0\\ 
\hline 
Power ($T_{\mathrm{FAN}}$) & 0.270& 0.328& 0.450& 0.570& 0.696 & 0.770& 0.906& 0.954& 0.988& 1\\ 
Power ($T_{\mathrm{ASE}}$) & 0.280& 0.322& 0.448& 0.564& 0.696 & 0.776& 0.906& 0.954& 0.988& 1\\        
Power ($T_{\mathrm{LSE}}$) & 0.292& 0.328& 0.448& 0.564& 0.702 & 0.780& 0.910& 0.954& 0.990& 1\\ 
\hline
\end{tabular}
\label{T5.31}
\end{center}
\end{table}

\begin{table}[htbp]
\begin{center}
\caption{Empirical estimates for the power of the test
statistics $G_{\mathrm{FAN}}$, $G_{\mathrm{ASE}}$ and $G_{\mathrm{LSE}}$ for degree-corrected mixed membership stochastic
block model graphs under local alternative hypothesis corresponding to $\pi_i =
(1,0,0)$ and $\pi_j = (1 - 10n^{-1/2}, 5n^{-1/2}, 5n^{-1/2})$ with various
choices of degree heterogeneity parameter $k$. The estimates are based on $500$
Monte Carlo replicates.}
\begin{tabular}{ccccccccccc}
\hline  
$k$& 1.1& 1.2& 1.3& 1.4& 1.5&1.6& 1.7&1.8&1.9&2.0\\
\hline 
Power ($G_{\mathrm{FAN}}$) &0.304 &0.356 &0.430 & 0.414&0.452 & 0.564& 0.642& 0.640& 0.444&0.834\\ 
Power ($G_{\mathrm{ASE}}$) &0.356 &0.404 &0.466 & 0.452&0.484 & 0.594& 0.664& 0.678& 0.480&0.840\\          
Power ($G_{\mathrm{LSE}}$) &0.354 &0.408 &0.470 & 0.456&0.504 & 0.592& 0.678& 0.666& 0.492&0.838\\  
\hline
\end{tabular}
\label{T5.32}
\end{center}
\end{table}

\section{Real Data Analysis}
\label{sec5}
\subsection{U.S. Political Blogs data}
\label{sec:pol}
We now analyze a network of U.S. Political Blogs as compiled in
\citep{adamic2005political}. This directed network contains snapshots of 1494 web blogs on US politics recorded in 2005. Each blog is represented by a node and a (directed) link between two nodes indicates the presence of a hyperlink between them. Blogs are labelled as being either liberal or conservative by self-reported or automated categorizations or by manually looking at the incoming and outgoing links and posts of each blog around the time of the 2004 presidential election. While the resulting labels are not $100\%$ exact, they are still
reasonably accurate and will serve as the assumed ground truth for the
current analysis. 

We first consider its undirected version. We do some pre-processing on the data, namely (1) we keep only
the largest connected component (2) we convert directed edges to
be undirected and (3) we remove all multi-edges and loops. We then embed the resulting network into $a = d = 2$ dimensions. The choice of $a = d = 2$ is determined by looking at a scree plot of the eigenvalues. We then apply the test statistics in
Section~\ref{sec3} and Section~\ref{sec:LSE} to test the hypothesis
that a given pair of nodes have the same latent positions. 
Due to the large number of possible pairs of
nodes, we randomly
choose 1000 pairs of nodes within the same community and 1000 pairs in 
different communities to perform our test. The resulting sensitivity
and specificity are reported in Table~\ref{T10}.
Once again, the threshold for classifying a pair of vertices as
  having the same latent positions is based on the $95\%$ percentile
  of the chi-square distribution with the appropriate degrees of freedom.
Table~\ref{T10}
indicates that $T_{\mathrm{ASE}}$ and
$G_{\mathrm{ASE}}$ perform reasonably well; nevertheless $G_{\mathrm{ASE}}$ is
preferable to $T_{\mathrm{ASE}}$ as it has both high sensitivity and
high specificity.

\begin{table*}[htbp]
\caption{Sensitivity and specificity of different test statistics for
  the U.S. Political Blogs data. The threshold for classifying a pair of vertices as
  having the same latent positions is based on the $95\%$ percentile
  of the $\chi^2_{2}$ distribution (for $T_{\mathrm{ASE}}$ and
  $T_{\mathrm{LSE}}$) and $\chi^2_{1}$ distribution (for
  $G_{\mathrm{ASE}}$ and $G_{\mathrm{LSE}}$).}
\label{T10}
\begin{tabular}{@{}ccccc@{}}
\hline
&$T_{\mathrm{ASE}}$& $T_{\mathrm{LSE}}$& $G_{\mathrm{ASE}}$& $G_{\mathrm{LSE}}$\\
\hline
Sensitivity&0.839& 0.643& 0.653& 0.007\\
Specificity&0.335& 0.354& 0.749& 0.989\\
\hline
\end{tabular}
\end{table*}

The sensitivity and specificity values for $G_{\mathrm{ASE}}$
suggests that a degree corrected SBM is
a better model than a vanilla SBM for this political blogs network.
We verify this hypothesis by performing the model selection
procedure discussed in Section~\ref{sec:model_selection}. More
specifically we embed the graph into $\mathbb{R}^2$ using adjacency
spectral embedding and then cluster these embedded vertices into $K =
2$ communities. We then select $m = 10$ pairs of nodes
from each cluster and apply the test statistic $T_{\mathrm{ASE}}$ to each of these pairs
and get the $p$-values. Then the test statistic $S_1$ as defined in
Theorem~\ref{thm6} is computed using these $p$-values. We repeat the
above procedure $1000$ times, each time choosing a different random set of $m =
10$ pairs of nodes from each cluster. Among these 1000 Monte
Carlo replicates, we reject the null hypothesis that the stochastic
block model is a proper fit for the political blogs network more than
600 times. On the other hand, if we set the degree-corrected model as the
null model and repeat the above procedure except that we use $G_{\mathrm{ASE}}$
instead of $T_{\mathrm{ASE}}$ to get the $p$-values, we find that we reject the
null hypothesis less than $100$ times. We thus conclude that the
degree-corrected SBM is a more appropriate model for the political blogs. This
conclusion is consistent with earlier findings in
\citep{chen2018network,karrer2011stochastic,yan2014model}.

To further illustrate our test statistics, we randomly select
5 blogs each from the top 20 liberal and top 20 conservative blogs as
indicated in \citep{adamic2005political}. The information of these 10
blogs are presented in Table~\ref{Timprove}. We apply
$G_{\mathrm{ASE}}$ to each pair of nodes and the resulting $p$-values
are reported in Table~\ref{Timprovere}. 
Table~\ref{Timprovere} indicates that our test statistic is highly
accurate for predicting whether or not two blogs are similarly
labelled. For instance, blog $1$ and
blog $2$ in Table~\ref{Timprove} are both labelled as ``conservative''
and the $p$-value of our test statistic is close to $1$. Similarly,
blogs $3$ and $7$ have different labels and the $p$-value
of our test statistic is now almost $0$. 
Note however, that there are a few $p$-values that are possibly unexpected. 
For example the $p$-values for blog $5$ and blog $3$ are quite small
and the $p$-values for blog $5$ and blog $1$ are also quite small,
even though these three blogs are all ``conservative'' blogs. One
possible explanation is that blog $5$ is an aggregate blog and thus was written by multiple authors with
possibly different political leanings. Similarly, the $p$-values
between blog $6$ and other liberal blogs are also small, and this may
be due to the fact that blog $6$ has a substantially high proportion of links
to conservative blogs. 

\begin{table*}
\caption{5 of the top 20 conservative blogs and 5 of the top 20 liberal blogs. The two right columns show for comparison how many conservative and liberal blogs from the larger set linked to the blog in February 2005.}
\label{Timprove}
\begin{tabular}{@{}ccccc@{}}
\hline
ID&Weblog&Label& \# links to conservative blogs& \# links to liberal blogs\\
\hline
1&timblair.spleenville.com&Conservative& 80& 7\\
2&windsofchange.net&Conservative& 65& 16\\
3&vodkapundit.com&Conservative& 97& 9\\
4&rogerlsimon.com&Conservative& 74& 6\\
5&deanesmay.com&Conservative& 79& 8\\
6&wonkette.com&Liberal& 30& 83\\
7&j-bradford-delong.net/movable\_type&Liberal& 11& 98\\
8&prospect.org/weblog&Liberal& 11& 102\\
9&americablog.blogspot.com&Liberal& 5& 64\\
10&jameswolcott.com&Liberal& 6& 74\\
\hline
\end{tabular}
\end{table*}%
\begin{table*}
\caption{$p$-values based on the test statistic $G_{\mathrm{ASE}}$ for selected blogs in Table~\ref{Timprove}.}
\label{Timprovere}
\begin{tabular}{@{}cccccccccccc@{}}
\hline
ID& 1& 2& 3& 4& 5&6&7&8&9&10 \\
\hline
1 &1.000& 0.811& 0.702& 0.263&0.024&0.000& 0.000&0.000 &0.000&0.000\\
2 & 0.811&1.000& 0.882& 0.298& 0.013&0.000&0.000&0.000&0.000&0.000 \\ 
3 & 0.702& 0.882& 1.000& 0.303 &0.005&0.000&0.000&0.000&0.000&0.000 \\ 
4 & 0.263& 0.298& 0.303& 1.000 &0.091&0.000&0.000&0.000&0.000&0.000 \\ 
5& 0.024& 0.013 &0.005&0.091 &1.000 &0.000&0.000&0.000&0.000&0.000 \\
6&0.000&0.000&0.000&0.000&0.000 &1.000&0.001&0.001&0.001&0.000 \\
7 &0.000&0.000&0.000&0.000&0.000&0.001 &1.000 &0.830&0.300&0.077 \\ 
8 &0.000&0.000&0.000&0.000&0.000&0.001 &0.830 &1.000&0.364&0.087\\ 
9 &0.000&0.000&0.000&0.000&0.000&0.001 &0.300 &0.364&1.000&0.522\\   
10 &0.000&0.000&0.000&0.000&0.000&0.000 &0.077 &0.087&0.522&1.000\\   
\hline
\end{tabular}
\end{table*}%
Finally we also analyze the political blogs networks as a
directed network using the test statistics described in
Appendix~\ref{sec:directed1}. More specifically we keep only the largest
connected component, remove all multi-edges and loops, but keep the
orientation of the directed edges. We embed the resulting
directed graph into $\mathbb{R}^{2}$; we chose $d = 2$ to be
consistent with the above analysis of the undirected
case. Using this embedding, we test the null hypothesis that two nodes
have the same outgoing latent positions up to scaling. Our rationale
for looking only at the outgoing latent positions is that the author
of a blog can control the outgoing links (from their blog to other
blogs) but cannot control the incoming links. 
We apply $G_{\mathrm{out}}$ which is the degree-corrected version of $T_{\mathrm{out}}$ mentioned in Remark~\ref{rem:directed}. Once again, due to the large number of possible pairs of
nodes, we randomly
choose 1000 pairs of nodes within the same community and 1000 pairs in
different communities to perform our test.
We use the $95\%$ percentile of the $\chi^2_1$ distribution as the
threshold for classifying two blogs to be from the same
community. The resulting sensitivity and specificity are then $0.43$
and $0.92$, respectively. The pairwise $p$-values between the $10$
selected blogs are presented in Table~\ref{TF1}; once again we can
predict quite accurately whether or not two blogs are similarly
labelled. 
\begin{table*}
\caption{$p$-values based on the test statistic $G_{\mathrm{out}}$ for selected blogs in Table~\ref{Timprove}.}
\label{TF1}
\begin{tabular}{@{}cccccccccccc@{}}
\hline
ID& 1& 2& 3& 4& 5&6&7&8&9&10 \\
\hline
1 &1.0000& 0.8301& 0.1460& 0.9314& 0.1587& 0.0014 &0.0017 &0.0000& 0.0000& 0.0404\\
2 & 0.8301 &1.0000 &0.0946& 0.8848& 0.2231& 0.0011& 0.0017& 0.0000& 0.0000 &0.0406 \\ 
3 & 0.1460& 0.0946 &1.0000& 0.1090 &0.0334& 0.0065& 0.0019 &0.0000& 0.0000& 0.0396 \\ 
4 & 0.9314& 0.8848 &0.1090& 1.0000& 0.1446& 0.0011& 0.0016& 0.0000& 0.0000 &0.0399 \\ 
5& 0.1587& 0.2231& 0.0334& 0.1446 &1.0000& 0.0002& 0.0013 &0.0000 &0.0000& 0.0400 \\
6&0.0014 &0.0011 &0.0065& 0.0011 &0.0002& 1.0000 &0.0608 &0.0487& 0.0168& 0.0861 \\
7 &0.0017& 0.0017& 0.0019& 0.0016 &0.0013& 0.0608& 1.0000 &0.4952 &0.8831 &0.5430\\ 
8 &0.0000& 0.0000 &0.0000& 0.0000& 0.0000 &0.0487 &0.4952 &1.0000& 0.2994 &0.2005\\ 
9 &0.0000& 0.0000 &0.0000& 0.0000& 0.0000& 0.0168& 0.8831& 0.2994& 1.0000& 0.5311\\   
10 &0.0404& 0.0406& 0.0396& 0.0399& 0.0400& 0.0861& 0.5430& 0.2005& 0.5311& 1.0000\\   
\hline
\end{tabular}
\end{table*}

\subsection{Leeds Butterfly Dataset}
\label{leeds_butterfly}
We consider in this section the problem of testing for equality of community
assignments in Popularity Adjusted Block Models (PABM) and apply the
resulting test statistic to the Leeds Butterfly
dataset of \citep{wang2018network}. We start by describing the PABM 
proposed in \cite{sengupta2018block}. 
Let $K \geq 1$ be an integer and let $\bm{\Lambda}$ be a $n \times K$
matrix whose entries $\lambda_{ik} \in [0,1]$ for all $i \in
\{1,2,\dots,n\}$ and $k \in \{1,2,\dots,K\}$. 
Let $\tau=(\tau_1,\ldots,\tau_n) \in
\{1,\ldots,K\}^n$ be the community assignments where $\tau_i = k$ if
node $i$ belongs to community $k$. A graph $G$ with adjacency matrix
$\mathbf{A}$ is said to be a popularity adjusted block model (PABM) with $K$ communities, popularity vectors $\mathbf{\Lambda}$, and sparsity parameter $\rho_{n}$ if the $\mathbf{A}_{ij}$'s are independent Bernoulli variables satisfying
$$
P(\mathbf{A}_{ij}=1)=\mathbf{P}_{i j}=\rho_{n} \lambda_{i \tau_{j}} \lambda_{j \tau_{i}}
$$
The entries $\lambda_{ik}$ represents the popularity of node $i$ in
community $k$; that is to say, larger values of $\lambda_{ik}$ are
associated with more edges between node $i$ and other nodes in
community $k$.

The motivation for the PABM model is as follows. 
Recall that the degree-corrected SBM (DCSBM) is a generalization of the SBM and
allows for heterogeneous degrees for nodes in the same
community. A PABM also allows for degree heterogeneity
of nodes from the same community, but this heterogeneity 
is more flexible than that of a DCSBM. 
More specifically, suppose $i$ and $j$ are two nodes assigned to
the same community in a PABM. Then it is possible 
that node $i$ is more likely than
node $j$ to connect with nodes in some community $k$ (so that $\lambda_{ik} > \lambda_{jk}$),
while node $j$ is more likely than node $i$ to connect with nodes in some other
community $\ell \not = k$ (so that $\lambda_{j\ell} > \lambda_{i \ell}$). Now
suppose that $i$ and $j$ belong to the
same community in a degree-corrected SBM. Then as $i$ and $j$ are
associated with (scalar-valued) degree correction factors $\theta_i$
and $\theta_j$, if $\theta_i > \theta_j$ then node $i$ is more likely
than node $j$ to connect with any arbitrary node $v$. In other words, if a given
node $i$ is more popular than node $j$ in some community $k$, then
$i$ is also more popular than $j$ in all community $\ell \not = k$. 

A PABM is also a special case of a GRDPG. More specifically,
assume without loss of generality that the rows of $\bm{\Lambda}$ are arranged in
increasing order of the community assignment $\tau$, i.e., if $i < j$
then $\tau_i \leq \tau_j$. Denote the $i$th row of $\bm{\Lambda}$ as
$\lambda_i$. Now let $\bm{\Lambda}^{(k)}$ be the submatrix of
$\bm{\Lambda}$ obtained by keeping only the
$\lambda_i$'s for which $\tau_i =
k$. A PABM graph with parameters $\bm{\Lambda}$ given above
is equivalent to a GRPDG with signatures $a = K(K+1)/2$ and $b =
K(K-1)/2$, and latent positions matrix $\mathbf{X}$ given by
$\mathbf{X} = \bm{\Lambda}^{(1)} \oplus \bm{\Lambda}^{(2)} \oplus
\cdots \oplus \bm{\Lambda}^{(K)}$; here $\oplus$ denote the direct sum for
matrices, i.e., $\mathbf{M}_1 \oplus \mathbf{M}_2$ is the block matrix
of the form $\Bigl(\begin{smallmatrix} \mathbf{M}_1 & \bm{0} \\ \bm{0}
    & \mathbf{M}_2 \end{smallmatrix} \Bigr)$. 
The above structure for $\mathbf{X}$ implies that the $K$
communities of a PABM correspond to $K$ mutually orthogonal
subspaces in $\mathbb{R}^{K^2}$. 
See Theorem~1 and Theorem~2 in
\citep{koo2021popularity} for more details. 

Given two nodes $i$ and $j$ in a PABM we can test the hypothesis that
they have the same latent positions, possibly up to scaling, 
by using the test statistics provided in
Section~\ref{sec3} and Section~\ref{sec:LSE}. However, if our main
interest is in testing whether or not two given nodes $i$ and $j$
belong to the same community in a PABM then the above test
statistics no longer apply. 
Indeed, in contrast to the SBM or DCSBM where there exists a
$1$-to-$1$ correspondence between the community assignments $\tau_i$
and the point masses $\nu_{\tau_i}$ (see Remark~\ref{rem:SBM_MMSBM}),
two nodes from the same community in a PABM can have
drastically different latent positions. 

Let $U_i$ denote the $i$th row of the matrix
$\mathbf{U}$ where $\mathbf{U}\mathbf{S}
\mathbf{U}^{\top}$ is the eigen-decomposition of $\mathbf{P}$. Theorem 2 of
\citep{koo2021popularity} show that $U_i^{\top}U_j=0$ if and only if node
$i$ and $j$ belong to different communities. Leveraging this fact we
propose a test statistic for testing $\tau_i = \tau_j$ but, unlike the
test procedures in Section~\ref{sec3} and Section~\ref{sec:LSE}, the 
null hypothesis now is that two node $i$ and node $j$ belong to different
communities i.e., we are interested in testing the
hypothesis
\begin{equation}
  \label{eq:hypothesis_pabm}
\mathbb{H}_{0} \colon \tau_{i} \neq \tau_{j} \quad \text{against}
\quad \mathbb{H}_{A} \colon \tau_{i}= \tau_{j}
\end{equation}
We then have the following result.
\begin{theorem}\label{thmPABM}
Let $\mathbf{A}$ be a graph on $n$ vertices generated from a PABM with $K$ communities and sparsity factor $\rho_n$. Let $\hat{\mathbf{X}}$ be the adjacency spectral embedding of
$\mathbf{A}$ into $\mathbb{R}^{K^2}$. Define the test statistic
$$
T_{\mathrm{PABM}}(\hat{U}_i, \hat{U}_j)=\frac{n\rho_n^{1/2}\hat{U}_{i}^{\top}\hat{U}_{j}}{\bigl(\hat{U}_i^{\top}\hat{\boldsymbol{\Sigma}}(\hat{U}_{j})\hat{U}_i+\hat{U}_j^{\top}\hat{\boldsymbol{\Sigma}}(\hat{U}_{i})\hat{U}_j\bigr)^{1/2}}
$$
where $\hat{\bm{\Sigma}}(\hat{U}_i)$ and
$\hat{\bm{\Sigma}}(\hat{U}_j)$ are as defined in
Corollary~\ref{cor1}, with $a = K(K+1)/2$ and $b = K(K-1)/2$.  
Then under the null hypothesis $\mathbb{H}_0 \colon \tau_i \neq \tau_{j}$ and for
$n \rightarrow \infty$ with $n \rho_n = \omega(\log n)$, we have
$$T_{\mathrm{PABM}}(\hat{U}_i, \hat{U}_j) \rightsquigarrow \mathcal{N}(0,1)$$
\end{theorem} 
The above result indicates that, for a given significance level
$\alpha$, we reject $\mathbb{H}_{0}$ if
$|T_{\mathrm{PABM}}(\hat{U}_i, \hat{U}_j)|>Z_{1-\alpha/2}$ where
$Z_{1-\alpha/2}$ is the $100 \times(1-\alpha/2)$ th percentile of $\mathcal{N}(0,1)$. Simulation results for Theorem~\ref{thmPABM}
are provided in Section~\ref{app:pabm_simu} of the appendix.

We now apply the proposed test statistic to the Leeds Butterfly
dataset of \citep{wang2018network}. This dataset contains similarity
measurements between $832$ butterfly images; these images are labeled
into $10$ different classes. 
Following \cite{noroozi2019estimation}, we select a subset of $373$ images corresponding to the
$K = 4$ largest classes and form an adjacency matrix by thresholding
these pairwise similarities so that each image is mapped to a vertex
and two vertices are connected if their similarity measure is
positive. The resulting (undirected) graph has $20566$ edges. 
We use $T_{\mathrm{PABM}}$ to test, for each
of the $\tbinom{373}{2}$ pairs of nodes, the hypothesis in
Eq.~\eqref{eq:hypothesis_pabm}. Choosing the $95\%$ of the standard normal
distribution as a threshold, we achieve a specificity of $0.93$ and sensitivity of $0.67$.

\section{Discussion}
\label{sec6} In this paper we developed Mahalanobis distance based
test statistics to determine whether or not two vertices have the same
latent positions, or the same latent positions up to scaling (in the
degree-corrected case).  We established limiting chi-square
distributions for the test statistics under both the null and local
alternative hypothesis; furthermore, our expressions for the non-centrality
parameters under the local alternative are invariant with respect to
the non-identifiability of the latent positions.  Leveraging these
limit results, we also propose test statistics for deciding between
the standard stochastic block model (SBM) and a degree-corrected
stochastic block model (DCSBM), and choosing between the Erdős–Rényi model and stochastic block model.



We note that the values of the non-centrality
parameters for $T_{\mathrm{ASE}}$ and $T_{\mathrm{LSE}}$ in
Table~\ref{T2} are almost identical; similarly, the values of the non-centrality parameters for
$G_{\mathrm{ASE}}$ and $G_{\mathrm{LSE}}$ in Table~\ref{T5} are also
almost identical. This suggests that the test statistics constructed using the different
embeddings are, asymptotically, almost equivalent. Indeed, we were not able to find simulation
settings for which either the non-centrality parameters of $T_{\mathrm{ASE}}$
and $T_{\mathrm{LSE}}$, or the non-centrality parameters of
$G_{\mathrm{ASE}}$ and $G_{\mathrm{LSE}}$, are well separated.
Nevertheless, for the real data analysis in
Section~\ref{sec5}, the test statistics associated with different embeddings
do have significantly different error rates.
A more precise understanding of why these differences arise is
therefore of some practical interests. 

Finally, as we allude to in the introduction, a GRDPG is a special
case of a latent position graph. It is thus natural to pose the
question of testing the hypothesis $\mathbb{H}_0 \colon X_i = X_j$ for
general latent position graphs, and in particular to study test
statistics based on the Mahalanobis distance between the rows of the
embeddings as is done in the current paper. This problem is, however,
highly non-trivial. Indeed, the edge probabilities matrix of a latent position graph is
generally not low-rank; in contrast, limit results for spectral
embeddings for random graphs, such as those in
\citep{fan_eigenvectors,rubin2017statistical}, almost always
assume that the edge probabilities matrix is low-rank.
Theoretical results for testing $\mathbb{H}_0 \colon X_i = X_j$ in general
latent position graphs require new and far-reaching extensions of existing results for
spectral embeddings.
\bibliographystyle{imsart-number} 
\bibliography{bibliography}   

\begin{thebibliography}{54}

\bibitem{abbe2017community}
\begin{barticle}[author]
\bauthor{\bsnm{Abbe},~\bfnm{E.}\binits{E.}}
(\byear{2017}).
\btitle{Community detection and stochastic block models: recent developments}.
\bjournal{Journal of Machine Learning Research}
\bvolume{18}
\bpages{6446--6531}.
\end{barticle}
\endbibitem

\bibitem{adamic2005political}
\begin{binproceedings}[author]
\bauthor{\bsnm{Adamic},~\bfnm{L.~A.}\binits{L.~A.}} \AND
  \bauthor{\bsnm{Glance},~\bfnm{N.}\binits{N.}}
(\byear{2005}).
\btitle{The political blogosphere and the 2004 US election: divided they blog}.
In \bbooktitle{Proceedings of the 3rd international workshop on Link discovery}
\bpages{36--43}.
\end{binproceedings}
\endbibitem

\bibitem{agterberg2019vertex}
\begin{barticle}[author]
\bauthor{\bsnm{Agterberg},~\bfnm{J.}\binits{J.}},
  \bauthor{\bsnm{Park},~\bfnm{Y.}\binits{Y.}},
  \bauthor{\bsnm{Larson},~\bfnm{J.}\binits{J.}},
  \bauthor{\bsnm{White},~\bfnm{C.}\binits{C.}},
  \bauthor{\bsnm{Priebe},~\bfnm{C.~E.}\binits{C.~E.}} \AND
  \bauthor{\bsnm{Lyzinski},~\bfnm{V.}\binits{V.}}
(\byear{2020}).
\btitle{Vertex Nomination, consistent estimation, and adversarial
  modification}.
\bjournal{Electronic Journal of Statistics}
\bpages{3230--3267}.
\end{barticle}
\endbibitem

\bibitem{airoldi2008mixed}
\begin{barticle}[author]
\bauthor{\bsnm{Airoldi},~\bfnm{E.~M.}\binits{E.~M.}},
  \bauthor{\bsnm{Blei},~\bfnm{D.~M.}\binits{D.~M.}},
  \bauthor{\bsnm{Fienberg},~\bfnm{S.~E.}\binits{S.~E.}} \AND
  \bauthor{\bsnm{Xing},~\bfnm{E.~P.}\binits{E.~P.}}
(\byear{2008}).
\btitle{Mixed membership stochastic blockmodels}.
\bjournal{Journal of Machine Learning Research}
\bvolume{9}
\bpages{1981--2014}.
\end{barticle}
\endbibitem

\bibitem{athreya2016limit}
\begin{barticle}[author]
\bauthor{\bsnm{Athreya},~\bfnm{A.}\binits{A.}},
  \bauthor{\bsnm{Priebe},~\bfnm{C.~E.}\binits{C.~E.}},
  \bauthor{\bsnm{Tang},~\bfnm{M.}\binits{M.}},
  \bauthor{\bsnm{Lyzinski},~\bfnm{V.}\binits{V.}},
  \bauthor{\bsnm{Marchette},~\bfnm{D.~J.}\binits{D.~J.}} \AND
  \bauthor{\bsnm{Sussman},~\bfnm{D.~L.}\binits{D.~L.}}
(\byear{2016}).
\btitle{A limit theorem for scaled eigenvectors of random dot product graphs}.
\bjournal{Sankhya A}
\bvolume{78}
\bpages{1--18}.
\end{barticle}
\endbibitem

\bibitem{bandeira_vanhandel}
\begin{barticle}[author]
\bauthor{\bsnm{Bandeira},~\bfnm{A.~S.}\binits{A.~S.}} \AND
  \bauthor{\bsnm{Handel},~\bfnm{R.~Van}\binits{R.~V.}}
(\byear{2016}).
\btitle{Sharp nonasymptotic bounds on the norm of random matrices with
  independent entries}.
\bjournal{Annals of Probability}
\bvolume{44}
\bpages{2479--2506}.
\end{barticle}
\endbibitem

\bibitem{belkin03:_laplac}
\begin{barticle}[author]
\bauthor{\bsnm{Belkin},~\bfnm{M.}\binits{M.}} \AND
  \bauthor{\bsnm{Niyogi},~\bfnm{P.}\binits{P.}}
(\byear{2003}).
\btitle{Laplacian eigenmaps for dimensionality reduction and data
  representation}.
\bjournal{Neural Computation}
\bvolume{15}
\bpages{1373-1396}.
\end{barticle}
\endbibitem

\bibitem{bhatia_qnorm}
\begin{barticle}[author]
\bauthor{\bsnm{Bhatia},~\bfnm{R.}\binits{R.}}
(\byear{1987}).
\btitle{Some Inequalities for Norm Ideals}.
\bjournal{Communications in Mathematical Physics}
\bvolume{111}
\bpages{33--39}.
\end{barticle}
\endbibitem

\bibitem{bhatia}
\begin{bbook}[author]
\bauthor{\bsnm{Bhatia},~\bfnm{R.}\binits{R.}}
(\byear{1997}).
\btitle{Matrix Analysis}.
\bpublisher{Springer}.
\end{bbook}
\endbibitem

\bibitem{bickel2016hypothesis}
\begin{barticle}[author]
\bauthor{\bsnm{Bickel},~\bfnm{P.~J.}\binits{P.~J.}} \AND
  \bauthor{\bsnm{Sarkar},~\bfnm{P.}\binits{P.}}
(\byear{2016}).
\btitle{Hypothesis testing for automated community detection in networks}.
\bjournal{Journal of the Royal Statistical Society: Series B}
\bpages{253--273}.
\end{barticle}
\endbibitem

\bibitem{cai_outliers}
\begin{barticle}[author]
\bauthor{\bsnm{Cai},~\bfnm{T.}\binits{T.}} \AND
  \bauthor{\bsnm{Li},~\bfnm{X.}\binits{X.}}
(\byear{2015}).
\btitle{Robust and computationally feasible community detection in the presence
  of arbitrary outlier nodes}.
\bjournal{Annals of Statistics}
\bvolume{45}
\bpages{1027--1059}.
\end{barticle}
\endbibitem

\bibitem{chaudhuri}
\begin{binproceedings}[author]
\bauthor{\bsnm{Chaudhuri},~\bfnm{K.}\binits{K.}},
  \bauthor{\bsnm{Chung},~\bfnm{F.}\binits{F.}} \AND
  \bauthor{\bsnm{Tsiatas},~\bfnm{A.}\binits{A.}}
(\byear{2012}).
\btitle{Spectral partitioning of graphs with general degrees and the extended
  planted partition model}.
In \bbooktitle{Proceedings of the 25th conference on learning theory}.
\end{binproceedings}
\endbibitem

\bibitem{chen2018network}
\begin{barticle}[author]
\bauthor{\bsnm{Chen},~\bfnm{K.}\binits{K.}} \AND
  \bauthor{\bsnm{Lei},~\bfnm{J.}\binits{J.}}
(\byear{2018}).
\btitle{Network cross-validation for determining the number of communities in
  network data}.
\bjournal{Journal of the American Statistical Association}
\bvolume{113}
\bpages{241--251}.
\end{barticle}
\endbibitem

\bibitem{coifman06:_diffus_maps}
\begin{barticle}[author]
\bauthor{\bsnm{Coifman},~\bfnm{R.}\binits{R.}} \AND
  \bauthor{\bsnm{Lafon},~\bfnm{S.}\binits{S.}}
(\byear{2006}).
\btitle{Diffusion maps}.
\bjournal{Applied and Computational Harmonic Analysis}
\bvolume{21}
\bpages{5--30}.
\end{barticle}
\endbibitem

\bibitem{davis70}
\begin{barticle}[author]
\bauthor{\bsnm{Davis},~\bfnm{C.}\binits{C.}} \AND
  \bauthor{\bsnm{Kahan},~\bfnm{W.}\binits{W.}}
(\byear{1970}).
\btitle{The rotation of eigenvectors by a pertubation. {III}.}
\bjournal{Siam Journal on Numerical Analysis}
\bvolume{7}
\bpages{1--46}.
\end{barticle}
\endbibitem

\bibitem{fan_eigenvectors}
\begin{barticle}[author]
\bauthor{\bsnm{Fan},~\bfnm{J.}\binits{J.}},
  \bauthor{\bsnm{Fan},~\bfnm{Y.}\binits{Y.}},
  \bauthor{\bsnm{Han},~\bfnm{X.}\binits{X.}} \AND
  \bauthor{\bsnm{Lv},~\bfnm{J.}\binits{J.}}
(\byear{2021}+).
\btitle{Asymptotic Theory of Eigenvectors for Random Matrices with Diverging
  Spikes}.
\bjournal{Journal of the American Statistical Association}.
\end{barticle}
\endbibitem

\bibitem{fan2019simple}
\begin{barticle}[author]
\bauthor{\bsnm{Fan},~\bfnm{J.}\binits{J.}},
  \bauthor{\bsnm{Fan},~\bfnm{Y.}\binits{Y.}},
  \bauthor{\bsnm{Han},~\bfnm{X.}\binits{X.}} \AND
  \bauthor{\bsnm{Lv},~\bfnm{J.}\binits{J.}}
(\byear{2022}+).
\btitle{SIMPLE: Statistical Inference on Membership Profiles in Large
  Networks}.
\bjournal{Journal of the Royal Statistical Society, Series B}.
\end{barticle}
\endbibitem

\bibitem{fishkind2015vertex}
\begin{barticle}[author]
\bauthor{\bsnm{Fishkind},~\bfnm{D.~E.}\binits{D.~E.}},
  \bauthor{\bsnm{Lyzinski},~\bfnm{V.}\binits{V.}},
  \bauthor{\bsnm{Pao},~\bfnm{H.}\binits{H.}},
  \bauthor{\bsnm{Chen},~\bfnm{L.}\binits{L.}} \AND
  \bauthor{\bsnm{Priebe},~\bfnm{C.~E.}\binits{C.~E.}}
(\byear{2015}).
\btitle{Vertex nomination schemes for membership prediction}.
\bjournal{Annals of Applied Statistics}
\bvolume{9}
\bpages{1510--1532}.
\end{barticle}
\endbibitem

\bibitem{gao_optimal}
\begin{barticle}[author]
\bauthor{\bsnm{Gao},~\bfnm{C.}\binits{C.}},
  \bauthor{\bsnm{Ma},~\bfnm{Z.}\binits{Z.}},
  \bauthor{\bsnm{Zhang},~\bfnm{A.~Y.}\binits{A.~Y.}} \AND
  \bauthor{\bsnm{Zhou},~\bfnm{H.~H.}\binits{H.~H.}}
(\byear{2017}).
\btitle{Achieving optimal mis-classification proportion in stochastic
  blockmodels}.
\bjournal{Journal of Machine Learning Research}
\bvolume{18}
\bpages{1--45}.
\end{barticle}
\endbibitem

\bibitem{ghoshdastidar2017two}
\begin{barticle}[author]
\bauthor{\bsnm{Ghoshdastidar},~\bfnm{D.}\binits{D.}},
  \bauthor{\bsnm{Gutzeit},~\bfnm{M.}\binits{M.}},
  \bauthor{\bsnm{Carpentier},~\bfnm{A.}\binits{A.}} \AND
  \bauthor{\bparticle{von} \bsnm{Luxburg},~\bfnm{U.}\binits{U.}}
(\byear{2020}).
\btitle{Two-sample hypothesis testing for inhomogeneous random graphs}.
\bjournal{Annals of Statistics}
\bvolume{48}
\bpages{2208--2229}.
\end{barticle}
\endbibitem

\bibitem{eliassi}
\begin{binproceedings}[author]
\bauthor{\bsnm{Gilpin},~\bfnm{S.}\binits{S.}},
  \bauthor{\bsnm{Eliassi-Rad},~\bfnm{T.}\binits{T.}} \AND
  \bauthor{\bsnm{Davidson},~\bfnm{I.}\binits{I.}}
(\byear{2013}).
\btitle{Guided learning for role discovery (GLRD): Framework, Algorithms, and
  Applications}.
In \bbooktitle{Proceedings of the 19th ACM SIGKDD Conference on Knowledge
  Discovery and Data Mining}.
\end{binproceedings}
\endbibitem

\bibitem{ginestet2017hypothesis}
\begin{barticle}[author]
\bauthor{\bsnm{Ginestet},~\bfnm{C.~E.}\binits{C.~E.}},
  \bauthor{\bsnm{Li},~\bfnm{J.}\binits{J.}},
  \bauthor{\bsnm{Balachandran},~\bfnm{P.}\binits{P.}},
  \bauthor{\bsnm{Rosenberg},~\bfnm{S.}\binits{S.}} \AND
  \bauthor{\bsnm{Kolaczyk},~\bfnm{E.~D.}\binits{E.~D.}}
(\byear{2017}).
\btitle{Hypothesis testing for network data in functional neuroimaging}.
\bjournal{Annals of Applied Statistics}
\bvolume{11}
\bpages{725--750}.
\end{barticle}
\endbibitem

\bibitem{girvan2002community}
\begin{barticle}[author]
\bauthor{\bsnm{Girvan},~\bfnm{M.}\binits{M.}} \AND
  \bauthor{\bsnm{Newman},~\bfnm{M.~EJ.}\binits{M.~E.}}
(\byear{2002}).
\btitle{Community structure in social and biological networks}.
\bjournal{Proceedings of the national academy of sciences}
\bvolume{99}
\bpages{7821--7826}.
\end{barticle}
\endbibitem

\bibitem{hoff2002latent}
\begin{barticle}[author]
\bauthor{\bsnm{Hoff},~\bfnm{P.~D.}\binits{P.~D.}},
  \bauthor{\bsnm{Raftery},~\bfnm{A.~E.}\binits{A.~E.}} \AND
  \bauthor{\bsnm{Handcock},~\bfnm{M.~S.}\binits{M.~S.}}
(\byear{2002}).
\btitle{Latent space approaches to social network analysis}.
\bjournal{Journal of the American Statistical Association}
\bvolume{97}
\bpages{1090--1098}.
\end{barticle}
\endbibitem

\bibitem{holland1983stochastic}
\begin{barticle}[author]
\bauthor{\bsnm{Holland},~\bfnm{P.~W.}\binits{P.~W.}},
  \bauthor{\bsnm{Laskey},~\bfnm{K.~B.}\binits{K.~B.}} \AND
  \bauthor{\bsnm{Leinhardt},~\bfnm{S.}\binits{S.}}
(\byear{1983}).
\btitle{Stochastic blockmodels: First steps}.
\bjournal{Social networks}
\bvolume{5}
\bpages{109--137}.
\end{barticle}
\endbibitem

\bibitem{hu2020corrected}
\begin{barticle}[author]
\bauthor{\bsnm{Hu},~\bfnm{J.}\binits{J.}},
  \bauthor{\bsnm{Qin},~\bfnm{H.}\binits{H.}},
  \bauthor{\bsnm{Yan},~\bfnm{T.}\binits{T.}} \AND
  \bauthor{\bsnm{Zhao},~\bfnm{Y.}\binits{Y.}}
(\byear{2020}).
\btitle{Corrected Bayesian information criterion for stochastic block models}.
\bjournal{Journal of the American Statistical Association}
\bvolume{115}
\bpages{1771--1783}.
\end{barticle}
\endbibitem

\bibitem{jin2015fast}
\begin{barticle}[author]
\bauthor{\bsnm{Jin},~\bfnm{Jiashun}\binits{J.}}
(\byear{2015}).
\btitle{Fast community detection by SCORE}.
\bjournal{Annals of Statistics}
\bvolume{43}
\bpages{57--89}.
\end{barticle}
\endbibitem

\bibitem{karrer2011stochastic}
\begin{barticle}[author]
\bauthor{\bsnm{Karrer},~\bfnm{B.}\binits{B.}} \AND
  \bauthor{\bsnm{Newman},~\bfnm{M.~EJ.}\binits{M.~E.}}
(\byear{2011}).
\btitle{Stochastic blockmodels and community structure in networks}.
\bjournal{Physical review E}
\bvolume{83}
\bpages{016107}.
\end{barticle}
\endbibitem

\bibitem{koo2021popularity}
\begin{barticle}[author]
\bauthor{\bsnm{Koo},~\bfnm{John}\binits{J.}},
  \bauthor{\bsnm{Tang},~\bfnm{Minh}\binits{M.}} \AND
  \bauthor{\bsnm{Trosset},~\bfnm{Michael~W}\binits{M.~W.}}
(\byear{2021}).
\btitle{Popularity Adjusted Block Models are Generalized Random Dot Product
  Graphs}.
\bjournal{arXiv preprint arXiv:2109.04010}.
\end{barticle}
\endbibitem

\bibitem{le2015estimating}
\begin{barticle}[author]
\bauthor{\bsnm{Le},~\bfnm{C.~M.}\binits{C.~M.}} \AND
  \bauthor{\bsnm{Levina},~\bfnm{E.}\binits{E.}}
(\byear{2015}).
\btitle{Estimating the number of communities in networks by spectral methods}.
\bjournal{arXiv preprint arXiv:1507.00827}.
\end{barticle}
\endbibitem

\bibitem{lei2016goodness}
\begin{barticle}[author]
\bauthor{\bsnm{Lei},~\bfnm{J.}\binits{J.}}
(\byear{2016}).
\btitle{A goodness-of-fit test for stochastic block models}.
\bjournal{Annals of Statistics}
\bvolume{44}
\bpages{401--424}.
\end{barticle}
\endbibitem

\bibitem{graph_root}
\begin{barticle}[author]
\bauthor{\bsnm{Lei},~\bfnm{J.}\binits{J.}}
(\byear{2021}).
\btitle{Network representation using graph root distributions}.
\bjournal{Annals of Statistics}
\bvolume{49}
\bpages{745--768}.
\end{barticle}
\endbibitem

\bibitem{lei2015consistency}
\begin{barticle}[author]
\bauthor{\bsnm{Lei},~\bfnm{J.}\binits{J.}} \AND
  \bauthor{\bsnm{Rinaldo},~\bfnm{A.}\binits{A.}}
(\byear{2015}).
\btitle{Consistency of spectral clustering in stochastic block models}.
\bjournal{Annals of Statistics}
\bvolume{43}
\bpages{215--237}.
\end{barticle}
\endbibitem

\bibitem{li2020network}
\begin{barticle}[author]
\bauthor{\bsnm{Li},~\bfnm{T.}\binits{T.}},
  \bauthor{\bsnm{Levina},~\bfnm{E.}\binits{E.}} \AND
  \bauthor{\bsnm{Zhu},~\bfnm{J.}\binits{J.}}
(\byear{2020}).
\btitle{Network cross-validation by edge sampling}.
\bjournal{Biometrika}
\bvolume{107}
\bpages{257--276}.
\end{barticle}
\endbibitem

\bibitem{ma2021determining}
\begin{barticle}[author]
\bauthor{\bsnm{Ma},~\bfnm{S.}\binits{S.}},
  \bauthor{\bsnm{Su},~\bfnm{L.}\binits{L.}} \AND
  \bauthor{\bsnm{Zhang},~\bfnm{Y.}\binits{Y.}}
(\byear{2021}).
\btitle{Determining the Number of Communities in Degree-corrected Stochastic
  Block Models}.
\bjournal{Journal of Machine Learning Research}
\bvolume{22}
\bpages{1--63}.
\end{barticle}
\endbibitem

\bibitem{marshall_olkin}
\begin{bbook}[author]
\bauthor{\bsnm{Marshall},~\bfnm{A.~W.}\binits{A.~W.}},
  \bauthor{\bsnm{Olkin},~\bfnm{I.}\binits{I.}} \AND
  \bauthor{\bsnm{Arnold},~\bfnm{B.~C.}\binits{B.~C.}}
(\byear{2011}).
\btitle{Inequalities: Theory of Majorization and its Applications},
\bedition{2nd} ed.
\bpublisher{Springer}.
\end{bbook}
\endbibitem

\bibitem{mcnemar}
\begin{barticle}[author]
\bauthor{\bsnm{McNemar},~\bfnm{Q.}\binits{Q.}}
(\byear{1947}).
\btitle{Note on the sampling error of the difference between correlated
  proportions or percentages}.
\bjournal{Psychometrika}
\bvolume{12}
\bpages{153--157}.
\end{barticle}
\endbibitem

\bibitem{ng_spectral}
\begin{binproceedings}[author]
\bauthor{\bsnm{Ng},~\bfnm{A.}\binits{A.}},
  \bauthor{\bsnm{Jordan},~\bfnm{M.}\binits{M.}} \AND
  \bauthor{\bsnm{Weiss},~\bfnm{Y.}\binits{Y.}}
(\byear{2001}).
\btitle{On spectral clustering: analysis and an algorithm}.
In \bbooktitle{Advances in Neural Information Processing Systems}
\bvolume{14}.
\end{binproceedings}
\endbibitem

\bibitem{nickel2008random}
\begin{bphdthesis}[author]
\bauthor{\bsnm{Nickel},~\bfnm{C.~L.~M.}\binits{C.~L.~M.}}
(\byear{2008}).
\btitle{Random dot product graphs a model for social networks},
\btype{PhD thesis},
\bpublisher{Johns Hopkins University}.
\end{bphdthesis}
\endbibitem

\bibitem{noroozi2019estimation}
\begin{barticle}[author]
\bauthor{\bsnm{Noroozi},~\bfnm{M.}\binits{M.}},
  \bauthor{\bsnm{Rimal},~\bfnm{R.}\binits{R.}} \AND
  \bauthor{\bsnm{Pensky},~\bfnm{M.}\binits{M.}}
(\byear{2021}).
\btitle{Estimation and clustering in popularity adjusted block model}.
\bjournal{Journal of the Royal Statistical Society, Series B.}
\bvolume{83}
\bpages{293--317}.
\end{barticle}
\endbibitem

\bibitem{oliveira2009concentration}
\begin{barticle}[author]
\bauthor{\bsnm{Oliveira},~\bfnm{R.~I.}\binits{R.~I.}}
(\byear{2009}).
\btitle{Concentration of the adjacency matrix and of the Laplacian in random
  graphs with independent edges}.
\bjournal{arXiv preprint arXiv:0911.0600}.
\end{barticle}
\endbibitem

\bibitem{ostrowski}
\begin{barticle}[author]
\bauthor{\bsnm{Ostrowski},~\bfnm{A.~M.}\binits{A.~M.}}
(\byear{1963}).
\btitle{On positive matrices}.
\bjournal{Mathematische Annalen}
\bvolume{150}
\bpages{276--284}.
\end{barticle}
\endbibitem

\bibitem{rohe2011spectral}
\begin{barticle}[author]
\bauthor{\bsnm{Rohe},~\bfnm{K.}\binits{K.}},
  \bauthor{\bsnm{Chatterjee},~\bfnm{S.}\binits{S.}} \AND
  \bauthor{\bsnm{Yu},~\bfnm{B.}\binits{B.}}
(\byear{2011}).
\btitle{Spectral clustering and the high-dimensional stochastic blockmodel}.
\bjournal{Annals of Statistics}
\bvolume{39}
\bpages{1878--1915}.
\end{barticle}
\endbibitem

\bibitem{rubin2017statistical}
\begin{barticle}[author]
\bauthor{\bsnm{Rubin-Delanchy},~\bfnm{P.}\binits{P.}},
  \bauthor{\bsnm{Cape},~\bfnm{J.}\binits{J.}},
  \bauthor{\bsnm{Tang},~\bfnm{M.}\binits{M.}} \AND
  \bauthor{\bsnm{Priebe},~\bfnm{C.~E.}\binits{C.~E.}}
(\byear{2022}+).
\btitle{A statistical interpretation of spectral embedding: the generalised
  random dot product graph}.
\bjournal{Journal of the Royal Statistical Society, Series B.}
\end{barticle}
\endbibitem

\bibitem{sengupta2018block}
\begin{barticle}[author]
\bauthor{\bsnm{Sengupta},~\bfnm{Srijan}\binits{S.}} \AND
  \bauthor{\bsnm{Chen},~\bfnm{Yuguo}\binits{Y.}}
(\byear{2018}).
\btitle{A block model for node popularity in networks with community
  structure}.
\bjournal{Journal of the Royal Statistical Society: Series B (Statistical
  Methodology)}
\bvolume{80}
\bpages{365--386}.
\end{barticle}
\endbibitem

\bibitem{serre}
\begin{bbook}[author]
\bauthor{\bsnm{Serre},~\bfnm{D.}\binits{D.}}
(\byear{2002}).
\btitle{Matrices: Theory and Applications}.
\bpublisher{Springer}.
\end{bbook}
\endbibitem

\bibitem{Stewart_ginv}
\begin{barticle}[author]
\bauthor{\bsnm{Stewart},~\bfnm{G.~W.}\binits{G.~W.}}
(\byear{1977}).
\btitle{On the perturbation of pseudo-inverses, projections, and linear least
  squares problems}.
\bjournal{SIAM Review}
\bvolume{19}
\bpages{634--662}.
\end{barticle}
\endbibitem

\bibitem{sussman12}
\begin{barticle}[author]
\bauthor{\bsnm{Sussman},~\bfnm{D.~L.}\binits{D.~L.}},
  \bauthor{\bsnm{Tang},~\bfnm{M.}\binits{M.}},
  \bauthor{\bsnm{Fishkind},~\bfnm{D.~E.}\binits{D.~E.}} \AND
  \bauthor{\bsnm{Priebe},~\bfnm{C.~E.}\binits{C.~E.}}
(\byear{2012}).
\btitle{A consistent adjacency spectral embedding for stochastic blockmodel
  graphs.}
\bjournal{Journal of the American Statistical Association}
\bvolume{107}
\bpages{1119--1128}.
\end{barticle}
\endbibitem

\bibitem{tang2017semiparametric}
\begin{barticle}[author]
\bauthor{\bsnm{Tang},~\bfnm{M.}\binits{M.}},
  \bauthor{\bsnm{Athreya},~\bfnm{A.}\binits{A.}},
  \bauthor{\bsnm{Sussman},~\bfnm{D.~L.}\binits{D.~L.}},
  \bauthor{\bsnm{Lyzinski},~\bfnm{V.}\binits{V.}},
  \bauthor{\bsnm{Park},~\bfnm{Y.}\binits{Y.}} \AND
  \bauthor{\bsnm{Priebe},~\bfnm{C.~E.}\binits{C.~E.}}
(\byear{2017}).
\btitle{A semiparametric two-sample hypothesis testing problem for random
  graphs}.
\bjournal{Journal of Computational and Graphical Statistics}
\bvolume{26}
\bpages{344--354}.
\end{barticle}
\endbibitem

\bibitem{tang2018limit}
\begin{barticle}[author]
\bauthor{\bsnm{Tang},~\bfnm{M.}\binits{M.}} \AND
  \bauthor{\bsnm{Priebe},~\bfnm{C.~E.}\binits{C.~E.}}
(\byear{2018}).
\btitle{Limit theorems for eigenvectors of the normalized Laplacian for random
  graphs}.
\bjournal{Annals of Statistics}
\bvolume{46}
\bpages{2360--2415}.
\end{barticle}
\endbibitem

\bibitem{luxburg08:_consis}
\begin{barticle}[author]
\bauthor{\bparticle{von} \bsnm{Luxburg},~\bfnm{U.}\binits{U.}},
  \bauthor{\bsnm{Belkin},~\bfnm{M.}\binits{M.}} \AND
  \bauthor{\bsnm{Bousquet},~\bfnm{O.}\binits{O.}}
(\byear{2008}).
\btitle{Consistency of spectral clustering}.
\bjournal{Annals of Statistics}
\bvolume{36}
\bpages{555--586}.
\end{barticle}
\endbibitem

\bibitem{wang2018network}
\begin{barticle}[author]
\bauthor{\bsnm{Wang},~\bfnm{Bo}\binits{B.}},
  \bauthor{\bsnm{Pourshafeie},~\bfnm{Armin}\binits{A.}},
  \bauthor{\bsnm{Zitnik},~\bfnm{Marinka}\binits{M.}},
  \bauthor{\bsnm{Zhu},~\bfnm{Junjie}\binits{J.}},
  \bauthor{\bsnm{Bustamante},~\bfnm{Carlos~D}\binits{C.~D.}},
  \bauthor{\bsnm{Batzoglou},~\bfnm{Serafim}\binits{S.}} \AND
  \bauthor{\bsnm{Leskovec},~\bfnm{Jure}\binits{J.}}
(\byear{2018}).
\btitle{Network enhancement as a general method to denoise weighted biological
  networks}.
\bjournal{Nature communications}
\bvolume{9}
\bpages{1--8}.
\end{barticle}
\endbibitem

\bibitem{wang2017likelihood}
\begin{barticle}[author]
\bauthor{\bsnm{Wang},~\bfnm{YX.~R.}\binits{Y.~R.}} \AND
  \bauthor{\bsnm{Bickel},~\bfnm{P.~J.}\binits{P.~J.}}
(\byear{2017}).
\btitle{Likelihood-based model selection for stochastic block models}.
\bjournal{Annals of Statistics}
\bvolume{45}
\bpages{500--528}.
\end{barticle}
\endbibitem

\bibitem{yan2014model}
\begin{barticle}[author]
\bauthor{\bsnm{Yan},~\bfnm{X.}\binits{X.}},
  \bauthor{\bsnm{Shalizi},~\bfnm{C.}\binits{C.}},
  \bauthor{\bsnm{Jensen},~\bfnm{J.~E.}\binits{J.~E.}},
  \bauthor{\bsnm{Krzakala},~\bfnm{F.}\binits{F.}},
  \bauthor{\bsnm{Moore},~\bfnm{C.}\binits{C.}},
  \bauthor{\bsnm{Zdeborov{\'a}},~\bfnm{L.}\binits{L.}},
  \bauthor{\bsnm{Zhang},~\bfnm{P.}\binits{P.}} \AND
  \bauthor{\bsnm{Zhu},~\bfnm{Y.}\binits{Y.}}
(\byear{2014}).
\btitle{Model selection for degree-corrected block models}.
\bjournal{Journal of Statistical Mechanics: Theory and Experiment}
\bvolume{2014}
\bpages{P05007}.
\end{barticle}
\endbibitem

\end{thebibliography}
\newpage
\begin{appendix}
\section{Additional Numerical Experiments}
\subsection{Additional Plots}
\label{App:plots}
\begin{figure}[ht]
\centering
\subfigure[ASE]{
\label{figa2} 
\includegraphics[width=0.35\textwidth]{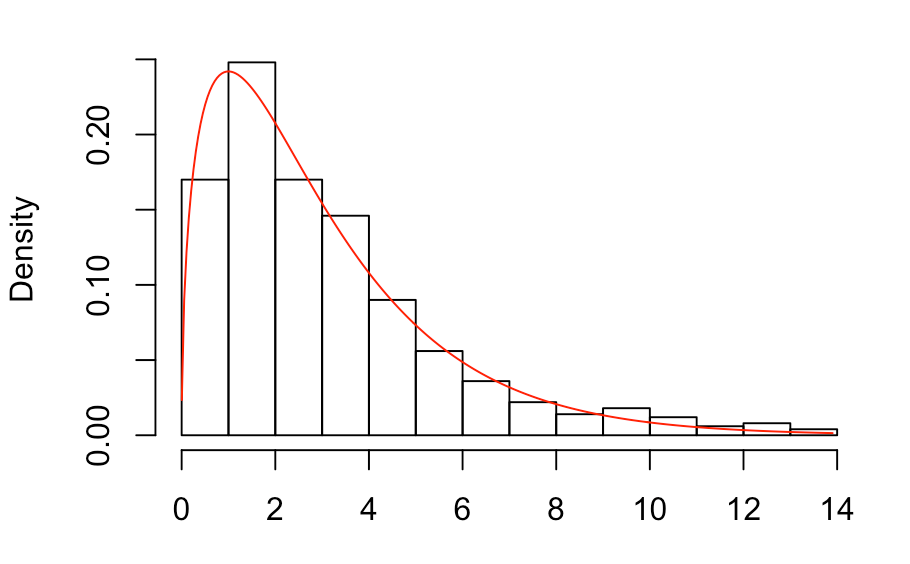}}
\subfigure[LSE]{
\label{figb2} 
\includegraphics[width=0.35\textwidth]{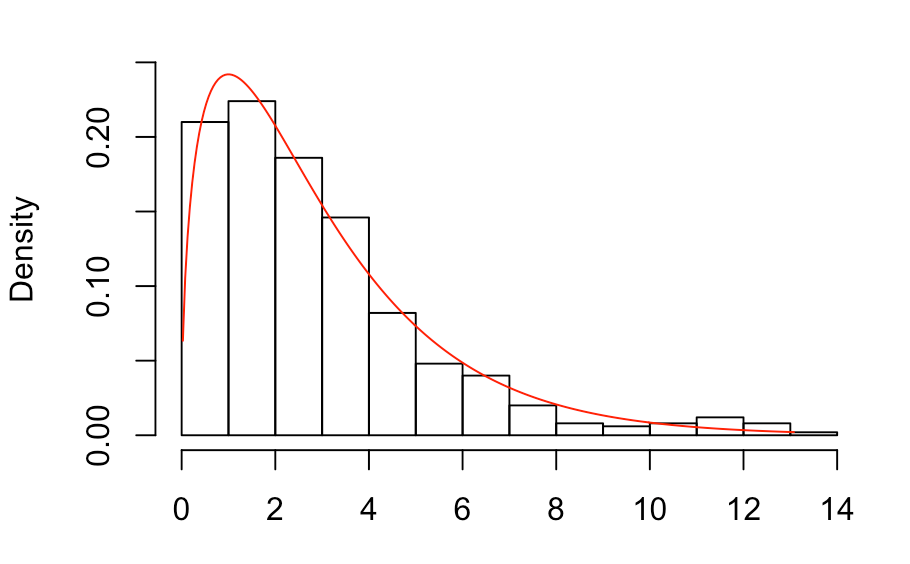}}
\caption{Empirical histograms for the test statistics
  $T_{\mathrm{ASE}}$ and $T_{\mathrm{LSE}}$ under the null hypothesis
  when $\rho=1.0$. The setting is that of mixed-membership graphs on
  $n = 3100$ vertices with
  parameters generated according to Model I. The red curve is the probability density function
  for the $\chi^2_3$ distribution.}
\label{f2} 
\end{figure}
\begin{figure}[ht]
\centering
\subfigure[ASE]{
\label{figa3} 
\includegraphics[width=0.35\textwidth]{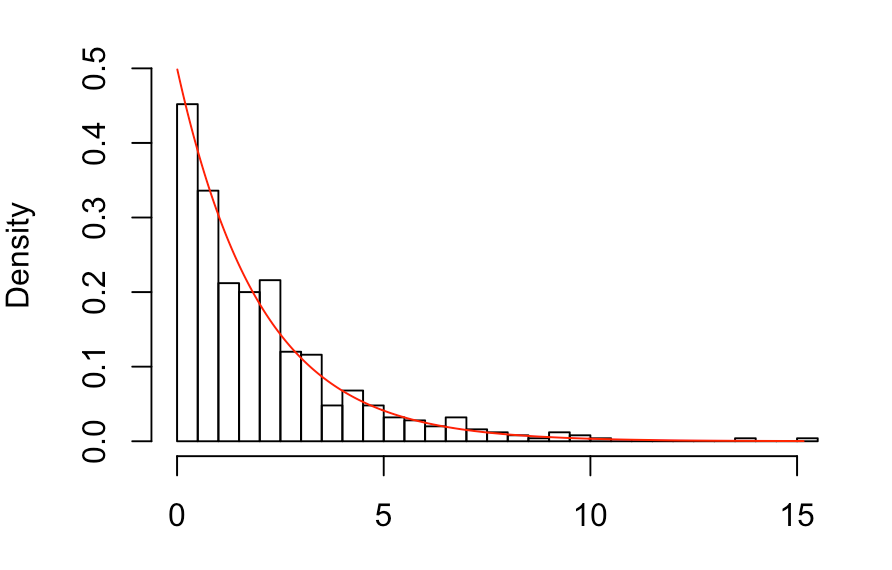}}
\subfigure[LSE]{
\label{figb3} 
\includegraphics[width=0.35\textwidth]{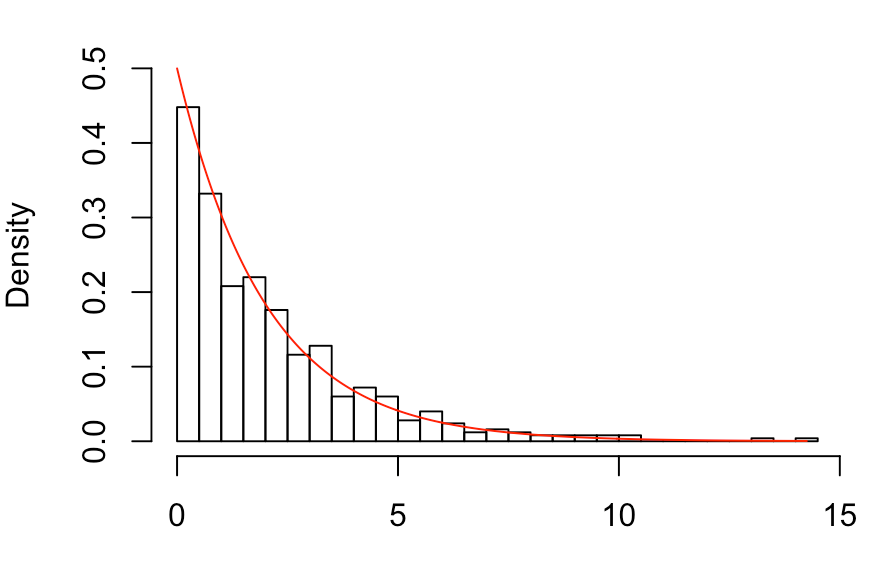}}
\caption{
  Empirical histograms for the test statistics $G_{\mathrm{ASE}}$ and $G_{\mathrm{LSE}}$ under the null hypothesis
  when the degree heterogeneity coefficients are uniformly distributed
  in the interval $[1,2]$.  The setting is that of degree-corrected
  mixed-membership graphs on $n = 3100$ vertices with sparsity factor
  $\rho = 0.25$ and parameters specified according to Model II.
  The red curve is the probability density function for the $\chi^2_2$ distribution.}
\label{f3} 
\end{figure}
\newpage
\subsection{Hypothesis Testing in PABM}
\label{app:pabm_simu}
In this section, we conduct simulations to investigate the finite sample performance of $T_{\mathrm{PABM}}$ proposed in Section~\ref{leeds_butterfly}. We generate graphs from PABM on $n=4800$ vertices with $K=2,3,4,5$. For each choice of $K$, $n$ vertices are equally assigned to the $K$ different communities. Then $\{\lambda^{(k \ell)}\}_{K}$ are generated according to the assigned community labels. Here $\{\lambda^{(k \ell)}\}_{K}$ are parameters of another view of PABM specified in \citep{koo2021popularity}. Specifically, we have within-group popularities $\lambda^{(k k)} \stackrel{\text { iid }}{\sim} \operatorname{Beta}(2,1)$ and between-group popularities $\lambda^{(k \ell)} \stackrel{\text { iid }}{\sim} \operatorname{Beta}(1,2)$ for $k \neq \ell$. $\mathbf{P}$ is constructed using the drawn $\{\lambda^{(k \ell)}\}_{K}$, and $\mathbf{A}$ is drawn from $\mathbf{P}$. We do $500$ Monte Carlo replicates for each choice of $K$ and $\alpha$ is set to be $0.05 $. The empirical histograms under the null hypothesis and empirical sizes and powers are presented below in Figure~\ref{fPABM} and Table~\ref{TPABM}. We see from Figure~\ref{fPABM} that the distributions of
$T_{ij}$ are well-approximated by the
standard normal distribution. The relatively large size and small power when $K=5$ might be due to the fact that we need larger sample size to ensure the finite sample convergence when the model becomes more complex.   

\begin{figure}[tp]
\centering
\subfigure[$K=2$]
{\label{figa} 
\includegraphics[width=0.4\textwidth]{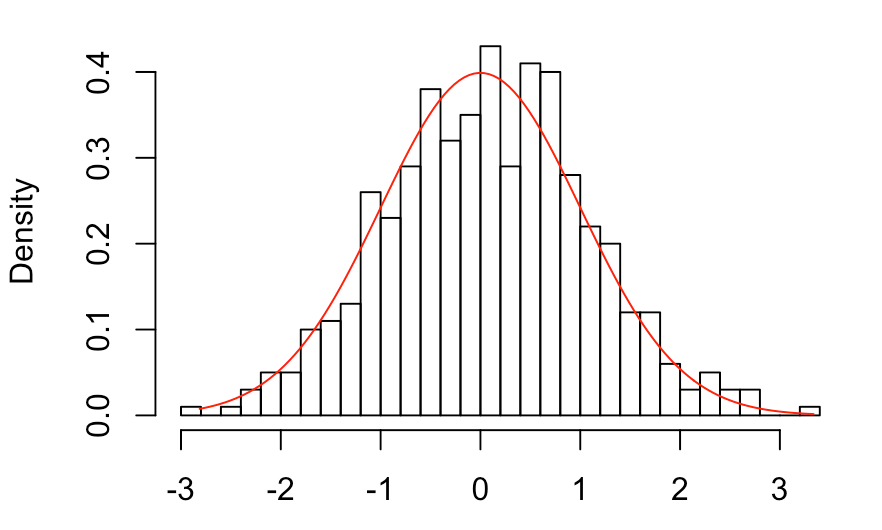}}
\subfigure[$K=3$]{
\label{figb} 
\includegraphics[width=0.4\textwidth]{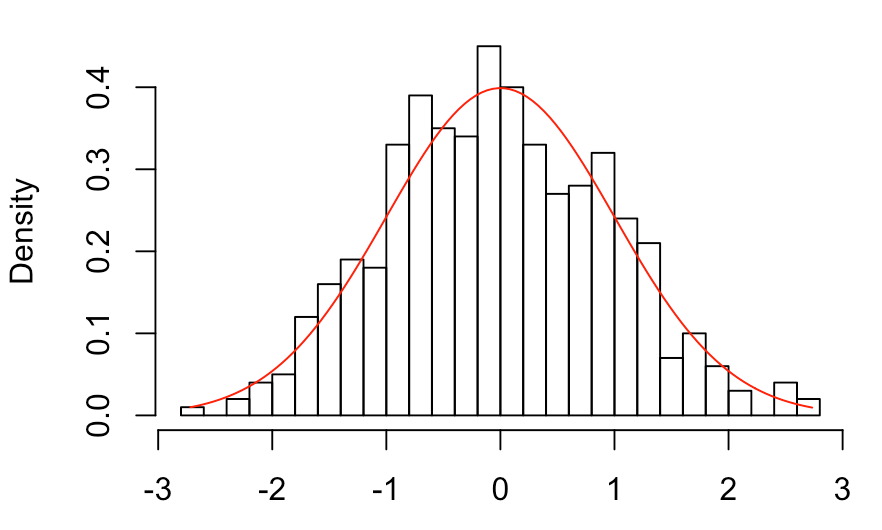}}
\subfigure[$K=4$]{
\label{figc} 
\includegraphics[width=0.4\textwidth]{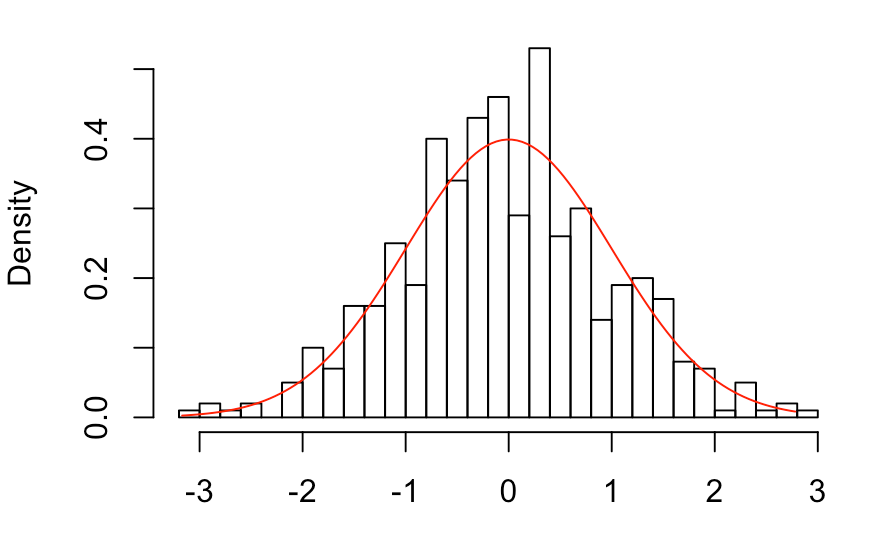}}
\subfigure[$K=5$]{
\label{figd} 
\includegraphics[width=0.4\textwidth]{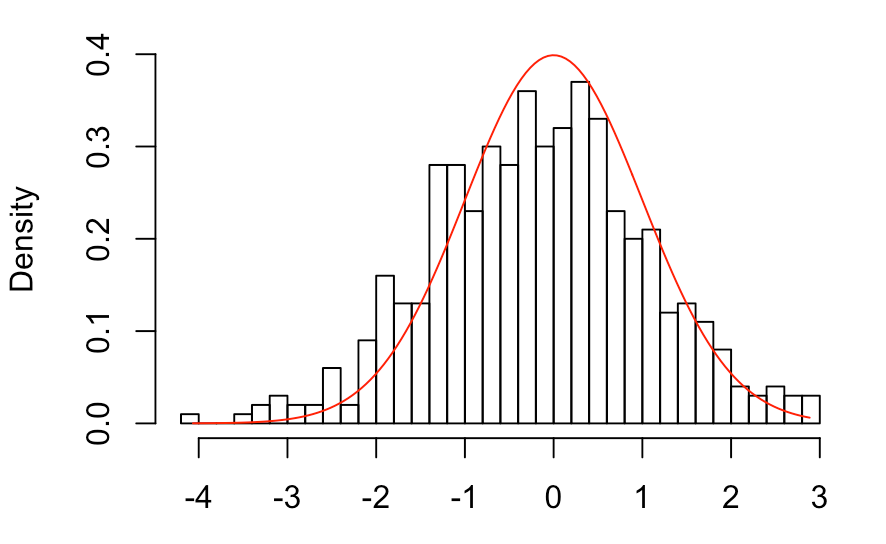}}
\caption{Empirical histograms for the test statistic $T_{\mathrm{PABM}}$ under the null hypothesis when $n=4800$ and $K=2,3,4,5$. The setting is that of PABM graphs on $n = 4800$ vertices with $K$ communities. The red curve is the probability density function
for the standard normal distribution.}
\label{fPABM} 
\end{figure}

\begin{table}[htbp]
\begin{center}
\caption{Empirical sizes for the test statistic under the null hypothesis when $n=4800$ and $K=2,3,4,5$.}
\begin{tabular}{@{}ccccc@{}}
\hline
$K$&$2$& $3$& $4$& $5$\\
\hline
Size&0.052& 0.034& 0.046& 0.086\\
Power&1&1&1&0.81\\
\hline
\end{tabular}
\label{TPABM}
\end{center}
\end{table}
\subsection{American Football data}

\begin{table*}[htbp]
\caption{Sensitivity and specificity of different test statistics for
the  American College Football data. The threshold for classifying a pair of vertices as
having the same latent positions is based on the $95\%$ percentile
of the $\chi^2_{4}$ distribution (for $T_{\mathrm{ASE}}$ and
$T_{\mathrm{LSE}}$) and $\chi^2_{3}$ distribution (for
$G_{\mathrm{ASE}}$ and $G_{\mathrm{LSE}}$).}
\label{T9}
\begin{tabular}{@{}ccccc@{}}
\hline
&$T_{\mathrm{ASE}}$& $T_{\mathrm{LSE}}$& $G_{\mathrm{ASE}}$& $G_{\mathrm{LSE}}$\\
\hline
Sensitivity&0.828& 0.486& 0.585& 0.578\\
Specificity&0.973& 0.952& 0.942& 0.934\\
\hline
\end{tabular}
\end{table*}

\begin{figure}[h]
\centering
\subfigure[$T_{\mathrm{ASE}}$]
{\label{figa4} 
\includegraphics[width=0.33\textwidth]{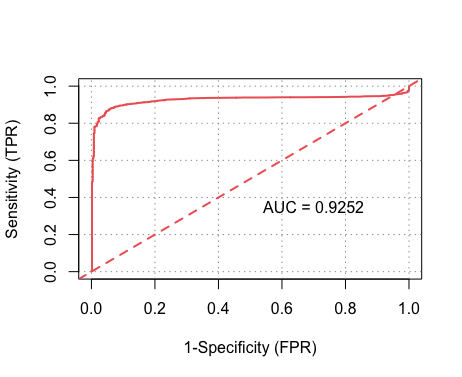}}
\subfigure[$T_{\mathrm{LSE}}$]{
\label{figb4} 
\includegraphics[width=0.33\textwidth]{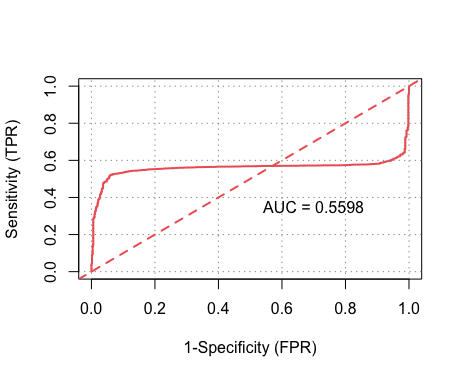}}
\subfigure[$G_{\mathrm{ASE}}$]{
\label{figc4} 
\includegraphics[width=0.33\textwidth]{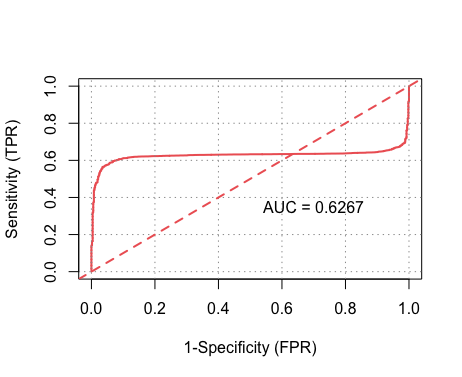}}
\subfigure[$G_{\mathrm{LSE}}$]{
\label{figd4} 
\includegraphics[width=0.33\textwidth]{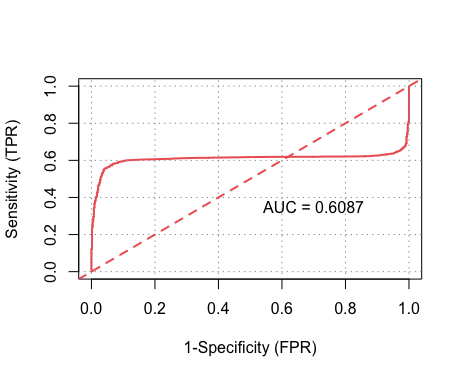}}
\caption{ROC curves and their corresponding AUC for the American College Football data.}
\label{f4} 
\end{figure}

We now examine a network based on the American College Football
data as considered in \citep{girvan2002community}. This network is a
representation of games between Division IA colleges during the regular
season of Fall 2000. Vertices represent teams and two teams are
connected by an edge whenever there is a game played between these teams.
There are 115 vertices and 616
edges. These 115 teams are divided into 12 conferences, each with 5-12
members, and each conference serves as a community/block. Games are more
frequent between teams in the same conference compared to teams in different
conferences. In addition, games between teams from different
conferences are more frequent if these teams are 
geographically close to each other.  For this network we choose the
embedding dimension as $d = 4$ by looking at a scree plot of the
eigenvalues, and among these $d = 4$ eigenvalues, all of them are
positive and thus we set $a = 4$ and $b = 0$. We apply all four test statistics we propose to
each pair of nodes and compute the sensitivity and specificity for
each test statistic for testing the hypothesis that two vertices
belong to the same community. These values are reported in
Table~\ref{T9}; the threshold for classifying a pair of vertices as
having the same latent positions is based on the $95\%$ percentile
of the $\chi^2_{4}$ distribution (for $T_{\mathrm{ASE}}$ and
$T_{\mathrm{LSE}}$) and $\chi^2_{3}$ distribution (for
$G_{\mathrm{ASE}}$ and $G_{\mathrm{LSE}}$).
Note that the number of ``true positives''
and ``true negatives'' are quite unbalanced as, among the
$\tbinom{115}{2}$ different pairs of nodes, there are $517$ pairs of nodes belonging
to the same community and $6038$ pairs of nodes belonging to different
communities.  We also plot, in Figure~\ref{f4}, the ROC curves and the
resulting AUCs for different choices of the test statistics. These ROC
curves are obtained by considering the values of a test
statistic as values of a ``score'' function, and ``classifying'' a given pair of
vertices as being labeled ``$+1$'' or ``$-1$'' depending on whether or not
the ``score'' exceeds some threshold. 
Figure~\ref{f4} indicates that the AUC of $0.925$ for
$T_{\mathrm{ASE}}$ is much higher compared to the
remaining test statistics, and that $T_{\mathrm{ASE}}$
performs well in terms of both the sensitivity and specificity. In
other words, by using $T_{\mathrm{ASE}}$, we are able to correctly infer the membership
relationships between almost all pair of nodes.  The incorrectness
might come from the nonuniform pattern of games between different
conferences and small sample size.

\section{Non-identifiability of latent positions for GRDPG}
\label{sec:appendix_example_Z}
We now provide a simple example for why the
representation $\mathbf{Z} = \mathbf{U} |\mathbf{S}|^{1/2}$ obtained
from the eigendecomposition of the edge probability matrix
$\mathbf{P}$ is generally not suitable for generating $\mathbf{P}$. 

Consider a mixed membership stochastic block model
   with block probabilities matrix
   $$\mathbf{B}= 0.4 \bm{1} \bm{1}^{\top} - 0.2 \mathbf{I} = \begin{bmatrix} 0.2 & 0.4 & 0.4 \\ 0.4 & 0.2 & 0.4 \\ 0.4 & 0.4 & 0.2 \end{bmatrix}.$$
The matrix $\mathbf{B}$ induces a generalized random dot product graph
with signature $a = 1$ and $b = 2$. Let $v_1, v_2, v_3$ be the vectors
in $\mathbb{R}^{3}$ given by
$$v_1 \approx (-0.577, 0.365, 0), \quad v_2 \approx (-0.577, -0.182,
-0.316), \quad v_3 \approx (-0.577, -0.182, 0.316).$$
The $\{v_1, v_2, v_3\}$ are the latent positions generating the block probabilities
matrix $\mathbf{B}$, i.e., $v_i^{\top} \mathbf{I}_{1,2} v_i = 0.2$ for all $i$ and $v_i^{\top}
\mathbf{I}_{1,2} v_j = 0.4$ for all $j \not = i$. Now given any node $i$ with mixed
membership vector $\pi_i = (\pi_{i1}, \pi_{i2}, \pi_{i3})$, we can
assign to $i$ the latent position representation $X_i = \sum_{j}
\pi_{ij} v_j$. Next consider two mixed membership SBM graphs on $4$
vertices with $\pi_1 = (1,0,0), \pi_2 = (0,1,0)$, and $\pi_3 = (0,0,1)$ for both graphs, $\pi_4 =
(0.25, 0.25, 0.5)$ for the first graph and $\pi_4 = (0.2, 0.6, 0.2)$
for the second graph. Let $\mathbf{P}^{(1)}$ and $\mathbf{P}^{(2)}$ be
the edge probabilities matrix for these graphs and let $\mathbf{Z}^{(1)}$ and $\mathbf{Z}^{(2)}$ be the
representations obtained from the eigendecompositions of
$\mathbf{P}^{(1)}$ and $\mathbf{P}^{(2)}$. The first
three rows of $\mathbf{Z}^{(1)}$ and $\mathbf{Z}^{(2)}$ correspond to
the same vertices; in particular the second and third row of $\mathbf{Z}^{(1)}$
and $\mathbf{Z}^{(2)}$ are the vertices with $\pi_2 = (0, 1, 0)$ and
$\pi_3 = (0, 0, 1)$. We expect their representations to be
identical in $\mathbf{Z}^{(1)}$ and $\mathbf{Z}^{(2)}$. This
is, however, not the case; the eigendecomposition of $\mathbf{P}^{(1)}$ and
$\mathbf{P}^{(2)}$ using \texttt{R} yields
\begin{gather*}
  Z^{(1)}_2 = (-0.584, -0.202, -0.316), \,\, \|Z^{(1)}_2\| \approx 0.69, \quad
  Z^{(2)}_{2} = (-0.558, 0.334, 0), \,\, \|Z^{(2)}_2\| \approx 0.65, \\
  Z^{(1)}_3 = (-0.565, 0.346, 0), \,\, \|Z^{(1)}_3\| \approx 0.66, \quad Z^{(2)}_3
  \approx (-0.588, -0.214, 0.316), \,\, \|Z^{(2)}_3\| \approx 0.70.
\end{gather*}
Here $Z^{(k)}_{i}$ denote the $i$th row of
$\mathbf{Z}^{(k)}$ and $\|\cdot\|$ denote the $\ell_2$ norm of a
vector. As $\|Z^{(1)}_2\| \not = \|Z^{(2)}_2\|$ and $\|Z^{(1)}_3\| \not
= \|Z^{(2)}_3\|$, these representations cannot be aligned using orthogonal transformations. 
In summary, by changing the mixed membership
vector of a single vertex $\pi_4$ we also change the latent vectors
representation of the remaining vertices in $\mathbf{Z}^{(1)}$ and
$\mathbf{Z}^{(2)}$. 

\subsection*{Proofs of Proposition~\ref{prop:non_identifiable1} and Proposition~\ref{prop:non_identifiable2}}
\label{sec:proofs_proposition}
We first prove Proposition~\ref{prop:non_identifiable1}.
Let $\mathbf{X}$ and $\mathbf{Y}$ be $n \times d$ matrices of
full-column ranks. Then both $\mathbf{X}^{\top} \mathbf{X}$ and
$\mathbf{Y}^{\top} \mathbf{Y}$ are invertible. The condition $\mathbf{X} \mathbf{I}_{a,b} \mathbf{X}^{\top}
= \mathbf{Y} \mathbf{I}_{a,b} \mathbf{Y}^{\top}$ then implies
$$\mathbf{X} = \mathbf{Y} \mathbf{I}_{a,b} \mathbf{Y}^{\top}
\mathbf{X} (\mathbf{X}^{\top} \mathbf{X})^{-1} \mathbf{I}_{a,b}.$$
Now let $\mathbf{Q} = \mathbf{I}_{a,b} \mathbf{Y}^{\top}
\mathbf{X} (\mathbf{X}^{\top} \mathbf{X})^{-1}
\mathbf{I}_{a,b}$. We then have
$$\mathbf{Q}^{\top} \mathbf{I}_{a,b} \mathbf{Q} = \mathbf{I}_{a,b}
(\mathbf{X}^{\top} \mathbf{X})^{-1} \mathbf{X}^{\top} \mathbf{Y}
\mathbf{I}_{a,b} \mathbf{Y}^{\top} \mathbf{X} (\mathbf{X}^{\top} \mathbf{X})^{-1}
\mathbf{I}_{a,b} = \mathbf{I}_{a,b}$$
and hence $\mathbf{Q}^{\top}$ is an indefinite orthogonal
matrix. Therefore $\mathbf{Q}$ is also indefinite orthogonal, and
Proposition~\ref{prop:non_identifiable1} is established.

We now prove Proposition~\ref{prop:non_identifiable2}. We will prove a
stronger result in that $\|\mathbf{Z}\|_{*} \leq
\|\mathbf{X}\|_{*}$ for any unitarily invariant $Q$-norm
$\|\cdot\|_{*}$. A norm $\|\cdot\|_{*}$ is said to be a $Q$-norm if
there exists a unitarily invariant norm $\|\cdot\|_{\mathrm{UI}}$ such that
$\|\mathbf{M}\|_{*}^2 = \|\mathbf{M}^{\top} \mathbf{M}\|_{\mathrm{UI}}$ for all
matrices $\mathbf{M}$. The Schatten $p$-norms for $p
\geq 2$, which include the spectral norm ($p = \infty$) and the
Frobenius norm ($p = 2$), are all $Q$-norms; see \citep{bhatia_qnorm} for further
discussion. 

Let $\mathbf{X}$ be any matrix such that $\mathbf{X} \mathbf{I}_{a,b}
\mathbf{X}^{\top} = \mathbf{P}$. Then from
Proposition~\ref{prop:non_identifiable1}, we have $\mathbf{X} =
\mathbf{Z} \mathbf{Q}$ for some indefinite orthogonal matrix
$\mathbf{Q}$. We first consider the singular values of $\mathbf{X}$; let
$\sigma_j(\cdot)$ and $\lambda_j(\cdot)$ denote the singular values
and eigenvalues of some matrix $(\cdot)$, respectively. We then have
$$\sigma_j(\mathbf{X})^2 = \lambda_j(\mathbf{Q}^{\top}
\mathbf{Z}^{\top} \mathbf{Z} \mathbf{Q}) = \lambda_j(\mathbf{Q}^{\top}
|\mathbf{S}| \mathbf{Q}) = 
\lambda_j(|\mathbf{S}|^{1/2} \mathbf{Q} \mathbf{Q}^{\top} |\mathbf{S}|^{1/2}).$$
Let $\mathbf{M} = \mathbf{Q} \mathbf{Q}^{\top}$. Then the diagonal entries of $|\mathbf{S}|^{1/2} \mathbf{Q}
\mathbf{Q}^{\top} |\mathbf{S}|^{1/2}$ are of the form $|\lambda_i|
m_{ii}$ where $|\lambda_i| > 0$ are the eigenvalues in
$|\mathbf{S}|$. 
Now $m_{ii} =
\sum_{j} q_{ij}^2$ where $q_{ij}$ is the $ij$th entry of
$\mathbf{Q}$. Since
$\mathbf{Q} \mathbf{I}_{a,b} \mathbf{Q}^{\top} = \mathbf{I}_{a,b}$, we have
\begin{equation}
  \label{eq:qqt_form}\sum_{j \leq a} q_{ij}^2 - \sum_{j \geq a + 1} q_{ij}^2
= \begin{cases} 1 & \text{if $i \leq a$}, \\
  -1 & \text{if $i \geq a + 1$}. \end{cases}
\end{equation}
In either cases, we have $\sum_{j} q_{ij}^2 \geq 1$ for all $i$, i.e.,
$m_{ii} \geq 1$. Hence
$$(|\lambda_1|, |\lambda_2|, \dots, |\lambda_d|) \leq \mathrm{diag}(|\mathbf{S}|^{1/2} \mathbf{Q}
\mathbf{Q}^{\top} |\mathbf{S}|^{1/2})$$
where $\mathrm{diag}(\cdot)$ refers to the diagonal elements of the
matrix $(\cdot)$ and $\leq$ denote elementwise ordering. Now by the Schur majorization theorem for
eigenvalues, we have
$$\mathrm{diag}(|\mathbf{S}|^{1/2} \mathbf{Q}
\mathbf{Q}^{\top} |\mathbf{S}|^{1/2}) \prec
(\lambda_1(|\mathbf{S}|^{1/2} \mathbf{Q} \mathbf{Q}^{\top}
|\mathbf{S}|^{1/2}), \dots, \lambda_d(|\mathbf{S}|^{1/2} \mathbf{Q} \mathbf{Q}^{\top}
|\mathbf{S}|^{1/2})),$$
and hence, from Fact~5.A.9 of \citep{marshall_olkin}, we have
$$(|\lambda_1|, \dots, |\lambda_d|)  \prec_{w} (\lambda_1(|\mathbf{S}|^{1/2} \mathbf{Q} \mathbf{Q}^{\top}
|\mathbf{S}|^{1/2}), \dots, \lambda_d(|\mathbf{S}|^{1/2} \mathbf{Q} \mathbf{Q}^{\top}
|\mathbf{S}|^{1/2})).$$
Here $u \prec v$ and $u \prec_{w} v$ denote that $u$ is majorized
by $v$ and $u$ is weakly majorized by $v$, respectively. Therefore, by Fan's domination theorem \citep[Theorem~IV.2.2]{bhatia},
we have
$$\|\mathbf{Z}\|_* = \|\mathbf{S}\|_* \leq \||\mathbf{S}|^{1/2} \mathbf{Q} \mathbf{Q}^{\top}
|\mathbf{S}|^{1/2}\|_{\mathrm{UI}}^{1/2} = \|\mathbf{X}\|_{*}. $$ 
as desired.

We now show that if $\|\mathbf{X}\|_{F} = \|\mathbf{Z}\|_{F}$ then
$\mathbf{Z} = \mathbf{X} \mathbf{W}$ for some block orthogonal matrix
  $\mathbf{W}$. Suppose $\|\mathbf{X}\|_{F}^2 = \sum_{i} m_{ii}
  |\lambda_{i}| = \sum_{i} |\lambda_i| = \|\mathbf{Z}\|_{F}^2$. Since $m_{ii} \geq 1, |\lambda_i| > 0$ for all $i$, we
  must have $m_{ii} = 1$ for all $i$. Now $m_{ii} = \sum_{j} q_{ij}^2$
  and, together with Eq.~\eqref{eq:qqt_form}, we have
\begin{gather*}\sum_{j \leq a} q_{ij}^2 = 1, \quad \sum_{j \geq a +1} q_{ij}^2 = 0,
  \quad \text{for $i \leq a$}, \\
  \sum_{j \leq a} q_{ij}^2 = 0, \quad \sum_{j \geq a +1} q_{ij}^2 = 1,
  \quad \text{for $i \geq a+1$}. 
\end{gather*}
Hence $q_{ij} = 0$ whenever $i \leq a, j \geq a + 1$ or $i \geq a + 1,
j \leq a$. The matrix $\mathbf{Q}$ can thus be decomposed into
diagonal blocks of the form $\mathbf{Q} = \Bigl[\begin{smallmatrix}
  \mathbf{Q}_{1} & 0 \\ 0 & \mathbf{Q}_2 \end{smallmatrix} \Bigr]$
where $\mathbf{Q}_1$ and $\mathbf{Q}_2$ are of size $a \times a$ and
$b \times b$, respectively. Finally, since $\mathbf{Q}
\mathbf{I}_{a,b} \mathbf{Q}^{\top} = \mathbf{I}_{a,b}$, we conclude
that $\mathbf{Q}_1 \mathbf{Q}_1^{\top} = \mathbf{I}$ and
$\mathbf{Q}_{2} \mathbf{Q}_2^{\top} = \mathbf{I}$ and hence
$\mathbf{Q}$ is a block-orthogonal matrix. 
\section{Testing the Degree-corrected Hypothesis using LSE}
\label{sec:app_lse}
We now discuss the test statistic for testing $\mathbb{H}_0 \colon X_i/\|X_i\| = X_j/\|X_j\|$ using $\breve{\mathbf{X}}.$
\begin{theorem}\label{thm4}
  Consider the setting in Theorem~\ref{thm3}. Let $s(\xi) =
  \xi/\|\xi\|$ and $\mathbf{J}(\xi) = \bigl(\|\xi\|^2
\mathbf{I}-\xi\xi^{\top}\bigr)/\|\xi\|^3$ be its Jacobian
transformation.  
Now define the test statistic
$$G_{\mathrm{LSE}}(\breve{X}_i,\breve{X}_j)=n^2\rho_n\bigl(s(\breve{X}_i)-s(\breve{X}_j)\bigr)^{\top}\Bigl(\mathbf{J}(\breve{X}_i)\Bigl[\breve{\mathbf{\Sigma}}(\breve{X}_i)+\tfrac{\|\breve{X}_i\|^2}{\|\breve{X}_j\|^2}
\breve{\mathbf{\Sigma}}(\breve{X}_j)\Bigr]\mathbf{J}(\breve{X}_i)\Bigr)^{\dagger}\bigl(s(\breve{X}_i)-s(\breve{X}_j)\bigr).$$
where $\breve{\bm{\Sigma}}(\breve{X}_i)$ and $\breve{\bm{\Sigma}}(\breve{X}_j)$ are as defined in 
Theorem~\ref{thm2}.
Then under $\mathbb{H}_0 \colon X_i/\|X_i\|=X_{j}/\|X_j\|$ and for
$n \rightarrow \infty$ with $n \rho_n = \omega(\log n)$, we have
$$G_{\mathrm{LSE}}(\breve{X}_i, \breve{X}_j) \rightsquigarrow \chi_{d-1}^2.$$
Recall the definition of $\tilde{\bm{\Sigma}}(X_i)$ in
Theorem~\ref{thm2}. Let $\tilde{\mathbf{X}} = \rho_n^{1/2}
\mathbf{T}^{-1/2} \mathbf{X}$ and let $\tilde{\mathbf{Q}}_{\mathbf{X}}$ be the indefinite
orthogonal transformation such that $\tilde{\mathbf{X}} =
\tilde{\mathbf{U}}|\tilde{\mathbf{S}}|^{1/2}
\tilde{\mathbf{Q}}_{\mathbf{X}}$ where $\tilde{\mathbf{U}}
\tilde{\mathbf{S}} \tilde{\mathbf{U}}^{\top}$ is the
eigendecomposition of $\tilde{\mathbf{X}} \mathbf{I}_{a,b}
\tilde{\mathbf{X}}^{\top}$.
Next define
$$
\tilde{Z}_i = (\tilde{\mathbf{Q}}_{\mathbf{X}}^{-1})^{\top}
\rho_n^{1/2} X_i/\sqrt{t_i}, \quad
\tilde{\bm{\Sigma}}(\tilde{Z}_i) = (\tilde{\mathbf{Q}}_{\mathbf{X}}^{-1})^{\top}
\tilde{\bm{\Sigma}}(X_i) \tilde{\mathbf{Q}}_{\mathbf{X}}^{-1}.
$$
Let $\tilde{\mu} > 0$ be a finite constant such that $\tfrac{X_i}{\|X_i\|} \not =
\tfrac{X_j}{\|X_j\|}$ satisfies a local alternative where
\begin{equation}
  \label{eq:ncp_G_LSE}
n^2 \rho_n \bigl(s(\tilde{Z}_i)-s(\tilde{Z}_j)\bigr)^{\top}\bigl(\mathbf{J}(\tilde{Z}_i) \tilde{\bm{\Sigma}}(\tilde{Z}_i) \mathbf{J}(\tilde{Z}_i) +\mathbf{J}(\tilde{Z}_j)\tilde{\bm{\Sigma}}(\tilde{Z}_j) \mathbf{J}(\tilde{Z}_j)\bigr)^{\dagger}
\bigl(s(\tilde{Z}_i)-s(\tilde{Z}_j)\bigr) \rightarrow \tilde{\mu}.
\end{equation}
Then $\mathrm{G}_{\mathrm{LSE}}(\breve{Y}_i, \breve{Y}_{j}) \rightsquigarrow \chi_{d-1}^2\left(\tilde{\mu}\right)$
where $\chi_{d-1}^2\left(\tilde{\mu}\right)$ is the
noncentral chi-square distribution with $d-1$ degrees of freedom and
    noncentrality parameter $\tilde{\mu}$.
\end{theorem} 
We note that, for ease of
exposition, the non-centrality
parameter $\tilde{\mu}$ in Eq.~\eqref{eq:ncp_G_LSE} depends on the choice of
representation $\{\tilde{Z}_i\}$; nevertheless, if desired, one can
derive a condition for $\tilde{\mu}$ that is similar to
Eq.~\eqref{eq:indefinite_alt_cond2} and is invariant with respect to the
non-identifiability of $\{X_i\}$.
\section{Proofs of main results}
\label{sec:proof_main}
We first discuss how the result in Theorem~\ref{thm0} relates to existing 
results in \citep{rubin2017statistical}. We emphasize that the normal approximation in
Theorem~4 of \citep{rubin2017statistical} assumes
that the rows of $\mathbf{X} = [X_1, \dots, X_n]^{\top}$ are sampled iid from some distribution $F$. This
iid assumption is not essential; rather
it is enforced so that we can obtain a simpler convergence statement
wherein the limiting covariance matrix for $\hat{X}_{i}$ 
depends only on the latent position $X_i$ and not on any other $X_j$
for $j \not =i$. In any case, the proofs in
\citep{rubin2017statistical} only require slight modification so that
the results still hold given a {\em deterministic} $\mathbf{X}^{(n)}$,
provided that $\mathbf{X}^{(n)}$ satisfies Condition~1 through Condition~3. More specifically,
Eq.~(15) in \citep{rubin2017statistical} yields the expansion
\begin{equation}
  \label{eq:sketch}
  \begin{split}
\sqrt{n} (\mathbf{Q}_{\mathbf{X}}^{\top} \mathbf{W}_n \hat{X}_i -
\rho_n^{1/2} X_i) &= \sqrt{n} \sum_{j} (a_{ij} - p_{ij}) \rho_n^{1/2}
X_j (\rho_n \mathbf{X}^{\top} \mathbf{X})^{-1} + o(1) \\ &=
\frac{1}{\sqrt{n \rho_n}} \sum_{j} (a_{ij} - p_{ij}) X_j
\Bigl(\frac{\mathbf{X}^{\top} \mathbf{X}}{n}\Bigr)^{-1} + o(1).
\end{split}
\end{equation}
Now the term $\sum_{j} (a_{ij} -
p_{ij}) X_j$ is a sum of {\em independent} mean $0$
random vectors. Eq.~\eqref{eq:convergence} then follows from the Lindeberg-Feller central limit
theorem together with the observation that
$$\mathrm{var}\Bigl[\frac{1}{\sqrt{n \rho_n}} \sum_{j} (a_{ij} -
p_{ij}) X_j\Bigr] = \frac{1}{n} \sum_{j} X_j X_j^{\top} X_i^{\top}
\mathbf{I}_{a,b} X_j (1 - \rho_n X _i^{\top} \mathbf{I}_{a,b} X_j).$$
If we also assume that the $X_1, \dots, X_n$ are iid 
samples from some distribution $F$, 
then we can simplify the covariance matrix in
Eq.~\eqref{eq:cov_X_theoretical} and obtain, in place of
Eq.~\eqref{eq:cov_X_theoretical}, the  covariance matrix 
  \begin{equation}
    \label{eq:theoretical2}
  \bm{\Sigma}(X_i) = \mathbf{I}_{a,b} (\mathbb{E}[\xi
  \xi^{\top}])^{-1} \mathbb{E}[\xi \xi^{\top} X_i^{\top}
  \mathbf{I}_{a,b} \xi (1 - \rho_n X_i^{\top} \mathbf{I}_{a,b} \xi)] (\mathbb{E}[\xi
  \xi^{\top}])^{-1} \mathbf{I}_{a,b}
  \end{equation}
  where all of the expectations in Eq.~\eqref{eq:theoretical2}
  are taken with respect to a random vector
  $\xi \sim F$. Indeed, the strong law of large
  numbers implies $\frac{\mathbf{X}^{\top} \mathbf{X}}{n} \overset{\mathrm{a.s}}{\longrightarrow} \mathbb{E}[\xi
  \xi^{\top}]$ and
  $$\frac{1}{n} \sum_{k=1}^{n} X_k X_k^{\top} X_i^{\top}
  \mathbf{I}_{a,b} X_k (1 - \rho_n X_i^{\top} \mathbf{I}_{a,b}
  X_k) \overset{\mathrm{a.s}}{\longrightarrow} \mathbb{E}[\xi \xi^{\top} X_i^{\top}
  \mathbf{I}_{a,b} \xi (1 - \rho_n X_i^{\top} \mathbf{I}_{a,b} \xi)].$$

  The covariance matrix $\bm{\Sigma}(X_i)$ in Eq.~\eqref{eq:theoretical2} appeared in
  Theorem~4 of \citep{rubin2017statistical} and is a {\em deterministic} function that depends
  only on $X_i$. The normal approximation for the iid setting can
  therefore be written as $r_i^{(n)}
  \rightsquigarrow \mathcal{N}(0, \bm{\Sigma}(X_i)).$
  In contrast, if we only assume that
  the $\{X_i\}$ satisfy Conditions~1 and
  Condition~2 then $\bm{\Sigma}(X_i)$ as defined in
  Eq.~\eqref{eq:cov_X_theoretical} need not converge and hence writing
  $\sqrt{n} \bigl(\mathbf{Q}_{\mathbf{X}}^{\top} \mathbf{W}_n \hat{X}_i -
  \rho_n^{1/2} X_i\bigr) \rightsquigarrow \mathcal{N}\bigl(0,
  \bm{\Sigma}(X_i)\bigr)$ is slightly inaccurate. Hence, for ease of intuition, we interpret Eq.~\eqref{eq:convergence} as saying that $r_i^{(n)}$ has the
  same limiting distribution as the random vector
  $\bm{\Sigma}(X_i)^{1/2} \mathcal{N}(0, \mathbf{I})$.

Now recall that, from the definition of $\mathbf{Z}^{(n)}$, we have
$\rho_n^{1/2} X_i = \mathbf{Q}_{\mathbf{X}^{(n)}}^{\top} Z_i^{(n)}$
and hence Eq.~\eqref{eq:convergence} can be rewritten as
\begin{equation}
  \label{eq:convergence_2}
\sqrt{n} \bm{\Sigma}(X_i^{(n)}; \mathbf{X}^{(n)})^{-1/2}
\mathbf{Q}_{\mathbf{X}^{(n)}}^{\top}
\bigl(\mathbf{W}_n \hat{X}_i^{(n)} - Z_i^{(n)}\bigr) \rightsquigarrow \mathcal{N}\bigl(0,
  \mathbf{I}).
\end{equation}
Eq.~\eqref{eq:convergence_2} and Eq.~\eqref{eq:convergence_Z} are
equivalent.
Indeed the definition of $\bm{\Sigma}(Z_i^{(n)})$ 
in Theorem~\ref{thm0} implies $\bm{\Sigma}(Z_i^{(n)}) = (\mathbf{Q}_{\mathbf{X}^{(n)}}^{-1})^{\top}
\bm{\Sigma}(X_i^{(n)}; \mathbf{X}^{(n)})
\mathbf{Q}_{\mathbf{X}^{(n)}}^{-1}$ and hence there exists an
orthogonal matrix $\mathbf{W}_*$ such that
$\bm{\Sigma}(Z_i^{(n)})^{1/2}  =
(\mathbf{Q}_{\mathbf{X}^{(n)}}^{-1})^{\top} \bm{\Sigma}(X_i^{(n)};
\mathbf{X}^{(n)})^{1/2} \mathbf{W}_*$. We can thus rewrite Eq.~\eqref{eq:convergence_2}
as
$$\sqrt{n} \mathbf{W}_* \bm{\Sigma}(Z_i^{(n)})^{-1/2}
\bigl(\mathbf{W}_n \hat{X}_i^{(n)} - Z_i^{(n)}\bigr) \rightsquigarrow \mathcal{N}\bigl(0,
  \mathbf{I}).$$
Eq.~\eqref{eq:convergence_Z} follows from the above display and
the fact that $\mathcal{N}(0, \mathbf{I})$ is spherically symmetric. 

We now proceed with the proofs in Section~\ref{sec3}. We start by stating an important technical lemma 
for bounding the maximum $\ell_2$ norm of the errors $\mathbf{W}_n
\hat{X}_i - Z_i, i = 1,2,\dots,n$. The proof of this lemma is given in \citep{rubin2017statistical}.

\begin{lem}
  \label{lem:norm_bounds}
  Let $\mathbf{A} \sim \mathrm{GRDPG}(\mathbf{X}, \rho_n)$ be a
  generalized random dot product graph on $n$ vertices with sparsity
  factor $\rho_n$ and signature $(a,b)$. Suppose that $\mathbf{X}$
  satisfies Conditions~1 through Conditions~3 in
  Section~\ref{sec:ase}. Then there exists a $(a,b)$ block orthogonal
  matrix $\mathbf{W}_n$ such that, with high probability,
    \begin{equation}
    \label{eq:2toinfty}
  \max_{i} \|\mathbf{W}_n \hat{X}_i - Z_i\| =
  O_{\mathbb{P}}(\frac{\log n}{\sqrt{n}}\Bigr).
  \end{equation}
\end{lem}
Eq.~\eqref{eq:2toinfty} implies a Frobenius norm bound of
\begin{equation}
  \label{eq:frob_bound}
  \|\hat{\mathbf{X}} \mathbf{W}_n^{\top} - \mathbf{Z}\|_{F} = O_{\mathbb{P}}(\log n).
\end{equation}
Note that a stronger bound than Eq.~\eqref{eq:frob_bound} is possible,
namely that $\|\hat{\mathbf{X}} \mathbf{W}_n - \mathbf{Z}\| =
O_{\mathbb{P}}(1)$; nevertheless Eq.~\eqref{eq:frob_bound} suffices for
the subsequent derivations in this paper. 
\subsection{Proof of Lemma~\ref{lem:conv1}}
We first define
\begin{align*}
  \boldsymbol{\Psi}(Z_i) &= \frac{1}{n \rho_n^2} \sum_{k=1}^{n} 
Z_k Z_k^{\top} Z_i^{\top}
\mathbf{I}_{a,b} Z_k (1 - Z_i^{\top} \mathbf{I}_{a,b}
Z_k), \\
  \hat{\boldsymbol{\Psi}}(\hat{X}_i) &= \frac{1}{n \rho_n^2} \sum_{k=1}^{n} \hat{X}_k \hat{X}_k^{\top}
  \hat{X}_i^{\top} \mathbf{I}_{a,b} \hat{X}_k (1 - \hat{X}_i^{\top}
  \mathbf{I}_{a,b} \hat{X}_k), \\
  \hat{\boldsymbol{\Sigma}}(\hat{X}_i) &=  
  \mathbf{I}_{a,b} \Bigl(\frac{\hat{\mathbf{X}}^{\top}
\hat{\mathbf{X}}}{n \rho_n}\Bigr)^{-1} 
\hat{\boldsymbol{\Psi}}(\hat{X}_i) \Bigl(\frac{\hat{\mathbf{X}}^{\top}
  \hat{\mathbf{X}}}{n \rho_n}\Bigr)^{-1}
\mathbf{I}_{a,b}. 
\end{align*}
Then by Eq.~\eqref{eq:2toinfty} and \eqref{eq:frob_bound}, we have
\begin{gather}
  \label{eq:thm1_proof4}
  \Bigl(\frac{\hat{\mathbf{X}}^{\top} \hat{\mathbf{X}}}{n \rho_n}\Bigr)^{-1}
  - \mathbf{W}_n^{\top} \Bigl(\frac{
    \mathbf{Z}^{\top} \mathbf{Z}}{n \rho_n}\Bigr)^{-1}\mathbf{W}_n
  \overset{\mathrm{a.s.}}{\longrightarrow} 0, \\
  \label{eq:thm1_proof4b}
  \frac{1}{\rho_n^2} \Bigl(\mathbf{W}_n^{\top} \hat{X}_k \hat{X}_k^{\top} \mathbf{W}_n \hat{X}_k^{\top} \mathbf{I}_{a,b}
  \hat{X}_i (1 - \hat{X}_k^{\top} \mathbf{I}_{a,b} \hat{X}_i) - Z_k
  Z_k^{\top} Z_k^{\top} \mathbf{I}_{a,b} Z_i (1 - Z_k^{\top}
  \mathbf{I}_{a,b} Z_i)\Bigr) \overset{\mathrm{a.s.}}{\longrightarrow} 0
\end{gather}
Eq.~\eqref{eq:thm1_proof4} and \eqref{eq:thm1_proof4b} then implies
\begin{equation}
  \label{eq:thm1_proof5}
\hat{\boldsymbol{\Psi}}(\hat{X}_i) - \mathbf{W}_n^{\top}
 \boldsymbol{\Psi}(Z_i) \mathbf{W}_n
\overset{\mathrm{a.s.}}{\longrightarrow} 0.
\end{equation}
Since both $\boldsymbol{\Psi}(Z_i)$ and
$\hat{\boldsymbol{\Psi}}(\hat{X}_i)$ are bounded in spectral norm, we
have
\begin{equation*}
  \begin{split}
\hat{\boldsymbol{\Sigma}}(\hat{X}_i) -
\mathbf{W}_n^{\top}\boldsymbol{\Sigma}(Z_i)\mathbf{W}_n &=\hat{\boldsymbol{\Sigma}}(\hat{X}_i) - \mathbf{W}_n^{\top} \mathbf{I}_{a,b} 
 \Bigl(\frac{
    \mathbf{Z}^{\top} \mathbf{Z}}{n \rho_n}\Bigr)^{-1} \boldsymbol{\Psi}(Z_i) \Bigl(\frac{
    \mathbf{Z}^{\top} \mathbf{Z}}{n \rho_n}\Bigr)^{-1}
 \mathbf{I}_{a,b} \mathbf{W}_n
\\&=
\hat{\boldsymbol{\Sigma}}(\hat{X}_i) -
  \mathbf{I}_{a,b} \mathbf{W}_n^{\top}
\Bigl(\frac{
    \mathbf{Z}^{\top} \mathbf{Z}}{n \rho_n}\Bigr)^{-1} \boldsymbol{\Psi}(Z_i) \Bigl(\frac{
    \mathbf{Z}^{\top} \mathbf{Z}}{n\rho_n}\Bigr)^{-1} \mathbf{W}_n \mathbf{I}_{a,b} \\ &
  \overset{\mathrm{a.s.}}{\longrightarrow} 0.
\end{split}
\end{equation*}
where the second equality follows from the fact that $\mathbf{W}_n$ is
{\em block-orthogonal} and the convergence to $0$ follows from
Eq.~\eqref{eq:thm1_proof4} and Eq.~\eqref{eq:thm1_proof5}.

\subsection{Proof of Theorem~\ref{thm1}}
We first recall the statement of Theorem~\ref{thm0}, i.e.,
\begin{equation}
   \label{eq:thm1_proof1}
 \sqrt{n} \bm{\Sigma}(Z_i)^{-1/2} \bigl(\mathbf{W}_n \hat{X}_i-
  Z_i\bigr) \rightsquigarrow N(0,\mathbf{I}).
\end{equation}
Furthermore, recall that $\sqrt{n}(\mathbf{W}_n \hat{X}_i -
Z_i)$ and $\sqrt{n}(\mathbf{W}_n \hat{X}_j - Z_j)$ are asymptotically independent whenever $i \not
= j$. Then under $\mathbb{H}_0 \colon X_i = X_j$, we have
\begin{equation}
  \sqrt{n} \bm{\Sigma}(Z_i)^{-1/2} \mathbf{W}_n (\hat{X}_i - \hat{X}_j) \rightsquigarrow
  \mathcal{N}(\bm{0}, 2 \mathbf{I}).
\end{equation}
and hence, for $X_i = X_j$, that
\begin{equation}
  \label{eq:chisq1}
n\bigl(\hat{X}_i -\hat{X}_j \bigr)^{\top} \mathbf{W}_n^{\top}\bigl(\boldsymbol{\Sigma}(Z_i) +\boldsymbol{\Sigma}(Z_j)\bigr)^{-1} \mathbf{W}_n \bigl(\hat{X}_i -
\hat{X}_j\bigr) \rightsquigarrow \chi^{2}_{d}
\end{equation}
as $n \rightarrow \infty$. Note that in Eq.~\eqref{eq:chisq1} we have used $\bm{\Sigma}(Z_i) +
\bm{\Sigma}(Z_j)$ as opposed to $2 \bm{\Sigma}(Z_i)$; indeed, for the 
construction of our test statistic we will not know whether
the null hypothesis is true and hence it is simpler to use
$\bm{\Sigma}(Z_i) + \bm{\Sigma}(Z_j)$ in the subsequent
derivations.
Now Lemma~\ref{lem:conv1} implies
\begin{equation}
  \label{eq:thm1_proof5b}
  \hat{\boldsymbol{\Sigma}}(\hat{X}_i) +
\hat{\boldsymbol{\Sigma}}(\hat{X}_j)  -
\mathbf{W}_n^{\top}
\bigl(\boldsymbol{\Sigma}(Z_i) +\boldsymbol{\Sigma}(Z_j)\bigr) \mathbf{W}_n
\overset{\mathrm{a.s.}}{\longrightarrow} 0.
\end{equation}
We therefore have
\begin{equation}
  \label{eq:thm1_proof6} 
\Bigl(\hat{\boldsymbol{\Sigma}}(\hat{X}_i) +
\hat{\boldsymbol{\Sigma}}(\hat{X}_j)\Bigr)^{-1}-\mathbf{W}_n^{\top}
\bigl(\boldsymbol{\Sigma}(Z_i) +\boldsymbol{\Sigma}(Z_j)\bigr)^{-1} \mathbf{W}_n
\overset{\mathrm{a.s.}}{\longrightarrow} 0
\end{equation}
We thus conclude, by Eq.~\eqref{eq:chisq1} and Slutsky's theorem that
$$n\bigl(\hat{X}_i - \hat{X}_j \bigr)^{\top} \bigl(\hat{\boldsymbol{\Sigma}}(\hat{X}_i) +
\hat{\boldsymbol{\Sigma}}(\hat{X}_j)\bigr)^{-1} \bigl(\hat{X}_i -  
\hat{X}_j\bigr) \rightsquigarrow \chi^{2}_{d}
$$
as desired.

We next derive the condition for the local alternative. Let $\sqrt{n
  \rho_n}(X_i - X_j)$ be bounded as $n \rightarrow \infty$. Then
$\|X_i - X_j\| = o(1)$ and hence $\|\bm{\Sigma}(Z_i) -
\bm{\Sigma}(Z_j)\| \rightarrow 0$. Then from
Eq.~\eqref{eq:thm1_proof1} and Slutsky's theorem together 
with the asymptotic independence of $\hat{X}_i$ and $\hat{X}_j$ for $i
\not = j$, we have
$$ \sqrt{n} (\bm{\Sigma}(Z_i) + \bm{\Sigma}(Z_j))^{-1/2} \left[\mathbf{W}_n (\hat{X}_i -
\hat{X}_j)-(Z_i - Z_j)\right] \rightsquigarrow \mathcal{N}( 0,
\mathbf{I})$$
Let $\zeta_{ij} = \sqrt{n} (\bm{\Sigma}(Z_i) + \bm{\Sigma}(Z_j))^{-1/2} \mathbf{W}_n (\hat{X}_i -
\hat{X}_j)$. Recall that the condition for the local alternative is
$n (Z_i - Z_j)^{\top}(\bm{\Sigma}(Z_i) +
\bm{\Sigma}(Z_j))^{-1} (Z_i - Z_j) \rightarrow \mu$. We therefore have
$$\|\zeta_{ij}\|^2 \rightsquigarrow \chi^2_{d}(\mu).$$
Recalling the definition of $Z_i
= \rho_n^{1/2} (\mathbf{Q}_{\mathbf{X}}^{-1})^{\top} X_i$ together
with the definition of $\bm{\Sigma}(Z_i)$ in Theorem~\ref{thm0}, we have
\begin{equation*}
  \begin{split}
     & \quad n (Z_i - Z_j)^{\top}(\bm{\Sigma}(Z_i) +
\bm{\Sigma}(Z_j))^{-1} (Z_i - Z_j) \\ &= n \rho_n (X_i - X_j)^{\top}
\mathbf{Q}_{\mathbf{X}}^{-1}
\Bigl((\mathbf{Q}_{\mathbf{X}}^{-1})^{\top} \bigl(\bm{\Sigma}(X_i) +
\bm{\Sigma}(X_j)\bigr) \mathbf{Q}_{\mathbf{X}}^{-1} \Bigr)^{-1}
(\mathbf{Q}_{\mathbf{X}}^{-1})^{\top} (X_i - X_j) \\
&= n \rho_n (X_i - X_j)^{\top} \Bigl(\bm{\Sigma}(X_i) +
\bm{\Sigma}(X_j)\Bigr)^{-1}(X_i - X_j).
\end{split}
\end{equation*}
Eq.~\eqref{eq:noncentral1} is thus established. Finally, from Eq.~\eqref{eq:thm1_proof6}, we have 
$T_{\mathrm{ASE}}(\hat{X}_i, \hat{X}_j) - \|\zeta_{ij}\|^2 \rightarrow
0$ and hence $T_{\mathrm{ASE}}(\hat{X}_i, \hat{X}_j) \rightsquigarrow
\chi^2_{d}(\mu)$ as desired.

\subsection{Proof of Theorem~\ref{thm3}}
First recall the limit result in Theorem~\ref{thm0}, namely
\begin{equation}
  \label{eq:thm3_proof1}
\sqrt{n} \bm{\Sigma}(Z_i)^{-1/2} (\mathbf{W}_n \hat{X}_i - Z_i) \rightsquigarrow
\mathcal{N}\bigl(0, \mathbf{I}\bigr)
\end{equation}
Now recall the definition of $s(\xi) = \xi/\|\xi\|$ and its Jacobian
$\mathbf{J}(\xi) = (\|\xi\|^2 \mathbf{I} - \xi
\xi^{\top})/\|\xi\|^3$.
Then for any vector $\xi \in \mathbb{R}^{d}$ and any $d \times d$ orthogonal matrix
$\mathbf{W}$, we have
\begin{equation}
  \label{eq:prop_s_jacobian}
  s(\mathbf{W} \xi) = \mathbf{W} s(\xi), \quad \mathbf{J}(\mathbf{W} \xi) = \frac{\|\mathbf{W} \xi\|^2 \mathbf{I} -
  \mathbf{W} \xi \xi^{\top} \mathbf{W}^{\top}}{\|\mathbf{W} \xi\|^3} =
\mathbf{W} \mathbf{J}(\xi) \mathbf{W}^{\top}.
\end{equation}
Furthermore, for any constant $c > 0$, we have $\mathbf{J}(c \xi) =
c^{-1} \mathbf{J}(\xi)$. 
Note, however, that $s(X_i) \not = \mathbf{Q}_{\mathbf{X}} s(Z_i)$
unless $b = 0$ so that $\mathbf{Q}_{\mathbf{X}}$ reduces to an
orthogonal matrix.
Next we note that if $\xi \in \mathbb{R}^{d}$ then $\mathbf{J}(\xi)$
is of rank $d - 1$. Indeed, $\mathbf{J}(\xi) \xi = \bm{0}$ for any
vector $\xi$.

From Eq.~\eqref{eq:thm3_proof1} together with the delta method, we have
\begin{equation}
  \label{eq:thm3_proof2}
  \sqrt{n \rho_n} \bigl( \mathbf{W}_n s( \hat{X}_i) - s(Z_i)\bigr)
- \rho_n^{1/2} \mathbf{J}(Z_i) \bm{\Sigma}(Z_i)^{1/2}
\mathcal{N}(0, \mathbf{I}) \rightarrow 0
\end{equation}
in probability.
We emphasize the difference in scaling between Eq.~\eqref{eq:thm3_proof1} and
Eq.~\eqref{eq:thm3_proof2}; indeed, if $\rho_n \rightarrow 0$ then
$\|\hat{X}_i\| = \Theta(\rho_n^{1/2})$ but $\|s(\hat{X}_i)\| = 1$. 

Therefore, for any $i \not =j$, by the asymptotic
independence of $\hat{X}_i$ and $\hat{X}_j$, we have
\begin{equation}
  \label{eq:thm3_proof2a}
  \begin{split}
  \sqrt{n \rho_n} \, \mathbf{W}_n \bigl(s(\hat{X}_i) - s(\hat{X}_j )\bigr) &=
   \sqrt{n \rho_n} \, (s(Z_i) - s(Z_j))
\\ &+ \rho_n^{1/2} \mathbf{J}(Z_i) \bm{\Sigma}(Z_i)^{1/2}
\zeta_1 + \rho_n^{1/2} \mathbf{J}(Z_j)
\bm{\Sigma}(Z_j)^{1/2} \zeta_2 + o_{\mathbb{P}}(1),
\end{split}
\end{equation}
where $\zeta_1$ and $\zeta_2$ are independent
$\mathcal{N}(0, \mathbf{I})$ random variables. 

Suppose that $\mathbb{H}_0$ is true. Then $X_i = c X_j$ for some
constant $c$. We then have $Z_i = c Z_j$ and hence
$$s(Z_i) = s(Z_j), \quad \mathbf{J}(Z_j) = \frac{\|Z_i\|}{\|Z_j\|}
\mathbf{J}(Z_i) = \frac{\|X_i\|}{\|X_j\|} \mathbf{J}(Z_i).$$
Now suppose that $ X_i/\|X_i\| = X_j/\|X_j\|$ holds. Then
Eq.~\eqref{eq:thm3_proof2a} simplifies to
\begin{equation}
  \label{eq:thm3_proof2b}
   \sqrt{n \rho_n} \, \mathbf{W}_n\bigl( s(\hat{X}_i) - s(\hat{X}_j
   )\bigr) - \rho_n^{1/2}
   \mathbf{J}(Z_i) \Bigl(\bm{\Sigma}(Z_i) +
   \frac{\|Z_i\|^2}{\|Z_j\|^2} \bm{\Sigma}(Z_j)\Bigr)^{1/2}
   \mathcal{N}(0, \mathbf{I}) \rightarrow 0
\end{equation}
in probability, and hence
\begin{equation}
  \label{eq:chisq_thm3}
  n \bigl(s(\hat{X}_i) - s(\hat{X}_j) \bigr)^{\top} \mathbf{W}_n^{\top}
\Bigl(\mathbf{J}( Z_i)
\Bigl(\boldsymbol{\Sigma}(Z_i) + \frac{\|Z_i\|^2}{\|Z_j\|^2} 
\boldsymbol{\Sigma}(Z_j)\Bigr)
\mathbf{J}(Z_i)\Bigr)^{\dagger} \mathbf{W}_n \bigl(s(\hat{X}_i) -  
s(\hat{X}_j)\bigr) \rightsquigarrow \chi^{2}_{d-1}
\end{equation}
as $n \rightarrow \infty$. Here $\mathbf{M}^{\dagger}$ denote the
Moore-Penrose pseudo-inverse of a matrix $\mathbf{M}$.
Now we have, by Eq.~\eqref{eq:2toinfty}, that
\begin{equation}
  \label{eq:conv_Jacobian}
\rho_n^{1/2} \bigl(\mathbf{J}(\hat{X}_i) - \mathbf{W}_n^{\top} \mathbf{J}(Z_i)
\mathbf{W}_n \bigr) = \frac{\rho_n^{1/2}}{\|\hat{X}_i\|}\Bigl( \mathbf{I} - \tfrac{\hat{X}_i
  \hat{X}_i^{\top}}{\|\hat{X}_i\|^2}\Bigr) - \frac{\rho_n^{1/2}}{\|Z_i\|}
\mathbf{W}_n^{\top} \Bigl( \mathbf{I} - \tfrac{Z_i
  Z_i^{\top}}{\| Z_i\|^2} \Bigr)
\mathbf{W}_n \overset{\mathrm{a.s.}}{\longrightarrow} 0.
\end{equation}
Let $c_{ij} = \|X_i\|/\|X_j\|$ and $\hat{c}_{ij} =
\|\hat{X}_i\|/\|\hat{X}_j\|$. Combining Eq.~\eqref{eq:conv_Jacobian}
and Eq.~\eqref{eq:thm1_proof5b}, we obtain
\begin{equation}
  \label{eq:thm3_proof3}
  \rho_n \Bigl(\mathbf{J}(\hat{X}_i)
  \bigl(\hat{\boldsymbol{\Sigma}}(\hat{X}_i) + \hat{c}_{ij}^2 
\hat{\boldsymbol{\Sigma}}(\hat{X}_j)\bigr) \mathbf{J}(\hat{X}_i) -
\mathbf{W}_n^{\top} \mathbf{J}(Z_i)
\bigl(\boldsymbol{\Sigma}(Z_i) + c_{ij}^2
\boldsymbol{\Sigma}(Z_j)\bigr)
\mathbf{J}(Z_i) \mathbf{W}_n\Bigr)
\overset{\mathrm{a.s.}}{\longrightarrow} 0.
\end{equation}
Now let $\mathbf{M}$ be a matrix with Moore-Penrose pseudoinverse
$\mathbf{M}^{\dagger}$. Then for {\em any} orthogonal 
$\mathbf{W}$,
$$(\mathbf{W} \mathbf{M} \mathbf{W}^{\top})^{\dagger} = \mathbf{W}
\mathbf{M}^{\dagger} \mathbf{W}^{\top}.$$
Indeed, $\mathbf{W} \mathbf{M}^{\dagger} \mathbf{W}^{\top}$ satisfies
the four conditions that uniquely define the Moore-Penrose
pseudoinverse. For example
\begin{gather*}
  \Bigl(\mathbf{W} \mathbf{M} \mathbf{W}^{\top} \mathbf{W}
\mathbf{M}^{\dagger} \mathbf{W}^{\top}\Bigr)^{\top} = \mathbf{W} (\mathbf{M}
\mathbf{M}^{\dagger})^{\top} \mathbf{W}^{\top} = \mathbf{W} \mathbf{M}
\mathbf{M}^{\dagger} \mathbf{W}^{\top} = \mathbf{W} \mathbf{M} \mathbf{W}^{\top} \mathbf{W}
\mathbf{M}^{\dagger} \mathbf{W}^{\top}, \\
(\mathbf{W} \mathbf{M} \mathbf{W}^{\top}) \mathbf{W}
\mathbf{M}^{\dagger} \mathbf{W}^{\top} \mathbf{W} \mathbf{M}
\mathbf{W}^{\top} = \mathbf{W} \mathbf{M} \mathbf{M}^{\dagger}
\mathbf{M} \mathbf{W}^{\top} = \mathbf{W} \mathbf{M} \mathbf{W}^{\top}.
\end{gather*}
We therefore have
$$\mathbf{W}_{n}^{\top} \Bigl(\mathbf{J}(Z_i)
\bigl(\boldsymbol{\Sigma}(Z_i) + c_{ij}^2
\boldsymbol{\Sigma}(Z_j)\bigr)
\mathbf{J}(Z_i)\Bigr)^{\dagger} \mathbf{W}_n = \Bigl(\mathbf{W}_n^{\top} \mathbf{J}(Z_i)
\bigl(\boldsymbol{\Sigma}(Z_i) + c_{ij}^2
\boldsymbol{\Sigma}(Z_j)\bigr)
\mathbf{J}(Z_i) \mathbf{W}_n\Bigr)^{\dagger}.$$
The convergence in Eq.~\eqref{eq:thm3_proof3} together with
perturbation bounds for the Moore-Penrose pseudoinverse (see e.g.,
Theorem~3.3 of \citep{Stewart_ginv}) then implies
$$\rho_n^{-1} \Bigl(\mathbf{J}(\hat{X}_i) \bigl(\hat{\boldsymbol{\Sigma}}(\hat{X}_i) + \hat{c}_{ij}^2
\hat{\boldsymbol{\Sigma}}(\hat{X}_j)\bigr) \mathbf{J}(\hat{X}_i)\Bigr)^{\dagger} -
\rho_n^{-1} \mathbf{W}_n^{\top} \Bigl(\mathbf{J}(Z_i)
\bigl(\boldsymbol{\Sigma}(Z_i) + c_{ij}^2
\boldsymbol{\Sigma}(Z_j)\bigr)
\mathbf{J}(Z_i)\Bigr)^{\dagger} \mathbf{W}_n
\overset{\mathrm{a.s.}}{\longrightarrow} 0$$ 
We conclude, by Eq.~\eqref{eq:chisq_thm3} and Slutsky's theorem that
$$n\bigl(s(\hat{X}_i) - s(\hat{X}_j) \bigr)^{\top} \Bigl(\mathbf{J}(\hat{X}_i) \bigl(\hat{\boldsymbol{\Sigma}}(\hat{X}_i) + \hat{c}_{ij}^2
\hat{\boldsymbol{\Sigma}}(\hat{X}_j)\bigr) \mathbf{J}(\hat{X}_i)\Bigr)^{\dagger}  \bigl(s(\hat{X}_i) -  
s(\hat{X}_j)\bigr) \rightsquigarrow \chi^{2}_{d-1}.
$$
under $\mathbb{H}_0 \colon X_i/\|X_i\| = X_j/\|X_j\|$. 
The condition for the local alternative follows directly from
Eq.~\eqref{eq:thm3_proof2a} and the observation that if $\sqrt{n
  \rho_n}(s(Z_i) - s(Z_j))$ is bounded as $n \rightarrow \infty$ then
$$\|\bm{\Sigma}(Z_i) - \bm{\Sigma}(Z_j)\| \longrightarrow 0, \quad
\text{and}, \quad \rho_n^{1/2}\bigl(\|\mathbf{J}(Z_i) - \tfrac{\|Z_i\|}{\|Z_j\|}
\mathbf{J}(Z_j)\|\bigr) \longrightarrow 0.$$
\subsection{Proof of Proposition~\ref{prop:equality_ncp}}
Let $\mathbf{M}_{ij}$ and $\mathbf{M}'_{ij}$ be the matrices
\begin{gather*}
  \mathbf{M}_{ij} = \bm{\Sigma}(Z_i) +
  \frac{\|Z_i\|^2}{\|Z_j\|^2} \bm{\Sigma}(Z_j), \quad   \mathbf{M}'_{ij} = \bm{\Sigma}(Z_i) +
  \frac{\|Z_i\|_{\mathbf{I}_{a,b}}^2}{\|Z_j\|_{\mathbf{I}_{a,b}}^2} \bm{\Sigma}(Z_j).
\end{gather*}
Eq.~\eqref{eq:local_alt_ase2} can now be written as
$$n (s(Z_i) - s(Z_j))^{\top} \bigl(\mathbf{J}(Z_i) \mathbf{M}_{ij}
\mathbf{J}(Z_i)\bigr)^{\dagger} (s(Z_i) - s(Z_j)) \longrightarrow \mu.$$ 
We first show that Eq.~\eqref{eq:local_alt_ase2} holds if and only if
\begin{equation}
  \label{eq:prop3_proof_1}
n (s'(Z_i) - s'(Z_j))^{\top} \bigl(\mathbf{J}'(Z_i) \mathbf{M}'_{ij}
\mathbf{J}'(Z_i)^{\top}\bigr)^{\dagger} (s'(Z_i) - s'(Z_j))
\longrightarrow \mu.
\end{equation}
Suppose Eq.~\eqref{eq:local_alt_ase2} holds. Then a Taylor
approximation for $s(Z_i) - s(Z_j)$ around $Z_i$ implies
$$n (Z_i - Z_j)^{\top} \mathbf{J}(Z_i) \bigl(\mathbf{J}(Z_i)
\mathbf{M}_{ij} \mathbf{J}(Z_i)\bigr)^{\dagger} \mathbf{J}(Z_i) (Z_i - Z_j)
\longrightarrow \mu.$$
Since $\mathbf{M}_{ij}$ is positive definite, this is equivalent to
\begin{equation}
  \label{eq:prop3_proof_2}
n (Z_i - Z_j)^{\top} \mathbf{M}_{ij}^{-1/2} \mathbf{M}_{ij}^{1/2} \mathbf{J}(Z_i) \bigl(\mathbf{J}(Z_i)
\mathbf{M}_{ij} \mathbf{J}(Z_i)\bigr)^{\dagger} \mathbf{J}(Z_i)
\mathbf{M}_{ij}^{1/2} \mathbf{M}_{ij}^{-1/2} (Z_i - Z_j)
\longrightarrow \mu.
\end{equation}
Now let $\mathbf{N} = \mathbf{M}_{ij}^{1/2} \mathbf{J}(Z_i)$. Then by
the properties of the Moore-Penrose pseudoinverse, we have
\begin{equation}
  \label{eq:prop3_proof_2b}
  \begin{split}\mathbf{M}_{ij}^{1/2} \mathbf{J}(Z_i) \bigl(\mathbf{J}(Z_i)
\mathbf{M}_{ij} \mathbf{J}(Z_i)\bigr)^{\dagger} \mathbf{J}(Z_i)
\mathbf{M}_{ij}^{1/2} &= \mathbf{N} (\mathbf{N}^{\top}
\mathbf{N})^{\dagger} \mathbf{N}^{\top} = \mathbf{N}
\mathbf{N}^{\dagger} (\mathbf{N}^{\dagger})^{\top} \mathbf{N}^{\top}
\\ &= \mathbf{N} \mathbf{N}^{\dagger} (\mathbf{N}
\mathbf{N}^{\dagger})^{\top} = \mathbf{N} \mathbf{N}^{\dagger}
\mathbf{N} \mathbf{N}^{\dagger} = \mathbf{N} \mathbf{N}^{\dagger}.
\end{split}
\end{equation}
Now $\mathbf{N} \mathbf{N}^{\dagger}$ is the {\em unique} orthogonal
projection onto the column space of $\mathbf{N}$.
Let $\mathbf{N}' = \mathbf{M}_{ij}^{1/2} \mathbf{J}'(Z_i)^{\top}$.
Next recall that
$$\mathbf{J}(Z_i) = \frac{1}{\|Z_i\|} \Bigl(\mathbf{I} - \frac{Z_i
  Z_i^{\top}}{\|Z_i\|^2}\Bigr), \quad \mathbf{J}'(Z_i)^{\top} = \frac{1}{\|Z_i\|}_{\mathbf{I}_{a,b}}
\Bigl(\mathbf{I} - \frac{\mathbf{I}_{a,b} Z_i Z_i^{\top}}{\|Z_i\|_{\mathbf{I}_{a,b}}^2} \Bigr).$$
Since $\|Z_i\|^2_{\mathbf{I}_{a,b}} = Z_i^{\top} \mathbf{I}_{a,b}
Z_i$, we have
\begin{gather*}
  \Bigl(\mathbf{I} - \frac{Z_i Z_i^{\top}}{\|Z_i\|^2}\Bigr) 
\Bigl(\mathbf{I} - \frac{\mathbf{I}_{a,b} Z_i
  Z_i^{\top}}{\|Z_i\|_{\mathbf{I}_{a,b}}^2}\Bigr) = \Bigl(\mathbf{I} - \frac{\mathbf{I}_{a,b} Z_i
  Z_i^{\top}}{\|Z_i\|_{\mathbf{I}_{a,b}}^2}\Bigr), \quad
\mathbf{J}'(Z_i)^{\top} = \|Z_i\| \mathbf{J}(Z_i) \mathbf{J}'(Z_i)^{\top}  \\ \Bigl(\mathbf{I} - \frac{\mathbf{I}_{a,b} Z_i
  Z_i^{\top}}{\|Z_i\|_{\mathbf{I}_{a,b}}^2}\Bigr) 
\Bigl(\mathbf{I} -
\frac{Z_i Z_i^{\top}}{\|Z_i\|^2}\Bigr)= \Bigl(\mathbf{I} -
\frac{Z_i Z_i^{\top}}{\|Z_i\|^2}\Bigr), \quad \mathbf{J}(Z_i) =
\|Z_i\|_{\mathbf{I}_{a,b}} \mathbf{J}'(Z_i)^{\top} \mathbf{J}(Z_i).
\end{gather*}
Hence, $\mathbf{N} = \|Z_i\|_{\mathbf{I}_{a,b}} \mathbf{N}' \mathbf{J}(Z_i)$ and
$\mathbf{N}' = \|Z_i\| \mathbf{N}
\mathbf{J}'(Z_i)^{\top}$. The column space of $\mathbf{N}$
and the column space of $\mathbf{N}'$ are therefore identical and
hence, by the uniqueness of orthogonal projection matrices,
$\mathbf{N} \mathbf{N}^{\dagger} = \mathbf{N}'
\mathbf{N}'^{\dagger}$. We therefore have
$$\mathbf{M}_{ij}^{1/2} \mathbf{J}(Z_i) \bigl(\mathbf{J}(Z_i)
\mathbf{M}_{ij} \mathbf{J}(Z_i)\bigr)^{\dagger} \mathbf{J}(Z_i)
\mathbf{M}_{ij}^{1/2} = \mathbf{M}_{ij}^{1/2} \mathbf{J}'(Z_i)^{\top} \bigl(\mathbf{J}'(Z_i)
\mathbf{M}_{ij} \mathbf{J}'(Z_i)^{\top}\bigr)^{\dagger} \mathbf{J}'(Z_i)
\mathbf{M}_{ij}^{1/2}.$$
Eq.~\eqref{eq:prop3_proof_2} is therefore equivalent to
\begin{equation}
  \label{eq:prop3_proof_3}
  n (Z_i - Z_j)^{\top} \mathbf{J}'(Z_i)^{\top} \bigl(\mathbf{J}'(Z_i)
\mathbf{M}_{ij} \mathbf{J}'(Z_i)^{\top}\bigr)^{\dagger} \mathbf{J}'(Z_i)
(Z_i - Z_j) \longrightarrow \mu.
\end{equation}
We can now do another Taylor series expansion for $s'(Z_i) - s'(Z_j)$
around $Z_i$ and thereby replace $\mathbf{J}'(Z_i)(Z_i - Z_j)$ in Eq.~\eqref{eq:prop3_proof_3}
with $s'(Z_i) - s'(Z_j)$ to obtain
$$n \bigl(s'(Z_i) - s'(Z_j)\bigr)^{\top} \bigl(\mathbf{J}'(Z_i)
\mathbf{M}_{ij} \mathbf{J}'(Z_i)^{\top}\bigr)^{\dagger} 
(s'(Z_i) - s'(Z_j)\bigr) \longrightarrow \mu.$$
Eq.~\eqref{eq:prop3_proof_1} now follows from the observation that, under a local alternative,
$$\frac{\|Z_i\|}{\|Z_j\|} -
\frac{\|Z_i\|_{\mathbf{I}_{a,b}}}{\|Z_j\|_{\mathbf{I}_{a,b}}}
\longrightarrow 0, \quad \text{and} \quad \mathbf{M}_{ij} - \mathbf{M}'_{ij} \longrightarrow 0.$$
Reversing the above steps yield the converse statement that Eq.~\eqref{eq:prop3_proof_1}
implies Eq.~\eqref{eq:local_alt_ase2}.

We now complete the proof of Proposition~\ref{prop:equality_ncp} by
showing that Eq.~\eqref{eq:prop3_proof_1} is identical to
Eq.~\eqref{eq:indefinite_alt_cond2}. 
We first make the observation that for any indefinite orthogonal matrix $\mathbf{Q}$,
\begin{gather*}
s'(\mathbf{Q} \xi) = \frac{\mathbf{Q} \xi}{\|\mathbf{Q} \xi\|_{\mathbf{I}_{a,b}}} =
  \frac{\mathbf{Q} \xi}{\|\xi\|_{\mathbf{I}_{a,b}}} = \mathbf{Q}
  s'(\xi)  \\
\mathbf{J}'(\mathbf{Q} \xi) = \frac{1}{\|\mathbf{Q}
  \xi\|_{\mathbf{I}_{a,b}}} \Bigl(\mathbf{I} - \frac{\mathbf{Q} \xi \xi^{\top} \mathbf{Q}^{\top}
  \mathbf{I}_{a,b}}{\|\mathbf{Q} \xi\|^2_{\mathbf{I}_{a,b}}}\Bigr) =
\frac{1}{\|\xi\|_{\mathbf{I}_{a,b}}} \Bigl(\mathbf{I} - \frac{\mathbf{Q} \xi \xi^{\top}
  \mathbf{I}_{a,b} \mathbf{Q}^{-1}}{\|\xi\|_{\mathbf{I}_{a,b}}^2}\Bigr) =
\mathbf{Q} \mathbf{J}'(\xi) \mathbf{Q}^{-1}.
\end{gather*}
Furthermore, for any constant $c > 0$, we also have $s'(c \xi) = s'(\xi)$
and $\mathbf{J}'(c \xi) = \tfrac{1}{c} \mathbf{J}'(\xi)$. 

Recall that $\rho_n^{1/2} \mathbf{X} = \mathbf{Z} \mathbf{Q}_{\mathbf{X}}$
for some indefinite orthogonal matrix $\mathbf{Q}_{\mathbf{X}}$. Then
$\|Z_i\|_{\mathbf{I}_{a,b}} = \rho_n^{1/2}
\|X_i\|_{\mathbf{I}_{a,b}}$; furthermore
by the definition of $\bm{\Sigma}(Z_i)$ in Theorem~\ref{thm0}, we
have
$$\mathbf{J}'(X_i) \bm{\Sigma}(X_i)
\mathbf{J}'(X_i)^{\top} =
\mathbf{Q}_{\mathbf{X}}^{\top} \mathbf{J}'(\rho_n^{-1/2} Z_i)
\bm{\Sigma}(Z_i) \mathbf{J}'(\rho_n^{-1/2} Z_i)^{\top}
\mathbf{Q}_{\mathbf{X}} = \rho_n(\mathbf{Q}_{\mathbf{X}}^{\top}) \mathbf{J}'(Z_i)
\bm{\Sigma}(Z_i) \mathbf{J}'(Z_i)^{\top} \mathbf{Q}_{\mathbf{X}}.$$
Eq.~\eqref{eq:indefinite_alt_cond2} can now be written as
\begin{equation}
  \label{eq:indefinite_alt_cond2_proof}
n \bigl(s'(Z_i) - s'(Z_j)\bigr)^{\top} \mathbf{Q}_{\mathbf{X}} \Bigl(\mathbf{Q}_{\mathbf{X}}^{\top} \mathbf{J}'(Z_i)
\mathbf{M}'_{ij} \mathbf{J}'(Z_i)^{\top} \mathbf{Q}_{\mathbf{X}}\Bigr)^{\dagger}
\mathbf{Q}_{\mathbf{X}}^{\top} \bigl(s'(Z_i) - s'(Z_j)\bigr)
\longrightarrow \mu.
\end{equation}
We thus need to show that 
Eq.~\eqref{eq:prop3_proof_1} and \eqref{eq:indefinite_alt_cond2_proof} are identical. By
replacing $s'(Z_i) - s'(Z_j)$ with $\mathbf{J}'(Z_i)(Z_i - Z_j)$, this
is equivalent to showing that the following conditions are identical
\begin{gather*}
n(Z_i - Z_j)^{\top} \mathbf{J}'(Z_i)^{\top} \Bigl(\mathbf{J}'(Z_i)
\mathbf{M}'_{ij} \mathbf{J}'(Z_i)^{\top} \Bigr)^{\dagger}
\mathbf{J}'(Z_i) (Z_i - Z_j) \longrightarrow \mu, \\
n(Z_i - Z_j)^{\top} \mathbf{J}'(Z_i)^{\top} \mathbf{Q}_{\mathbf{X}}
\Bigl( \mathbf{Q}_{\mathbf{X}}^{\top} \mathbf{J}'(Z_i)
\mathbf{M}'_{ij} \mathbf{J}'(Z_i)^{\top} \mathbf{Q}_{\mathbf{X}} \Bigr)^{\dagger}
\mathbf{Q}_{\mathbf{X}}^{\top} \mathbf{J}'(Z_i) (Z_i - Z_j) \longrightarrow \mu.
\end{gather*}
Let $\mathbf{N}'' = (\mathbf{M}'_{ij})^{1/2} \mathbf{J}'(Z_i)^{\top}$
and $\mathbf{N}''' = (\mathbf{M}'_{ij})^{1/2} \mathbf{J}'(Z_i)^{\top}
\mathbf{Q}_{\mathbf{X}}$. Using a similar argument to that for
deriving Eq.~\eqref{eq:prop3_proof_2} and \eqref{eq:prop3_proof_2b},
we have
\begin{gather*}
\mathbf{J}'(Z_i)^{\top} \Bigl(\mathbf{J}'(Z_i)
\mathbf{M}'_{ij} \mathbf{J}'(Z_i)^{\top} \Bigr)^{\dagger}
\mathbf{J}'(Z_i) = (\mathbf{M}'_{ij})^{-1/2} \mathbf{N''} \mathbf{N''}^{\dagger}
(\mathbf{M}'_{ij})^{-1/2}, \\
\mathbf{J}'(Z_i)^{\top} \mathbf{Q}_{\mathbf{X}} 
\Bigl(\mathbf{Q}_{\mathbf{X}}^{\top} \mathbf{J}'(Z_i)
\mathbf{M}'_{ij} \mathbf{J}'(Z_i)^{\top} \mathbf{Q}_{\mathbf{X}} \Bigr)^{\dagger}
\mathbf{Q}_{\mathbf{X}}^{\top} \mathbf{J}'(Z_i) = (\mathbf{M}'_{ij})^{-1/2} \mathbf{N'''} \mathbf{N'''}^{\dagger} (\mathbf{M}'_{ij})^{-1/2}. 
\end{gather*}
Now $\mathbf{N}'' \mathbf{N}''^{\dagger}$ and $\mathbf{N'''} \mathbf{N}'''^{\dagger}$ are the {\em unique}
orthogonal projection matrices onto the column spaces of $\mathbf{N}'' =
(\mathbf{M}'_{ij})^{1/2} \mathbf{J}'(Z_i)^{\top}$ and $\mathbf{N}''' =
(\mathbf{M}'_{ij})^{1/2} \mathbf{J}'(Z_i)^{\top}
\mathbf{Q}_{\mathbf{X}}$, respectively. However, since
$\mathbf{Q}_{\mathbf{X}}$ is {\em invertible}, the column spaces of
$\mathbf{N}''$ and $\mathbf{N}'''$ coincide. We therefore have
$$\mathbf{J}'(Z_i)^{\top} \Bigl(\mathbf{J}'(Z_i)
\mathbf{M}'_{ij} \mathbf{J}'(Z_i)^{\top} \Bigr)^{\dagger}
\mathbf{J}'(Z_i) = \mathbf{J}'(Z_i)^{\top} \mathbf{Q}_{\mathbf{X}}
\Bigl(\mathbf{Q}_{\mathbf{X}}^{\top} \mathbf{J}'(Z_i)
\mathbf{M}'_{ij} \mathbf{J}'(Z_i)^{\top} \mathbf{Q}_{\mathbf{X}} \Bigr)^{\dagger}
\mathbf{Q}_{\mathbf{X}}^{\top} \mathbf{J}'(Z_i),$$
and thus Eq.~\eqref{eq:prop3_proof_1} and
Eq.~\eqref{eq:indefinite_alt_cond2} are equivalent, as desired.

\subsection{Proof of Corollary~\ref{cor2}}
We will first show that the two expressions for
$\tilde{G}_{\mathrm{ASE}}(\hat{U}_i, \hat{U}_j)$ in 
Eq.~\eqref{eq:G_def_U} are the same. Given this equivalence, the limiting distribution for
$\tilde{G}_{\mathrm{ASE}}(\hat{U}_i, \hat{U}_j)$ follows by a careful
application of the delta method to $\tilde{s}(\hat{X}_i) -
\tilde{s}(\hat{X}_j)$. 

From the definition of $\tilde{s}(\xi)$, we have
\begin{equation}
  \label{eq:proof_cor2_1}
  \tilde{s}(\hat{U}_i) = \frac{\hat{U}_{i,2:d}}{\hat{U}_{i1}} =
\hat{\lambda}_1^{1/2} |\hat{\mathbf{S}}_{2:d}|^{-1/2}
\frac{\hat{X}_{i,2:d}}{\hat{X}_{i1}} = \hat{\lambda}_1^{1/2}
|\hat{\mathbf{S}}_{2:d}|^{-1/2} \tilde{s}(\hat{X}_i)
\end{equation}
Here $\hat{\lambda}_1 > 0$ denote the largest eigenvalue in modulus of
$|\hat{\mathbf{S}}|$ and $\hat{\mathbf{S}}_{2:d}$ denote the $(d-1)
\times (d-1)$ diagonal matrix obtained by removing $\hat{\lambda}_1$
from $\hat{\mathbf{S}}$. Using the above relationship, we have
\begin{equation}
  \begin{split}
    \label{eq:proof_cor2_2}
  \tilde{\mathbf{J}}(\hat{U}_i) |\hat{\mathbf{S}}|^{-1/2} &=
\frac{1}{\hat{U}_{i1}} \Bigl[-\tilde{s}(\hat{U}_i) \mid
\mathbf{I} \Bigr] \begin{bmatrix} \hat{\lambda}_1^{-1/2} & 0 \\ 0 &
  |\hat{\mathbf{S}}_{2:d}|^{-1/2} \end{bmatrix} \\ &=
\frac{\hat{\lambda}_1^{1/2}}{\hat{X}_{i1}}
\Bigl[-\hat{\lambda}_1^{1/2} |\hat{\mathbf{S}}_{2:d}|^{-1/2} \tilde{s}(\hat{X}_i) \mid
\mathbf{I} \Bigr] \begin{bmatrix} \hat{\lambda}_1^{-1/2} & 0 \\ 0 &
  |\hat{\mathbf{S}}_{2:d}|^{-1/2} \end{bmatrix} \\ & = \hat{\lambda}_1^{1/2}
|\hat{\mathbf{S}}_{2:d}|^{-1/2} \tilde{\mathbf{J}}(\hat{X}_i).
\end{split}
\end{equation}
Eq.~\eqref{eq:G_def_U} then follows from
Eq.~\eqref{eq:proof_cor2_1}, Eq.~\eqref{eq:proof_cor2_2}
and the expression $\hat{\bm{\Sigma}}(\hat{U}_i) = n \rho_n
|\hat{\mathbf{S}}|^{-1/2} \hat{\bm{\Sigma}}(\hat{X}_i)
|\hat{\mathbf{S}}|^{-1/2}$ provided in Corollary~\ref{cor1}. 

We now do Taylor series expansions for $\tilde{s}(\mathbf{W}_n \hat{X}_i) -
\tilde{s}(Z_i)$ around $Z_i$ and $\tilde{s}(\mathbf{W}_n \hat{X}_j) -
\tilde{s}(Z_j)$ around $Z_j$. We emphasize that the orthogonal
transformation $\mathbf{W}_n$ is present in this step.
First note that $\tilde{\mathbf{J}}(c \xi) =
c^{-1} \tilde{\mathbf{J}}(\xi)$.
We then have
\begin{equation}
  \label{eq:proof_cor2_decomp1}
  \begin{split}
  \sqrt{n \rho_n} \bigl(\tilde{s}(\mathbf{W}_n \hat{X}_i) -
  \tilde{s}(\mathbf{W}_n \hat{X}_j)\bigr) &=
  \sqrt{n \rho_n} \bigl(\tilde{s}(Z_i) -
  \tilde{s}(Z_j)\bigr) \\ &+
  \rho_n^{1/2} \tilde{\mathbf{J}}(Z_i) \sqrt{n} (\mathbf{W}_n \hat{X}_i -
    Z_i) \\ &+  \rho_n^{1/2} \tilde{\mathbf{J}}(Z_j) \sqrt{n}(\mathbf{W}_n \hat{X}_j - Z_j) + o_{\mathbb{P}}(1).
    \end{split}
    \end{equation}
Recall the definition of $\bm{\Sigma}(Z_i)$
  from the statement of Theorem~\ref{thm0}.
  Now define
  \begin{equation}
    \label{eq:proof_cor2_2b}
  \mathrm{var}[\tilde{s}(\mathbf{W}_n \hat{X}_i) -
  \tilde{s}(\mathbf{W}_n \hat{X}_j)] = \rho_n \tilde{\mathbf{J}}(Z_i)
\bm{\Sigma}(Z_i)
\tilde{\mathbf{J}}(Z_i)^{\top} + \rho_n \tilde{\mathbf{J}}(Z_j) \bm{\Sigma}(Z_j)
\tilde{\mathbf{J}}(Z_j)^{\top}.
\end{equation}
  We then have, under $\mathbb{H}_0 \colon X_i/\|X_i\| = X_j/\|X_j\|$,
  \begin{equation}
    \label{eq:proof_cor2_3}
 n \rho_n \bigl(\tilde{s}(\mathbf{W}_n \hat{X}_i) -
\tilde{s}(\mathbf{W}_n \hat{X}_j)\bigr)^{\top} \Bigl(
\mathrm{var}[\tilde{s}(\mathbf{W}_n \hat{X}_i) -
  \tilde{s}(\mathbf{W}_n \hat{X}_j)] \Bigr)^{-1}
\bigl(\tilde{s}(\mathbf{W}_n
\hat{X}_i) - \tilde{s}(\mathbf{W}_n \hat{X}_j)\bigr) \rightsquigarrow
\chi^{2}_{d-1}.
\end{equation}
Note that the sparsity factor $\rho_n$ appears in
both Eq.~\eqref{eq:proof_cor2_3} and Eq.~\eqref{eq:proof_cor2_2b} and
thus canceled out. Nevertheless, by having the factor $\rho_n$ in
Eq.~\eqref{eq:proof_cor2_2b}, we guarantee that
Eq.~\eqref{eq:proof_cor2_2b} remains bounded, and this simplifies the
logic in the subsequent derivation. Indeed, as $\rho_n \rightarrow 0$, $Z_i \rightarrow 0$ and
thus $\|\tilde{\mathbf{J}}(Z_i)\| \rightarrow \infty$ due to the
scaling factor $1/Z_{i1}$ in the definition of $\tilde{\mathbf{J}}(Z_i)$.

In order to convert Eq.~\eqref{eq:proof_cor2_3} into an appropriate
test statistic, we need to (1) relate
$\tilde{s}(\mathbf{W}_n \hat{X}_i)$ to $\tilde{s}(\hat{X}_i)$ and (2)
find an estimate for $\mathrm{var}[\tilde{s}(\mathbf{W}_n \hat{X}_i) - \tilde{s}(\mathbf{W}_n \hat{X}_j)]$.
Recall that in our definition of $\hat{\mathbf{X}} = \hat{\mathbf{U}}
|\hat{\mathbf{S}}|^{1/2}$, the first column of $\hat{\mathbf{U}}$
is the eigenvector corresponding to the largest eigenvalue (in
modulus) of $\mathbf{A}$; similarly, for $\mathbf{Z} = \mathbf{U}
|\mathbf{S}|^{1/2}$, the first column of $\mathbf{U}$ is the
eigenvector corresponding to the largest eigenvalue of
$\mathbf{P}$. Let $\hat{\lambda}_1$ and $\lambda_1$ denote these
eigenvalues for $\mathbf{A}$ and $\mathbf{P}$, respectively. 
Then by the Perron Frobenius theorem, $\hat{\lambda}_1$ is a simple
eigenvalue of $\mathbf{A}$ and $\lambda_1$ is a simple eigenvalue of
$\mathbf{P}$; furthermore the entries in the corresponding column of
$\hat{\mathbf{U}}$ and $\mathbf{U}$ are all positive. Hence
$\hat{X}_{i1} > 0$ and $Z_{i1} > 0$ for all $i$.
Since $\mathbf{W}_n$ aligns $\hat{\mathbf{X}}$ to
$\mathbf{Z}$, and $\hat{\lambda}_1$ and $\lambda_1$ are both simple
eigenvalues, we also expect $\mathbf{W}_n$ to be of the form
\begin{equation}
  \label{eq:block_Wn}
\mathbf{W}_n = \begin{bmatrix} 1 & \bm{0} \\ \bm{0} &
  \bar{\mathbf{W}}_n \end{bmatrix}
\end{equation}
where $\bar{\mathbf{W}}_n$ is a $(d-1) \times (d-1)$ orthogonal
matrix. Lemma~\ref{lem:simple_eigenvalue} below guarantees that,
asymptotically almost surely, we can choose a sequence of orthogonal matrices
$\mathbf{W}_n$ satisfying this constraint.
\begin{lem}
  \label{lem:simple_eigenvalue}
  Let $\mathbf{A} \sim \mathrm{GRDPG}(\mathbf{X}, \rho_n)$ be a graph
  on $n$ vertices where $\mathbf{X}$ and $\rho_n$ satisfies
  Conditions~1 through Condition~3 in Section~\ref{sec:ase}. Then as
  $n \rightarrow \infty$, the largest eigenvalue of $\mathbf{P} =
  \rho_n \mathbf{X} \mathbf{I}_{a,b} \mathbf{X}^{\top}$ is
  well-separated from the remaining eigenvalues, i.e., there exists a
  {\em fixed} constant $c_0 > 0$ such that, for $j \not = 1$ and sufficiently large $n$,
  $\bigl|\tfrac{\lambda_1}{\lambda_j}\bigr| \geq (1 + c_0)$. Then,
  with high probability, the largest eigenvalue of $\mathbf{A}$ is also
  well-separated from the remaining eigenvalues, i.e., 
  $\bigl|\tfrac{\hat{\lambda}_1}{\hat{\lambda}_j}\bigr| \geq (1 + c_0)$ for
  $j \not = 1$. The $(a,b)$ block orthogonal matrix $\mathbf{W}_n$ mapping
  from $\hat{\mathbf{X}}$ to $\mathbf{Z}$ is therefore of the form 
$\mathbf{W}_n = \Bigl[\begin{smallmatrix} 1 & \bm{0} \\ \bm{0} &
  \bar{\mathbf{W}}_n \end{smallmatrix}\Bigr]$ where $\bar{\mathbf{W}}_n$
is $(a-1,b)$ block orthogonal.
\end{lem}
We now give a brief sketch of the proof of
Lemma~\ref{lem:simple_eigenvalue}. First consider the matrix
$\mathbf{P} = \rho_n \mathbf{X} \mathbf{I}_{a,b}
\mathbf{X}^{\top}$. Let $M = \max_{ij} p_{ij}$ and $m = \min_{ij} p_{ij}$.
Then Condition~2 in Section~\ref{sec:ase}
stipulates that $m \geq c \rho_n$ for some constant $c
> 0$ not depending on $n$. From the previous discussion, we know that
$\lambda_1$ is a simple eigenvalue;
furthermore, \citep[Theorem~V]{ostrowski} implies
$$\max_{j \not  =1} \Bigl|\frac{\lambda_j}{\lambda_1}\Bigr| \leq
\frac{M^2 - m^2}{M^2 + m^2} = 1 - \frac{2m^2}{M^2 + m^2} \leq 1 -
\frac{2}{c^{-2} + 1}.$$
Hence, there exists a constant $c_0 > 0$ not depending on $n$ such
that $\max_{j \not = 1}|\lambda_j|/\lambda_1 \leq 1 - c_0$, i.e., the
largest eigenvalue of $\mathbf{P}$ is well-separated from the
remaining eigenvalues of $\mathbf{P}$. Now by
Condition~1 in Section~\ref{sec:ase}, the $d$ largest eigenvalues of $\mathbf{P}$
are of order $\Theta(n \rho_n)$. Furthermore, matrix concentration
inequalities imply $\|\mathbf{A} - \mathbf{P}\| = O((n \rho_n)^{1/2})$
with high probability; see e.g.,
\citep{oliveira2009concentration,bandeira_vanhandel}. The
largest eigenvalue of $\mathbf{A}$ is therefore also well-separated from the
remaining eigenvalues of $\mathbf{A}$. 

Now the matrix $\mathbf{W}_n$ that appears in the statement of
Theorem~\ref{thm0} is the orthogonal matrix closest to
$\mathbf{U}^{\top} \hat{\mathbf{U}}$ in Frobenius norm. 
The $(1,1)$th entry of $\mathbf{U}^{\top} \hat{\mathbf{U}}$ is of the
form $u_1^{\top} \hat{u}_1 = 1 - \tfrac{1}{2} \|u_1 -
\hat{u}_1\|^2$. The Davis-Kahan theorem \citep{davis70} implies
$\|u_1 - \hat{u}_1\|^2 = O((n \rho_n)^{-1})$; we can thus set the
$(1,1)$th entry of $\mathbf{W}_n$ to be $1$ without changing the limit
result in Theorem~\ref{thm0}. For a more rigorous justification of
this claim, see Section~B.1 of \citep{rubin2017statistical}. 

We thus assume, without loss of generality, that $\mathbf{W}_n$
satisfies Eq.~\eqref{eq:block_Wn}. We then have
\begin{equation}
  \label{eq:tilde_s_Wn}
\tilde{s}(\mathbf{W}_n \hat{X}_i) = \bar{\mathbf{W}}_n
\tilde{s}(\hat{X}_i).
\end{equation}
Eq.~\eqref{eq:tilde_s_Wn} then implies
\begin{equation}
  \label{eq:commute_Wn_proof_cor2}
\tilde{\mathbf{J}}(\mathbf{W}_n \hat{X}_i) \mathbf{W}_n =
\frac{1}{\hat{X}_{i1}}\Bigl[-\bar{\mathbf{W}}_n \tilde{s}(\hat{X}_i) \mid
\mathbf{I} \Bigr] \begin{bmatrix} 1 & 0 \\ 0 &
  \bar{\mathbf{W}}_n \end{bmatrix} =
\bar{\mathbf{W}}_n \tilde{\mathbf{J}}(\hat{X}_i).
\end{equation}
Now recall Lemma~\ref{lem:conv1}. Then, by approximating $Z_i$ with
$\mathbf{W}_n \hat{X}_i$ and invoking
Eq.~\eqref{eq:commute_Wn_proof_cor2}, we obtain
\begin{equation}
  \label{eq:conv2_proof_cor2}
  \rho_n \bar{\mathbf{W}}_n \tilde{\mathbf{J}}(\hat{X}_i)
  \hat{\bm{\Sigma}}(\hat{X}_i) \tilde{\mathbf{J}}(\hat{X}_i)^{\top}
  \bar{\mathbf{W}}_n^{\top} - \rho_n \tilde{\mathbf{J}}(Z_i)
  \bm{\Sigma}(Z_i)
\tilde{\mathbf{J}}(Z_i)^{\top}
\overset{\mathrm{a.s}}{\longrightarrow} 0.
\end{equation}
Combining Eq.~\eqref{eq:proof_cor2_3}, Eq.~\eqref{eq:tilde_s_Wn}, and
Eq.~\eqref{eq:conv2_proof_cor2}, we obtain 
$$n \bigl(\tilde{s}(\hat{X}_i) - \tilde{s}(\hat{X}_j)\bigr)^{\top}
\Bigl(\widehat{\mathrm{var}}[\tilde{s}(\hat{X}_i) -
\tilde{s}(\hat{X}_j)]\Bigr)^{-1} \bigl(\tilde{s}(\hat{X}_i) -
\tilde{s}(\hat{X}_j)\bigr)^{\top} \rightsquigarrow \chi^2_{d-1}$$
under the null hypothesis. Here $\widehat{\mathrm{var}}[\tilde{s}(\hat{X}_i) -
\tilde{s}(\hat{X}_j)]$ is defined in Eq.~\eqref{eq:G_tilde_U_cov2} of
Corollary~\ref{cor2}. 

We now derive the non-centrality parameter for
$\tilde{G}_{ASE}(\hat{X}_i, \hat{X}_j)$ under a local alternative.
Note that Eq.~\eqref{eq:local_alt_ase_cor2b} follows directly
from the expansion in Eq.~\eqref{eq:proof_cor2_decomp1} and
thus we will only show that Eq.~\eqref{eq:local_alt_ase2} and
Eq.~\eqref{eq:local_alt_ase_cor2b} are equivalent. First observe that under a local alternative, we have
$$\rho_n^{1/2} \Bigl(\tilde{\mathbf{J}}(Z_j) - \tfrac{\|Z_i\|}{\|Z_j\|}
\tilde{\mathbf{J}}(Z_i)\Bigr) \rightarrow 0$$
and hence, under a local alternative,
$$\rho_n \Bigl(\tilde{\mathbf{J}}(Z_i) \bm{\Sigma}(Z_i)
\tilde{\mathbf{J}}(Z_i)^{\top} + \tilde{\mathbf{J}}(Z_j) \bm{\Sigma}(Z_j)
\tilde{\mathbf{J}}(Z_j)^{\top}\Bigr) - \rho_n
\Bigl(\tilde{\mathbf{J}}(Z_i) \Bigl(\bm{\Sigma}(Z_i) +
\tfrac{\|Z_i\|^2}{\|Z_j\|^2} \bm{\Sigma}(Z_j)\Bigr)
\tilde{\mathbf{J}}(Z_i)^{\top}\Bigr) \rightarrow 0.$$
Eq.~\eqref{eq:local_alt_ase_cor2b} is thus equivalent to the condition
that
\begin{equation}
  \label{eq:proof_cor3_4}
n \bigl(\tilde{s}(Z_i) - \tilde{s}(Z_j)\bigr)^{\top}
\Bigl(\tilde{\mathbf{J}}(Z_i) \Bigl(\bm{\Sigma}(Z_i) +
\tfrac{\|Z_i\|^2}{\|Z_j\|^2} \bm{\Sigma}(Z_j) \Bigr)
\tilde{\mathbf{J}}(Z_i)^{\top} \Bigr)^{-1} (\tilde{s}(Z_i) -
\tilde{s}(Z_j)) \rightarrow \mu.
\end{equation}
Now let $\mathbf{M}_{ij} = \bm{\Sigma}(Z_i) +
\tfrac{\|Z_i\|^2}{\|Z_j\|^2} \bm{\Sigma}(Z_j)$. Following an analogous
argument to that in the proof of Proposition~\ref{prop:equality_ncp}
(see the derivation of Eq.~\eqref{eq:prop3_proof_2}), 
we first replace $s(Z_i) - s(Z_j)$ and $\tilde{s}(Z_i) -
\tilde{s}(Z_j)$ with $\mathbf{J}(Z_i)(Z_i - Z_j)$ and
$\tilde{\mathbf{J}}(Z_i)(Z_i - Z_j)$, respectively. This yields, in
place of the conditions in Eq.~\eqref{eq:local_alt_ase2}
and Eq.~\eqref{eq:proof_cor3_4}, the conditions
\begin{gather*}
  n(Z_i - Z_j)^{\top} \mathbf{M}_{ij}^{-1/2} \mathbf{M}_{ij}^{1/2} \mathbf{J}(Z_i) \Bigl(\mathbf{J}(Z_i)
  \mathbf{M}_{ij} \mathbf{J}(Z_i)\Bigr)^{\dagger} \mathbf{J}(Z_i) \mathbf{M}_{ij}^{1/2}
  \mathbf{M}_{ij}^{-1/2} (Z_i - Z_j) \rightarrow \mu, \\
   n(Z_i - Z_j)^{\top} \mathbf{M}_{ij}^{-1/2} \mathbf{M}_{ij}^{1/2}
   \tilde{\mathbf{J}}(Z_i)^{\top}
   \Bigl(\tilde{\mathbf{J}}(Z_i)
  \mathbf{M}_{ij} \tilde{\mathbf{J}}(Z_i)^{\top}\Bigr)^{-1} \tilde{\mathbf{J}}(Z_i) \mathbf{M}_{ij}^{1/2}
  \mathbf{M}_{ij}^{-1/2} (Z_i - Z_j)  \rightarrow \mu.
\end{gather*}
Let $\mathcal{P}_{\mathbf{J}(Z_i)}$ and
  $\mathcal{P}_{\tilde{\mathbf{J}}(Z_i)^{\top}}$
  denote the orthogonal projection matrices onto the column spaces of
  $\mathbf{M}_{ij}^{1/2} \mathbf{J}(Z_i)$ and $\mathbf{M}_{ij}^{1/2}
  \tilde{\mathbf{J}}(Z_i)^{\top}$, respectively. We then have (see
  also the derivation of Eq.~\eqref{eq:prop3_proof_2b})
  \begin{gather*}
    \mathbf{M}_{ij}^{1/2} \mathbf{J}(Z_i) \Bigl(\mathbf{J}(Z_i)
  \mathbf{M}_{ij} \mathbf{J}(Z_i)\Bigr)^{\dagger} \mathbf{J}(Z_i)
  \mathbf{M}_{ij}^{1/2} = \mathcal{P}_{\mathbf{J}(Z_i)}, \\
  \mathbf{M}_{ij}^{1/2}
   \tilde{\mathbf{J}}(Z_i)^{\top}
   \Bigl(\tilde{\mathbf{J}}(Z_i)
  \mathbf{M}_{ij} \tilde{\mathbf{J}}(Z_i)^{\top}\Bigr)^{-1}
  \tilde{\mathbf{J}}(Z_i) \mathbf{M}_{ij}^{1/2} = \mathcal{P}_{\tilde{\mathbf{J}}(Z_i)^{\top}}.
  \end{gather*}
  The equivalence of Eq.~\eqref{eq:local_alt_ase2}
and Eq.~\eqref{eq:proof_cor3_4} then reduces to showing that
$\mathcal{P}_{\mathbf{J}(Z_i)} =
\mathcal{P}_{\tilde{\mathbf{J}}(Z_i)^{\top}}$. 
Now recall the definition of $\mathbf{J}(\xi)$ and
$\tilde{\mathbf{J}}(\xi)$, i.e.,
$$\mathbf{J}(\xi) = \frac{1}{\|\xi\|}\Bigl(\mathbf{I} - \frac{\xi
  \xi^{\top}}{\|\xi^2\|}\Bigr), \quad \tilde{\mathbf{J}}(\xi)^{\top} =
\frac{1}{\xi_{1}} \begin{bmatrix} - s(\xi)^{\top} \\ \mathbf{I} \end{bmatrix}.$$
Decomposing $\xi = (\xi_1, \dots, \xi_d)$ as $(\xi_1, \xi_{2:d})$, we
observe that
$$\xi \xi^{\top} \tilde{\mathbf{J}}(\xi)^{\top} = \frac{1}{\xi_{1}} 
\begin{bmatrix} \xi_1^2 & \xi_1
  \xi_{2:d}^{\top} \\ \xi_{2:d} \xi_1 & \xi_{2:d}
  \xi_{2:d}^{\top} \end{bmatrix} \begin{bmatrix}
  -\tfrac{1}{\xi_1} \xi_{2:d}^{\top} \\ \mathbf{I} \end{bmatrix} = \bm{0}.$$
Hence $\mathbf{M}_{ij}^{1/2} \tilde{\mathbf{J}}(Z_i)^{\top} =
\|Z_i\| \mathbf{M}_{ij}^{1/2} \mathbf{J}(Z_i)
\tilde{\mathbf{J}}(Z_i)^{\top}$ and the column space of
$\mathbf{M}_{ij}^{1/2} \tilde{\mathbf{J}}(Z_i)^{\top}$ is a subspace of
the column space of $\mathbf{M}_{ij}^{1/2} \mathbf{J}(Z_i)$.

We now show the converse statement, that is, the column space of
$\mathbf{M}_{ij}^{1/2} \mathbf{J}(Z_i)$ is a subspace of the column
space of $\mathbf{M}_{ij}^{1/2}
\tilde{\mathbf{J}}(Z_i)^{\top}$. Let $v$ be a vector in the
column space of $\mathbf{M}_{ij}^{1/2}
\mathbf{J}(Z_i)$. Then $v = \mathbf{M}_{ij}^{1/2}
\mathbf{J}(Z_i)w$ for some $w \in \mathbb{R}^{d}$ and,
from the definition of $\mathbf{J}(Z_i)$, $\mathbf{J}(Z_i) w$ is in
the null space of $Z_i^{\top}$. Let $u = \mathbf{J}(Z_i) w \in \mathbb{R}^{d}$. Writing
$Z_i = (Z_{i1}, Z_{i2:d})$ and $u = (u_1, u_{2:d})$, we have
$$Z_i^{\top} u = Z_{i1} u_1 + Z_{i2:d}^{\top} u_{2:d}= 0 \Longrightarrow u_1 =
\tfrac{-Z_{i2:d}^{\top} u_{2:d}}{Z_{i1}} \Longrightarrow u
= \begin{bmatrix} u_1 \\ u_{2:d} \end{bmatrix} =
\begin{bmatrix} -
  \frac{Z_{i2:d}^{\top}}{Z_{i1}} \\ \mathbf{I} \end{bmatrix} u_{2:d}
= Z_{i1} \tilde{\mathbf{J}}(Z_i)^{\top} u_{2:d}.$$
We therefore have $v = Z_{i1} \mathbf{M}_{ij}^{1/2}
\tilde{\mathbf{J}}(Z_i)^{\top} u_{2:d}$, i.e., $v$ also belongs to
the column space of $\mathbf{M}_{ij}^{1/2}
\tilde{\mathbf{J}}(Z_i)^{\top}$.

In summary, the column spaces of $\mathbf{M}_{ij}^{1/2}
\mathbf{J}(Z_i)$ and $\mathbf{M}_{ij}^{1/2}
\tilde{\mathbf{J}}(Z_i)^{\top}$ are identical. This then implies, by
the uniqueness of orthogonal projection matrices, that
$\mathcal{P}_{\mathbf{J}(Z_i)} =
\mathcal{P}_{\tilde{\mathbf{J}}(Z_i)^{\top}}$ as desired.

\subsection{Proof of Theorem~\ref{thm6}}
Assume without loss of generality that we use the test statistics in
Theorem~\ref{thm1}. Now suppose that the true model is a stochastic
block model. Then, asymptotically for all $k = 1,2,\dots,K$, we have
$p_{k 1}, p_{k 2}, \ldots, p_{k m}
\overset{\mathrm{iid}}{\sim}\operatorname{Uniform}(0,1)$. This
implies
\begin{equation*}
  \begin{split}
\mathbb{P}\bigl(-2m\log(P_k)<x\bigr) &= \mathbb{P}\bigl(P_k>\exp(-\frac{x}{2m})\bigr) \\ &=1-\left[\exp\Bigl(-\frac{x}{2m}\Bigr)\right]^m =1-\exp\Bigl(-\frac{x}{2}\Bigr).
\end{split}
\end{equation*}
We therefore have $-2m\log(P_k)\rightsquigarrow
\chi_{2}^2$ for all $i$. The $\{P_k\}_{k=1}^{K}$ are furthermore mutually
independent. Hence $S=\sum_{k=1}^K-2m\log(P_k) \rightsquigarrow \chi_{2K}^2$ as desired.

\subsection{Proof of Theorem~\ref{thm2}}
\label{appb.7}
We first define a few additional notations.
Let $\bar{X} = \tfrac{1}{n} \sum_{j=1}^{n} X_j$. Next recall $t_i =n \rho_n
X_i^{\top} \mathbf{I}_{a,b} \bar{X}$ as the expected degree of the
$i$th node and $\mathbf{T} = \operatorname{diag}(t_1, \dots, t_n)$. Now define
\begin{gather*}
  \tilde{X}_i = \frac{\rho_n^{1/2} X_i}{\sqrt{t_i}} = \frac{X_i}{(n
    X_i^{\top} \mathbf{I}_{a,b} \bar{X})^{1/2}}, \quad
  \tilde{\mathbf{X}} =\rho_n^{1/2} \mathbf{T}^{-1/2} \mathbf{X}, \\
  \rho_n \tilde{\mathbf{X}}^{\top} \tilde{\mathbf{X}} = \sum_{i=1}^{n} \tilde{X}_i \tilde{X}_i^{\top} = \sum_{i=1}^n\frac{X_iX_i^{\top}}{
  n X_i^{\top}\mathbf{I}_{a,b}\bar{X}}, \\
\quad Y_{ik} = 
\frac{(\tilde{\mathbf{X}}^{\top} \tilde{\mathbf{X}})^{-1}
        X_{k}}{X_{k}^{\top}\mathbf{I}_{a,b}
        \bar{X}}-\frac{\mathbf{I}_{a,b}X_i}{2
        X_i^{\top}\mathbf{I}_{a,b}
        \bar{X}}, \quad \text{for $i=1,2,\dots,n$ and $k = 1,2,\dots,n$}.
  \end{gather*}
Given the above notations, Theorem~8 in \citep{rubin2017statistical}
can be reformulated so that, {\em
  conditional} on the matrix $\mathbf{X}$ of latent positions, we have
\begin{equation}
  \label{eq:thm2_proof1}
  \begin{split}
n\rho_n^{1/2} \tilde{\bm{\Sigma}}(X_i)^{-1/2}
\Bigl(\tilde{\mathbf{Q}}_{\mathbf{X}}^{\top} \tilde{\mathbf{W}}_n \breve{X}_i - \tilde{X}_i \Bigr) & \rightsquigarrow
N(0,\mathbf{I}).
\end{split}
\end{equation}
where $\tilde{\mathbf{W}}_n$ is $(a,b)$ block orthogonal, $\tilde{\mathbf{Q}}_{\mathbf{X}}$
is indefinite orthogonal, and $\tilde{\bm{\Sigma}}(X_i)$ is defined as
\begin{gather}
  \label{eq:tilde_sigma}
  \tilde{\bm{\Sigma}}(X_i)=\frac{1}{n} \mathbf{I}_{a,b}
  \sum_{k=1}^{n}\left[Y_{ik} Y_{ik}^{\top} \frac{X_i^{\top}\mathbf{I}_{a,b} X_{k}\left(1-\rho_n
        X_i^{\top}\mathbf{I}_{a,b} X_{k}
      \right)}{X_i^{\top}\mathbf{I}_{a,b}
      \bar{X}}\right]\mathbf{I}_{a,b}.
\end{gather}
We note that the covariance matrix $\tilde{\bm{\Sigma}}(X_i)$ in
Eq.~\eqref{eq:tilde_sigma} is
defined in terms of the $\{X_i\}$ but the
Laplacian spectral embedding $\breve{X}_i$ is an estimate of the
$\tilde{X}_i = \rho_n^{1/2} X_i/\sqrt{t_i}$. Thus, to
use $\{\breve{X}_i\}$ to construct a test statistic, we  
now rewrite $\tilde{\bm{\Sigma}}(X_i)$ in terms of the
$\{\tilde{X}_i\}$ and $\{t_i\}$. We first have
\begin{equation}
  \label{eq:tilde_sigma_1}
  \begin{split}
  \frac{X_i^{\top}\mathbf{I}_{a,b} X_{k}\left(1-\rho_n
        X_i^{\top}\mathbf{I}_{a,b} X_{k}
      \right)}{n X_i^{\top}\mathbf{I}_{a,b}
      \bar{X}} &= 
\frac{X_i^{\top}\mathbf{I}_{a,b} X_{k}\left(1-\rho_n
        X_i^{\top}\mathbf{I}_{a,b} X_{k}
      \right)}{t_i/\rho_n} \\ &= \frac{\sqrt{t_k}}{\sqrt{t_i}} \tilde{X}_i^{\top}
    \mathbf{I}_{a,b} \tilde{X}_k 
    - t_k (\tilde{X}_i^{\top}
    \mathbf{I}_{a,b} \tilde{X}_k)^{2}.
    \end{split}
\end{equation}
Meanwhile, for the term $Y_{ik}$, we have
\begin{equation}
  \label{eq:tilde_sigma_2}
Y_{ik} = \frac{(\tilde{\mathbf{X}}^{\top} \tilde{\mathbf{X}})^{-1}
        X_{k}}{t_k/(n \rho_n)}-\frac{\mathbf{I}_{a,b}X_i}{2
        t_i/(n \rho_n)} = n \rho_n^{1/2} \Bigl( \frac{(\tilde{\mathbf{X}}^{\top} \tilde{\mathbf{X}})^{-1}
        \tilde{X}_{k}}{\sqrt{t_k}} - \frac{\mathbf{I}_{a,b}
        \tilde{X}_i}{2\sqrt{t_i}}\Bigr) = 
      n \rho_n^{1/2} \zeta_{ik}
    \end{equation}
where $\zeta_{ik}$ is as defined in the statement of Theorem~\ref{thm2}. 
Combining Eq.~\eqref{eq:tilde_sigma_1} and Eq.~\eqref{eq:tilde_sigma_2}
yields a representation for Eq.~\eqref{eq:tilde_sigma} in terms of the
$\{\tilde{X}_i\}$ and $t_i$ only. 

Now let $\tilde{Z}_i = (\tilde{\mathbf{Q}}_{\mathbf{X}}^{-1})^{\top}
\tilde{X}_i$. Then, conditional on
$\mathbf{X}$, the matrix $\tilde{\mathbf{Q}}_{\mathbf{X}}$ is 
{\em deterministic} and hence we can rewrite
Eq.~\eqref{eq:thm2_proof1} as
 \begin{equation}
  \label{eq:thm5_proof1}
  \begin{split}
n\rho_n^{1/2} \tilde{\bm{\Sigma}}(\tilde{Z}_i)^{-1/2}
\Bigl( \tilde{\mathbf{W}}_n \breve{X}_i - \tilde{Z}_i \Bigr) & \rightsquigarrow
N(0,\mathbf{I}).
\end{split}
\end{equation}
where $\tilde{\bm{\Sigma}}(\tilde{Z}_i)$ is the $d \times
d$ matrix of the form 
\begin{gather*}
\tilde{\bm{\Sigma}}(\tilde{Z}_i) =
(\tilde{\mathbf{Q}}_{\mathbf{X}}^{-1})^{\top} \tilde{\bm{\Sigma}}(X_i) \tilde{\mathbf{Q}}_{\mathbf{X}}^{-1}= n^2 \rho_n \mathbf{I}_{a,b} \Bigl[\sum_{k=1}^n \tilde{\zeta}_{ik}
  \tilde{\zeta}_{ik}^{\top} \Bigl(\frac{\sqrt{t_k}}{\sqrt{t_i}}
  \tilde{Z}_i^{\top} \mathbf{I}_{a,b} \tilde{Z}_k - t_k
  (\tilde{Z}_i^{\top} \mathbf{I}_{a,b} \tilde{Z}_k)^2 \Bigr) \Bigr]
  \mathbf{I}_{a,b}, \\
\tilde{\zeta}_{ik}=\Bigl( \frac{(\tilde{\mathbf{Z}}^{\top} \tilde{\mathbf{Z}})^{-1}
        \tilde{Z}_{k}}{\sqrt{t_k}} - \frac{\mathbf{I}_{a,b}
        \tilde{Z}_i}{2\sqrt{t_i}}\Bigr).
\end{gather*}
Finally, for any pair of indices $i \not = j$, the vectors $n\rho_n^{1/2}
\Bigl( \tilde{\mathbf{W}}_n \breve{X}_i - \tilde{Z}_i \Bigr) $ and $n\rho_n^{1/2}
\Bigl( \tilde{\mathbf{W}}_n \breve{X}_j - \tilde{Z}_j \Bigr) $ are
asymptotically independent.

We therefore have, under $\mathbb{H}_0 \colon X_i = X_j$, that
$\tilde{Z}_i = \tilde{Z}_j$ and hence
\begin{equation}
  \label{eq:chisq_proof5}
n^2 \rho_n (\breve{X}_i - \breve{X}_j)^{\top} \tilde{\mathbf{W}}_n^{\top} \bigl(\tilde{\bm{\Sigma}}(\tilde{Z}_i) +
\tilde{\bm{\Sigma}}(\tilde{Z}_j)\bigr)^{-1} \tilde{\mathbf{W}}_n (\breve{X}_i - \breve{X}_j) \rightsquigarrow \chi^2_{d}.
\end{equation}
In order to convert Eq.~\eqref{eq:chisq_proof5} into an appropriate
test statistic, we need to find an estimate for $\tilde{\bm{\Sigma}}(\tilde{Z}_i)$ in
terms of the $\{\breve{X}_i\}$. Let 
\begin{gather*}
  \breve{\zeta}_{ik} = \frac{(\breve{\mathbf{X}}^{\top} \breve{\mathbf{X}})^{-1}
  \breve{X}_k}{\sqrt{d_k}}-\frac{\mathbf{I}_{a,b}\breve{X}_i
  }{2\sqrt{d_i}}, \\
  \breve{\boldsymbol{\Sigma}}(\breve{X}_i) =  
  n^2 \rho_n \mathbf{I}_{a,b} \Bigl[\sum_{k=1}^n \breve{\zeta}_{ik}
  \breve{\zeta}_{ik}^{\top} \Bigl(\frac{\sqrt{d_k}}{\sqrt{d_i}}
  \breve{X}_i^{\top} \mathbf{I}_{a,b} \breve{X}_k - d_k
  (\breve{X}_i^{\top} \mathbf{I}_{a,b} \breve{X}_k)^2 \Bigr) \Bigr] \mathbf{I}_{a,b}.
\end{gather*}
We now show that $\breve{\bm{\Sigma}}(\breve{X}_i) -
\tilde{\mathbf{W}}_n^{\top} \tilde{\bm{\Sigma}}(\tilde{Z_i})
\tilde{\mathbf{W}}_n \rightarrow 0$. 
We start with the following bound 
    \begin{equation}
    \label{eq:2toinfty2}
  \max_{i} \|\tilde{\mathbf{W}}_n \breve{X}_i - \tilde{Z}_i\| =
  O_{\mathbb{P}}\Bigl(\frac{\log n}{n\sqrt{\rho_n}}\Bigr).
  \end{equation}
  Eq.~\eqref{eq:2toinfty2} follows from Theorem~6 in
  \citep{rubin2017statistical} and is analogous to the bound
  in Lemma~\ref{lem:norm_bounds} for the adjacency spectral embedding
  $\hat{X}_i$. Eq.~\eqref{eq:2toinfty2} then implies a Frobenius norm bound of
\begin{equation}
  \label{eq:frob_bound2}
  \|\breve{\mathbf{X}} \tilde{\mathbf{W}}_n^{\top} - \tilde{\mathbf{Z}}\|_{F} = O_{\mathbb{P}}\Bigl(\frac{\log n}{\sqrt{n\rho_n}}\Bigr).
\end{equation}
In addition, we also have
\begin{gather}
  \label{eq:proof_thm3_bd1}
\max
\limits_{i,k}\left|\frac{\sqrt{d_k}}{\sqrt{d_i}}-\frac{\sqrt{t_k}}{\sqrt{t_i}}\right|=
O_{\mathbb{P}}\Bigl(\frac{\sqrt{\log n}}{\sqrt{n\rho_n}}\Bigr) \\
\label{eq:proof_thm3_bd2}
\max
\limits_{i}\left|\frac{1}{\sqrt{d_i}}-\frac{1}{\sqrt{t_i}}\right|
=
O_{\mathbb{P}}\Bigl(\frac{\sqrt{\log n}}{n\rho_n}\Bigr).
\end{gather}
Eq.~\eqref{eq:proof_thm3_bd1} and Eq.~\eqref{eq:proof_thm3_bd2}
follows from standard concentration inequalities for sum of Bernoulli
random variables. For example, for Eq.~\eqref{eq:proof_thm3_bd1}, we
have
$$\Bigl|\frac{\sqrt{d_k}}{\sqrt{d_i}} - \frac{\sqrt{t_k}}{\sqrt{t_i}}\Bigr| =
\Bigl|\frac{\sqrt{d_k t_i} - \sqrt{d_i t_k}}{\sqrt{d_i t_i}}\Bigr|
\leq \frac{|d_k|^{1/2} |d_i - t_i|}{\sqrt{d_i t_i}(\sqrt{d_i} +
  \sqrt{t_i})} + \frac{|d_i|^{1/2} |d_k - t_k|}{\sqrt{d_i t_i}(\sqrt{d_k} +
  \sqrt{t_k})}.$$ 
Conditions~1 through Condition~3 in Section~\ref{sec:ase}
then stipulate that $t_i = \Theta(n \rho_n)$ and hence, by e.g.,
Hoeffding's inequality, we have $\max_{i} |d_i - t_i| = O_{\mathbb{P}}((n
\rho_n \log n)^{1/2})$. This implies, with high probability, that $|d_i| = \Theta(n \rho_n)$ for
all $i$, from which Eq.~\eqref{eq:proof_thm3_bd1} follows.

The bounds in Eq.~\eqref{eq:2toinfty2} through
Eq.~\eqref{eq:proof_thm3_bd2}, together with multiple applications of
the triangle inequality, yield
\begin{gather}
  \label{eq:thm5_proof5} n^2 \rho_n\breve{\zeta}_{ik}\breve{\zeta}_{ik}^{\top}
  - n^2 \rho_n\tilde{\mathbf{W}}_n^{\top} \tilde{\zeta}_{ik}\tilde{\zeta}_{ik}^{\top} \tilde{\mathbf{W}}_n
  \overset{\mathrm{a.s}}{\longrightarrow} 0, \\
  \label{eq:thm5_proof6}
n \Bigl(\frac{\sqrt{d_k}}{\sqrt{d_i}}
  \breve{X}_i^{\top} \mathbf{I}_{a,b} \breve{X}_k - d_k
  (\breve{X}_i^{\top} \mathbf{I}_{a,b} \breve{X}_k)^2 \Bigr)- n \Bigl(\frac{\sqrt{t_k}}{\sqrt{t_i}}
  \tilde{Z}_i^{\top} \mathbf{I}_{a,b} \tilde{Z}_k - t_k
  (\tilde{Z}_i^{\top} \mathbf{I}_{a,b} \tilde{Z}_k)^2 \Bigr)
\overset{\mathrm{a.s}}{\longrightarrow} 0.
\end{gather}
We therefore have
\begin{equation*}
  \begin{split}
\breve{\boldsymbol{\Sigma}}(\breve{X}_i) -
\tilde{\mathbf{W}}_n^{\top}\tilde{\boldsymbol{\Sigma}}(\tilde{Z}_i)\tilde{\mathbf{W}}_n &=\breve{\boldsymbol{\Sigma}}(\breve{X}_i) - n^2\rho_n\tilde{\mathbf{W}}_n^{\top}\mathbf{I}_{a,b} \Bigl[\sum_{k=1}^n \tilde{\zeta}_{ik}
  \tilde{\zeta}_{ik}^{\top} \Bigl(\frac{\sqrt{t_k}}{\sqrt{t_i}}
  \tilde{Z}_i^{\top} \mathbf{I}_{a,b} \tilde{Z}_k - t_k
  (\tilde{Z}_i^{\top} \mathbf{I}_{a,b} \tilde{Z}_k)^2 \Bigr) \Bigr] \mathbf{I}_{a,b} \tilde{\mathbf{W}}_n
\\&= 
\breve{\boldsymbol{\Sigma}}(\breve{X}_i) -n^2\rho_n
  \mathbf{I}_{a,b} \tilde{\mathbf{W}}_n^{\top}
\Bigl[\sum_{k=1}^n \tilde{\zeta}_{ik}
  \tilde{\zeta}_{ik}^{\top} \Bigl(\frac{\sqrt{t_k}}{\sqrt{t_i}}
  \tilde{Z}_i^{\top} \mathbf{I}_{a,b} \tilde{Z}_k - t_k
  (\tilde{Z}_i^{\top} \mathbf{I}_{a,b} \tilde{Z}_k)^2 \Bigr) \Bigr] \tilde{\mathbf{W}}_n \mathbf{I}_{a,b} \\ &
  \overset{\mathrm{a.s.}}{\longrightarrow} 0.
\end{split}
\end{equation*}
where the second equality follows from the fact that $\tilde{\mathbf{W}}_n$ is
{\em block-orthogonal} and the convergence to $0$ follows from
Eq.~\eqref{eq:thm5_proof5} and Eq.~\eqref{eq:thm5_proof6}. This indicates
\begin{equation}
  \label{eq:thm5_proof5b}
  \breve{\boldsymbol{\Sigma}}(\breve{X}_i) +
\breve{\boldsymbol{\Sigma}}(\breve{X}_j)  -
\tilde{\mathbf{W}}_n^{\top}
\bigl(\tilde{\boldsymbol{\Sigma}}(\tilde{Z}_i) +\tilde{\boldsymbol{\Sigma}}(\tilde{Z}_j)\bigr) \tilde{\mathbf{W}}_n
\overset{\mathrm{a.s.}}{\longrightarrow} 0.
\end{equation}
We therefore have
\begin{equation}
  \label{eq:thm5_proof6b}
 \bigl( \breve{\boldsymbol{\Sigma}}(\breve{X}_i) +
\breve{\boldsymbol{\Sigma}}(\breve{X}_j) \bigr)^{-1} -
\tilde{\mathbf{W}}_n^{\top}
\bigl(\tilde{\boldsymbol{\Sigma}}(\tilde{Z}_i) +\tilde{\boldsymbol{\Sigma}}(\tilde{Z}_j)\bigr)^{-1} \tilde{\mathbf{W}}_n
\overset{\mathrm{a.s.}}{\longrightarrow} 0.
\end{equation}
We thus conclude, by Eq.~\eqref{eq:chisq_proof5} and Slutsky's theorem that
$$n^2 \rho_n\bigl(\breve{X}_i - \breve{X}_j \bigr)^{\top} \bigl(\breve{\boldsymbol{\Sigma}}(\breve{X}_i) +
\breve{\boldsymbol{\Sigma}}(\breve{X}_j)\bigr)^{-1} \bigl(\breve{X}_i -  
\breve{X}_j\bigr) \rightsquigarrow \chi^{2}_{d}
$$
as desired. The proof for the local alternative follows from
Eq.~\eqref{eq:thm5_proof1} using an almost identical argument to that given
in Theorem~\ref{thm1} and is thus omitted.

\subsection{Proof Sketch for Remark~\ref{rem:ncp_eq}}
\label{appb.8}
Let $\mu_{\mathrm{ASE}}^{(n)}$ and $\mu_{\mathrm{LSE}}^{(n)}$ be the left sides of Eq.~\eqref{eq:noncentral1} and Eq.~\eqref{eq:alt_cond_lse}.
First note that when $\rho_n \rightarrow 0$, $X_i^{\top} \mathbf{I}_{a,b} X_k (1 - \rho_n X_i^{\top} \mathbf{I}_{a,b} X_k)$ is the same asymptotically as $X_i^{\top} \mathbf{I}_{a,b} X_k$. Secondly, in balanced SBM, we have $t_i=t$ for all $i=1,...,n$. With these two facts, we have 
$$
\tilde{\mathbf{\Sigma}}(X_i) = \frac{n^2 \rho^2_n}{t} 
    \mathbf{I}_{a,b} \Bigl(\sum_{k=1}^n
    \zeta_{ik} \zeta_{ik}^{\top} X_i^{\top} \mathbf{I}_{a,b} X_k
 \Bigr) \mathbf{I}_{a,b}$$ 
 and $$\mu_{\mathrm{LSE}}^{(n)}=n^2 \rho_n^2\Bigl(\frac{X_i}{\sqrt{t}}-
\frac{X_j}{\sqrt{t}}\Bigr)^{\top}\bigl(\tilde{\boldsymbol{\Sigma}}(X_i)
+ \tilde{\boldsymbol{\Sigma}}(X_j)\bigr)^{-1}\Bigl(\frac{X_i}{\sqrt{t}}-
\frac{X_j}{\sqrt{t}}\Bigr)
$$ If we let $\tilde{\mathbf{\Sigma}}^{\prime}(X_i) = n \rho_n \; 
    \mathbf{I}_{a,b} \Bigl(\sum_{k=1}^n
    \zeta_{ik} \zeta_{ik}^{\top} X_i^{\top} \mathbf{I}_{a,b} X_k
 \Bigr) \mathbf{I}_{a,b}$, then
\begin{equation}
\label{eq:tlide_prime_sigma}
\mu_{\mathrm{LSE}}^{(n)}=n\rho_n(X_i-X_j)^{\top}\bigl(\tilde{\mathbf{\Sigma}}^{\prime}(X_i) + \tilde{\mathbf{\Sigma}}^{\prime}(X_j)\bigr)^{-1}
(X_i-X_j)
\end{equation}
Now $\tilde{\mathbf{\Sigma}}^{\prime}(X_i)$ can be simplified to
$$\tilde{\mathbf{\Sigma}}^{\prime}(X_i)=\mathbf{\Sigma}(X_i)-\frac{3n\rho_n}{4t}X_iX_i^{\top}$$
Apply the Sherman Morrison Woodbury formula to $\tilde{\bm{\Sigma}}^{\prime}(X_i)$, we get
\begin{equation}
\label{eq:tilde_Sigma_balanced}
\tilde{\bm{\Sigma}}^{\prime}(X_i)^{-1} = \bm{\Sigma}(X_i)^{-1} + \frac{3n\rho_n\bm{\Sigma}(X_i)^{-1}X_iX_i^{\top}\bm{\Sigma}(X_i)^{-1}}{4t(1-\frac{3n\rho_n}{4t}X_i^{\top}\bm{\Sigma}(X_i)^{-1}X_i)}.
\end{equation}
Now, under the local alternative, we have $\mathbf{\Sigma}(X_i)=\mathbf{\Sigma}(X_j)+R_{i1}$ and $X_iX_i^{\top}=X_jX_j^{\top}+R_{i2}$ where $R_{i1}$ and $R_{i2}$ are lower order terms. Then by ignoring these lower-order terms,
substituting Eq.~\eqref{eq:tilde_Sigma_balanced} 
into Eq.~\eqref{eq:tlide_prime_sigma} and simplifying, we obtain
$$
\mu_{\mathrm{LSE}}^{(n)} = \mu_{\mathrm{ASE}}^{(n)}+\frac{3n^2\rho_n^2}{8t(1-\frac{3n\rho_n}{4t}X_i^{\top}\bm{\Sigma}(X_i)^{-1}X_i)} \cdot \left[(X_i-X_j)^{\top}\bm{\Sigma}(X_i)^{-1}X_i\right]^2
$$
as desired.
\section{Testing in Directed Graphs}
\label{sec:directed}
\subsection{Theoretical Result}
\label{sec:directed1}
In this section, we present an overview for extending our test
statistics to directed graphs. 
The main idea is similar to the directed case except that we now
assume $\mathbf{P}=\rho_n\mathbf{X} \mathbf{Y}^{\top}$ where
$\mathbf{X}$ is the $n \times d$ matrix of latent positions for the
outgoing edges and $\mathbf{Y}$ is the $n \times d$ matrix of latent
positions for the incoming edges, i.e., $p_{ij} = \rho_n X_i^{\top} Y_j$.
Given $\mathbf{P}$, we generate $\mathbf{A} \sim \text{Bernoulli}(\mathbf{P})$.
We shall assume that both $\mathbf{X}$ and $\mathbf{Y}$ satisfy
Condition 1 and 2 in Section~\ref{sec:ase}, and that $n \rho_n =
\omega(\log n)$ as $n \rightarrow \infty$. 

Denote the singular value decomposition of $\mathbf{P}$ as
$\mathbf{P} = \mathbf{U} \mathbf{S} \mathbf{V}^{\top}$ where 
$\mathbf{S} \in \mathbb{R}^{d \times d}$ is the diagonal matrix of the
non-zero singular values arranged in decreasing order. 
Let $\mathbf{A}=\hat{\mathbf{U}} \hat{\mathbf{S}}
\hat{\mathbf{V}}^{\top}+\hat{\mathbf{U}}_{\perp}
\hat{\mathbf{S}}_{\perp} \hat{\mathbf{V}}_{\perp}^{\top}$ where
$\hat{\mathbf{S}} \in \mathbb{R}^{d \times d}$ is the diagonal matrix
with entries given by the top $d$ singular values of $\mathbf{A}$ in
magnitude arranged in decreasing order, and the columns of
$\hat{\mathbf{U}}$, $\hat{\mathbf{V}}$ are the corresponding left and
right singular vectors. Define $\hat{\mathbf{X}}=\hat{\mathbf{U}}
\hat{\mathbf{S}}^{\frac{1}{2}}$ and $\hat{\mathbf{Y}}=\hat{\mathbf{V}}
\hat{\mathbf{S}}^{\frac{1}{2}}$. Now consider the symmetric dilations
of $\mathbf{A}$ and $\mathbf{P}$:
$$\widetilde{\mathbf{A}}=\left[\begin{array}{cc}\mathbf{0} & \mathbf{A} \\ \mathbf{A}^{\top} & \mathbf{0}\end{array}\right], \quad \quad \widetilde{\mathbf{P}}=\left[\begin{array}{cc}\mathbf{0} & \mathbf{P} \\ \mathbf{P}^{\top} & \mathbf{0}\end{array}\right].$$
The eigen-decompositions of $\widetilde{\mathbf{A}}$ and
$\widetilde{\mathbf{P}}$ are given by
$$
\begin{aligned} \widetilde{\mathbf{P}}&= \frac{1}{2}\left(\begin{array}{cc}
\mathbf{U} & \mathbf{U} \\
\mathbf{V} & -\mathbf{V}
\end{array}\right)\left(\begin{array}{cc}
\mathbf{S} & \mathbf{0} \\
\mathbf{0} & -\mathbf{S}
\end{array}\right) \left(\begin{array}{cc}
\mathbf{U} & \mathbf{U} \\
\mathbf{V} & -\mathbf{V}
\end{array}\right)^{\top} \\
\widetilde{\mathbf{A}}&=\frac{1}{2}\left(\begin{array}{cc}
\hat{\mathbf{U}} & \hat{\mathbf{U}} \\
\hat{\mathbf{V}} & -\hat{\mathbf{V}}
\end{array}\right)\left(\begin{array}{cc}
\hat{\mathbf{S}} & \mathbf{0} \\
\mathbf{0} & -\hat{\mathbf{S}}
\end{array}\right) \left(\begin{array}{cc}
\hat{\mathbf{U}} & \hat{\mathbf{U}} \\
\hat{\mathbf{V}} & -\hat{\mathbf{V}}
\end{array}\right)^{\top} + \frac{1}{2}\left(\begin{array}{cc}
\hat{\mathbf{U}}_{\perp} & \hat{\mathbf{U}}_{\perp} \\
\hat{\mathbf{V}}_{\perp} & -\hat{\mathbf{V}}_{\perp}
\end{array}\right)\left(\begin{array}{cc}
\hat{\mathbf{S}}_{\perp} & \mathbf{0} \\
\mathbf{0} & -\hat{\mathbf{S}}_{\perp}
\end{array}\right) \left(\begin{array}{cc}
\hat{\mathbf{U}}_{\perp} & \hat{\mathbf{U}}_{\perp} \\
\hat{\mathbf{V}}_{\perp} & -\hat{\mathbf{V}}_{\perp}
\end{array}\right)^{\top}.
\end{aligned}
$$
Note that $\tilde{\mathbf{P}}$ is of rank $2d$, with $d$ positive
eigenvalues and $d$ negative eigenvalues. 
Then using the same ideas as that presented in Appendix C of
\citep{rubin2017statistical}, we can show that
\begin{gather}
\label{directed_1}
\hat{\mathbf{U}}
         \hat{\mathbf{S}}^{\frac{1}{2}}=\mathbf{U}
         \mathbf{S}^{\frac{1}{2}}
         \mathbf{W}_{*}+(\mathbf{A}-\mathbf{P}) \mathbf{V}
         \mathbf{S}^{-\frac{1}{2}} \mathbf{W}_{*}+\mathbf{R}_{1}, \\
         \label{directed_1b}
         \hat{\mathbf{V}} \hat{\mathbf{S}}^{\frac{1}{2}}=\mathbf{V} \mathbf{S}^{\frac{1}{2}} \mathbf{W}_{*}+(\mathbf{A}-\mathbf{P})^{\top} \mathbf{U} \mathbf{S}^{-\frac{1}{2}} \mathbf{W}_{*}+\mathbf{R}_{2},
\end{gather}
where $\mathbf{W}_{*}$ is an orthogonal matrix and $\mathbf{R}_{1}$,
$\mathbf{R}_{2}$ are lower order terms, i.e., with high probability we
have $\|\mathbf{R}_1\|_{2 \to \infty} = o(n^{-1/2})$ and $\|\mathbf{R}_2\|_{2 \to \infty} =
o(n^{-1/2})$. Let $\mathbf{Z} = \mathbf{U} \mathbf{S}^{1/2}$ and
$\tilde{\mathbf{Z}} = \mathbf{V} \mathbf{S}^{1/2}$. 
Then for any index $i$, we have
from Eq.~\eqref{directed_1} and Eq.~\eqref{directed_1b} that
\begin{gather}
\label{directed:convergence1}
\sqrt{n} \bm{\Sigma}(Z_i)^{-1/2} \bigl( \mathbf{W}_* \hat{X}_i -
Z_i \bigr) \rightsquigarrow \mathcal{N}\bigl(0,
\mathbf{I}), \quad 
\sqrt{n} \bm{\Sigma}(\tilde{Z}_i)^{-1/2} \bigl( \mathbf{W}_* \hat{Y}_i -
\tilde{Z}_i \bigr) \rightsquigarrow \mathcal{N}\bigl(0,
\mathbf{I}),
\end{gather}
where $\bm{\Sigma}(Z_i)$ and $\bm{\Sigma}(\tilde{Z}_i)$ are $d \times d$ matrices of the form
\begin{gather*}
\bm{\Sigma}(Z_i) = n \mathbf{S}^{-1} \left[\sum_{k=1}^{n} \tilde{Z}_k \tilde{Z}_k^{\top} Z_i^{\top}\tilde{Z}_k
  (1 - Z_i^{\top} \tilde{Z}_k)\right] \mathbf{S}^{-1}, \\
\bm{\Sigma}(\tilde{Z}_i) = n \mathbf{S}^{-1} \left[\sum_{k=1}^{n} Z_k Z_k^{\top} \tilde{Z}_i^{\top}Z_k
  (1 - \tilde{Z}_i^{\top} Z_k)\right] \mathbf{S}^{-1}.
\end{gather*}
The above limit results allow us to conduct three different
hypothesis test of equality, namely equality of outgoing latent
positions ($\mathbb{H}_0 \colon X_i = X_j$), equality of incoming
latent positions ($\mathbb{H}_0 \colon Y_i =
Y_j$) and equality of both outgoing and incoming latent
positions. More specifically, suppose we are given two vertices $i$
and $j$ in $\mathbf{A}$, and we wish
to test the null hypothesis $\mathbb{H}_0 \colon X_i = X_j$
against the alternative hypothesis $\mathbb{H}_A \colon X_i \not =
X_j$. Then, similar to Theorem~\ref{thm1}, we can consider the test statistic
\begin{equation}
\label{thm5.51}
T_{\mathrm{out}}(\hat{X}_i, \hat{X}_j)
=n(\hat{X}_i - \hat{X}_j)^{\top}\Bigl(\hat{\boldsymbol{\Sigma}}(\hat{X}_i)
+ \hat{\boldsymbol{\Sigma}}(\hat{X}_j)\Bigr)^{-1} (\hat{X}_i - \hat{X}_j)
\end{equation}
where $\hat{\boldsymbol{\Sigma}}(\hat{X}_i) =  n \bigl(\hat{\mathbf{Y}}^{\top}\hat{\mathbf{Y}}\bigr)^{-1}
\Bigl[\sum_{k=1}^n \hat{Y}_k \hat{Y}_k^{\top}
\hat{X}_i^{\top}  \hat{Y}_k (1 - \hat{X}_i^{\top}  \hat{Y}_k)\Bigr]
\bigl(\hat{\mathbf{\mathbf{Y}}}^{\top}\hat{\mathbf{Y}}\bigr)^{-1}$. Then
under $\mathbb{H}_0$ we have, 
for $n \rightarrow \infty$ with $n \rho_n = \omega(\log n)$, that
$T_{\mathrm{out}}(\hat{X}_i, \hat{X}_j) \rightsquigarrow \chi_d^2.$
Next let $\mu > 0$ be a finite constant and suppose that $X_i \not =
X_j$ satisfies the local alternative
\begin{equation}
 \label{eq:noncentral1_directed}
n\rho_n(Z_i-Z_j)^{\top}\bigl(\boldsymbol{\Sigma}(Z_i) + \boldsymbol{\Sigma}(Z_j)\bigr)^{-1}
(Z_i-Z_j) \rightarrow \mu.
\end{equation}
We then have $T_{\mathrm{out}}(\hat{X}_i, \hat{X}_j) \rightsquigarrow \chi_d^2\left(\mu\right)$
where $\chi_d^2\left(\mu\right)$ is the
noncentral chi-square with $d$ degrees of freedom and
noncentrality parameter $\mu$.
To test $\mathbb{H}_0 \colon Y_i = Y_j$, we simply swap the
roles of $\{\hat{X}_i\}$ and $\{\hat{Y}_i\}$ in the above derivations.
Finally, for $\mathbb{H}_0 \colon (X_i, Y_i) = (X_{j}, Y_j)$, we
use the test statistic
\begin{equation}
\label{thm5.52}
T_{\mathrm{both}}(\hat{M}_i, \hat{M}_j)
=n(\hat{M}_i - \hat{M}_j)^{\top}\Bigl(\hat{\boldsymbol{\Sigma}}(\hat{M}_i)
+ \hat{\boldsymbol{\Sigma}}(\hat{M}_j)\Bigr)^{-1} (\hat{M}_i - \hat{M}_j)
\end{equation}
where $\hat{M}_i = (\hat{X}_i, \hat{Y}_i) \in \mathbb{R}^{2d}$ and $\hat{\boldsymbol{\Sigma}}(\hat{M}_i) =  \left(\begin{array}{cc}
\hat{\boldsymbol{\Sigma}}(\hat{X}_i) & \mathbf{0} \\
\mathbf{0} & \hat{\boldsymbol{\Sigma}}(\hat{Y}_i)
\end{array}\right)$. Then under $\mathbb{H}_0$ we have
$T_{\mathrm{both}}(\hat{M}_i, \hat{M}_j) \rightsquigarrow \chi_{2d}^2$
as $n \rightarrow \infty$. 
The non-centrality parameter for $T_{\mathrm{both}}$ has a similar expression to that in Eq.~\eqref{eq:noncentral1_directed}.
\begin{remark}
\label{rem:directed}
We can also test equality up to scaling, e.g., either $\mathbb{H}_0
\colon X_i/\|X_i\| = X_j/\|X_j\|$ or $\mathbb{H}_0 \colon
(X_i/\|X_i\|,Y_i/\|Y_i\|) = (X_j/\|X_j\|,Y_j/\|Y_j\|)$. 
Test statistics for these hypothesis, and their asymptotic properties,
follow {\em mutatis mutandis} from the arguments used in deriving 
Theorem~\ref{thm3} and the above discussions. We omit the 
details; see Section~\ref{sec:pol} for an application of these
hypothesis tests to the political blogs dataset of \cite{adamic2005political}. 
\end{remark}
\subsection{Numerical Simulation}
\label{sec:directed_simu}
\begin{figure}[htbp]
\centering
\subfigure[$T_{\mathrm{out}}$]{
\includegraphics[width=0.35\textwidth]{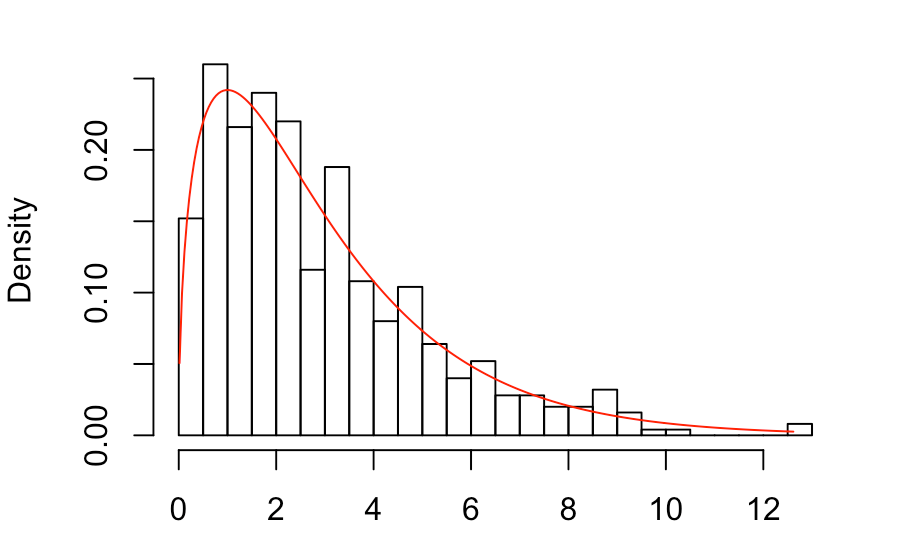}}
\subfigure[$T_{\mathrm{both}}$]{
\includegraphics[width=0.35\textwidth]{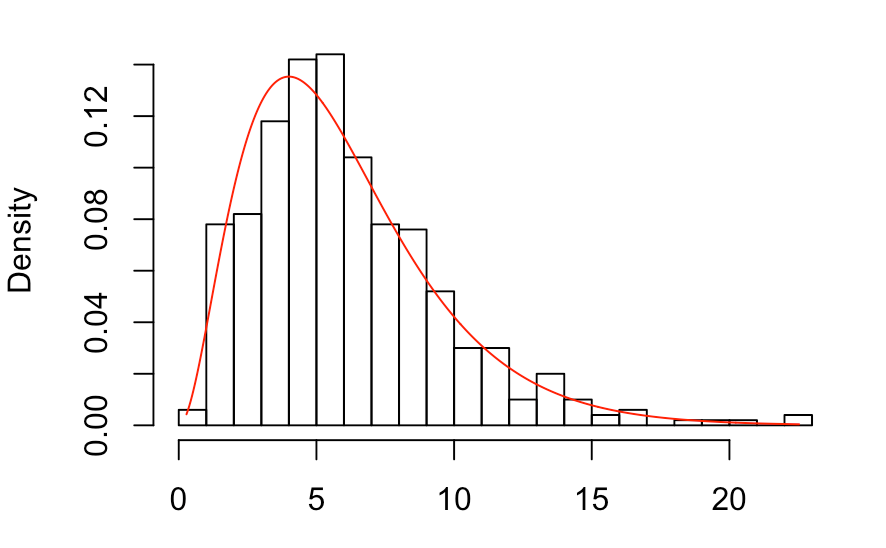}}
\caption{Empirical histograms for the test statistics
  $T_{\mathrm{out}}$ and $T_{\mathrm{both}}$ under the null hypothesis
  when $\rho=1.0$. The setting is that of directed mixed-membership graphs on
  $n = 4800$ vertices. The red curve on the left panel is the probability density function
  for the $\chi^2_3$ distribution and the one on the right panel is the probability density function
  for the $\chi^2_6$ distribution.}
\label{fdir} 
\end{figure}

In this section, we present simulation results about the empirical size and empirical power under the local alternative of our proposed test statistic. We consider a mixed membership SBM setting where the  block probabilities matrix $\mathbf{B}=\left(\begin{array}{ccc}
0.9 & 0.6 &0.5 \\
0.5 & 0.9 & 0.4\\
0.4 & 0.6 & 0.9
\end{array}\right)$. Note that $\mathbf{B}$ is not symmetric since the graph is directed here. Each node has two membership vectors $\bm{\pi}_1$ and $\bm{\pi}_2$, one for outgoing and one for incoming. They are chosen from the same seven possible membership vectors as those in Section~\ref{s5.1}. We generate graphs on $n=4800$ vertices. Among these $4800$ vertices, $300$ vertices are assigned to have membership vector $(1 - 2cn^{-1/2},
cn^{-1/2}, cn^{-1/2})$ with $c = 5$, and the remaining $4500$ vertices
are equally assigned to the remaining membership vectors with a random order. This assignment is done independently for both the outgoing and incoming membership vectors. To check the size of $T_{\mathrm{out}}$, we set the null hypothesis as $\bm{\pi}_{1i} = \bm{\pi}_{1j} =
(0.5,0.3,0.2)$ and we set $\bm{\pi}_{2i} = (1,0,0)$ and $\bm{\pi}_{2j} = (1 - 2cn^{-1/2}, cn^{-1/2}, cn^{-1/2})$ to check the power of $T_{\mathrm{in}}$ under local alternative hypothesis. For $T_{\mathrm{both}}$, we set the null hypothesis as $\bm{\pi}_{1i} = \bm{\pi}_{1j} =
(1,0,0)$ and $\bm{\pi}_{2i} = \bm{\pi}_{2j} =
(0.5,0.3,0.2)$ and the local alternative hypothesis as $\bm{\pi}_{1i} = \bm{\pi}_{2i} =
(1,0,0)$ and $\bm{\pi}_{1j} = (1 - 2cn^{-1/2}, cn^{-1/2}, cn^{-1/2})$, $\bm{\pi}_{2j} =
(1,0,0)$. The significant level $\alpha$ is to be $0.05$ and all the estimates of size and power are based on $500$ Monte Carlo replicates. Results under various choices of sparsity factors $\rho_n$ are reported in Table~\ref{Tdir}. Table~\ref{Tdir} also reports the large-sample, {\em theoretical} power. We see that the empirical estimates of the power are very close to the true theoretical values.
In addition, Figure~\ref{fdir} plots the empirical histograms
for $T_{\mathrm{out}}$ and $T_{\mathrm{both}}$ under the null
hypothesis for $\rho=1.0$.
We see that the distributions of
$T_{\mathrm{out}}$ and $T_{\mathrm{both}}$ are well-approximated by the
$\chi^2_3$ and $\chi^2_6$ distribution.
\begin{table*}
\caption{Empirical estimates for the size and power for the test
statistics $T_{\mathrm{out}}$, $T_{\mathrm{in}}$ and $T_{\mathrm{both}}$ with various
choices of sparsity parameter $\rho$. The
rows with labels $\mathrm{ncp}$ are the non-centrality parameters $\mu$ for the local alternative hypothesis.
}
\label{Tdir}
\begin{tabular}{@{}ccccccccc@{}}
\hline
$\rho$& 0.3& 0.4& 0.5& 0.6& 0.7&0.8&0.9&1.0 \\
\hline
Size ($T_{\mathrm{out}}$)  &0.066& 0.052& 0.050& 0.068&0.056&0.050& 0.060&0.048 \\
Power ($T_{\mathrm{in}}$)  & 0.158&0.192& 0.256& 0.350& 0.446&0.586&0.794&0.956 \\ 
ncp($T_{\mathrm{in}}$) & 1.15 & 1.72&2.46&3.44&4.82&6.92&10.51&18.15\\
Theoretical Power($T_{\mathrm{in}}$) & 0.127& 0.170& 0.230& 0.312 &0.426&0.584 &0.784 &0.961 \\ 
Size ($T_{\mathrm{both}}$)  &0.062& 0.074& 0.058& 0.054&0.068&0.056& 0.040&0.050 \\
Power ($T_{\mathrm{both}}$)  & 0.138&0.186& 0.226& 0.264& 0.356&0.474&0.646&0.844 \\ 
ncp($T_{\mathrm{both}}$) & 1.55 & 2.25&3.09&4.12&5.45&7.27&10.03&15.18\\
Theoretical Power($T_{\mathrm{both}}$) & 0.122& 0.161& 0.211& 0.278 &0.366&0.485 &0.646 &0.848 \\  
\hline
\end{tabular}
\end{table*}

\end{appendix}
\end{document}